\documentclass[tighten, twocolumn, twocolappendix]{aastex63}

\usepackage{newtxtext,newtxmath}
\usepackage{graphicx}	
\usepackage{amsmath}	
\usepackage{amssymb}	
\usepackage{multirow}
\usepackage{xcolor}

\defcitealias{Padovani2009AA}{P09}
\defcitealias{Webber1998ApJ}{W98}
\defcitealias{Moskalenko2002ApJ}{M02}
\defcitealias{McElroy2012}{RATE12}
\defcitealias{Millar1997}{RATE95}

\received{XXX}
\revised{YYY}
\accepted{ZZZ}
\submitjournal{ApJ}

\shorttitle{Observational signatures of CR interactions in molecular clouds}
\shortauthors{Owen et al.}
\graphicspath{{./}{figures/}}

\begin{document}

\title{Observational signatures of cosmic ray interactions in molecular clouds}

\correspondingauthor{Ellis R. Owen}
\email{erowen@gapp.nthu.edu.tw}

\author[0000-0003-1052-6439]{Ellis R. Owen}
\affiliation{Institute of Astronomy, Department of Physics, National Tsing Hua University, Hsinchu, Taiwan (ROC)}
\affiliation{Center for Informatics and Computation in Astronomy, National Tsing Hua University, Hsinchu, Taiwan (ROC)}

\author[0000-0003-4479-4415]{Alvina Y. L. On}
\affiliation{Institute of Astronomy, Department of Physics, National Tsing Hua University, Hsinchu, Taiwan (ROC)}
\affiliation{Center for Informatics and Computation in Astronomy, National Tsing Hua University, Hsinchu, Taiwan (ROC)}
\affiliation{Mullard Space Science Laboratory, University College London, Holmbury St. Mary, Dorking, Surrey, RH5 6NT, United Kingdom}

\author[0000-0001-5522-486X]{Shih-Ping Lai}
\affiliation{Institute of Astronomy, Department of Physics, National Tsing Hua University, Hsinchu, Taiwan (ROC)}

\author[0000-0002-7568-8765]{Kinwah Wu}
\affiliation{Mullard Space Science Laboratory, University College London, Holmbury St. Mary, Dorking, Surrey, RH5 6NT, United Kingdom}
\affiliation{Research center for Astronomy, Astrophysics and Astrophotonics, 
  Macquarie University, Sydney, NSW 2109, Australia}

\nocollaboration{4}



\begin{abstract}
We investigate ionization and heating of gas 
  in the dense, shielded clumps/cores of molecular clouds  
  bathed by an influx of energetic, charged cosmic rays (CRs). 
These molecular clouds  
  have complex structures, 
  with substantial variation in their physical properties 
  over a wide range of length scales. 
The propagation and distribution of the CRs is thus regulated 
  accordingly, 
  in particular, by the magnetic fields 
  threaded through the clouds 
  and into the dense regions within. 
We have found that a specific heating rate reaching  
  $10^{-26}$ erg\,cm$^{-3}$\,s$^{-1}$ 
  can be sustained in the dense clumps/cores for Galactic environments, 
  and this rate increases with CR energy density.  
The propagation of CRs and heating rates in some star-forming filaments identified 
  in IC 5146 
  are calculated,  
  with the CR diffusion coefficients in these structures determined 
  from magnetic field fluctuations 
  inferred from optical and near-infrared polarizations of starlight, 
  which is presumably a magnetic-field tracer. 
Our calculations indicate that 
  CR heating can vary by nearly three orders of magnitude 
  between different filaments within a cloud due to different levels of CR penetration.
  The CR ionization rate among these filaments is similar. 
The equilibrium temperature that could be maintained by CR heating alone
  is of order $1~{\rm K}$  
  in a Galactic environment, but this value would be higher 
  in strongly star-forming environments, 
  thus causing an increase in the Jeans mass of their molecular clouds. 
\end{abstract}


\section{Introduction}

The dense cores of molecular clouds (MCs) within our Galaxy
are expected to be shielded from much of the ionizing (particularly ultraviolet, UV) interstellar radiation by dust and molecular Hydrogen~\citep{Draine2011Book}.
However, observations
have revealed sustained ionization rates of up to $\zeta^{\rm H} = 10^{-17} - 10^{-15}~\text{s}^{-1}$ in dense cores~\citep{Caselli1998, Tak2000, Doty2002}, compared with $\zeta^{\rm H} = 10^{-16}~\text{s}^{-1}$ in diffuse interstellar clouds~\citep{Black1978ApJ, vanDishoeck1986ApJS, Federman1996ApJ}.
The cause of this ionization is widely attributed to cosmic rays -- CRs~\citep{Goldsmith1978, Goldsmith2001ApJ, Lequeux2005, Draine2011Book}, which would also act to regulate the temperature~\citep[e.g.][]{Spitzer1968ApJ} {and chemical evolution~\citep[e.g.][]{Desch2004ApJ, Dalgarno2006PNAS, Indriolo2015, Bisbas2017, Padovani2018, Albertsson2018, Gaches2019} of MCs.}
In this paper, we model the propagation of CRs in MC environments, accounting for their injection/absorption via hadronic interactions and ionizations in a self-consistent manner. We also demonstrate how this model can be applied to polarization observations of MCs and their cores, from which CR propagation parameters and ionization/heating patterns can be determined.

CRs are energetic, charged particles. 
They are able to penetrate into MCs, 
 causing ionization in dense MC cores 
 which are inaccessible to interstellar ionizing radiation.
CR protons in the MeV-GeV energy range 
  and CR electrons in the 10~keV-10~MeV range  
  are believed to contribute to the bulk of this ionization
   \citep[e.g.][]{Spitzer1968ApJ, Padovani2011, Yamamoto2017}.
Limits on CR ionization rates in interstellar clouds were first calculated by~\cite{Hayakawa1961PASJ} to be up to $10^{-15}~\text{s}^{-1}$. Later,
~\cite{Spitzer1968ApJ} 
  indicated a range from $\zeta^{\rm H} \approx 6.8\times 10^{-18}~\text{s}^{-1}$ 
  (when invoking a CR proton spectrum declining below 50 MeV), to values as high as $\zeta^{\rm H} \approx 1.2\times 10^{-15}~\text{s}^{-1}$ (when accounting for MeV protons injected by supernova, SN events). 
  The upper end of this range was disputed for some time,
   particularly by researchers inferring $\zeta^{\rm H}$ from the chemical balances of species influenced by CR ionization, however
later studies 
  showed CR ionization rates to be consistent with the lower end of this range~\citep[e.g.][]{Glassgold1974ApJ} with a consensus now having largely been reached, that the rate is around $10^{-16}~\text{s}^{-1}$ 
 for diffuse interstellar cloud environments (see, e.g.~\citealt{Hartquist1978MNRAS, Black1978ApJ, vanDishoeck1986ApJS, Federman1996ApJ, Geballe1999ApJ, Indriolo2007ApJ, Indriolo2012RSPTA, Indriolo2012, Geballe2007AAS}, and~\citealt{Padovani2009AA, Draine2011Book, Indriolo2013ASSP, Padovani2020SSRv} for overviews).
{CRs with energies above a GeV can also play a role~\citep{Bykov2020SSRv}}.
These are associated with
  star-forming activities 
  which yield massive stellar end products, e.g. 
  supernova remnants \citep[see][for discussion]{Blasi2011crpa}. Such environments can accelerate particles to relativistic energies 
  through, e.g. \citet{Fermi1949} acceleration 
  in diffusive shocks~\citep{Axford1977ICRC, Krymskii1977DoSSR, Blandford1978ApJ, Bell1978MNRASI, Bell1978MNRASII}. 
  At these high energies, pion-producing ($\pi^{0}, \pi^{\pm}$) pp interactions between CR protons and the dense MC gas can arise~\citep[cf.][]{Kafexhiu2014, Owen2018}. The decay of $\pi^{\pm}$ yields secondary MeV CR electrons, and these would be deposited locally (i.e. within the dense cloud).\footnote{We estimate that the contribution of these MeV CR electrons from pion decays
  to the ionization rate is negligible. See section~\ref{sec:secondary_elecs} for further discussion.} These higher-energy CRs can also engage with the ambient magnetic field, and could drive a gas heating rate via Alfv\'{e}n wave excitation~\citep{Wentzel1971, Wiener2013_b}.

The ionization level of a dense cloud governs the degree to which it is coupled to its ambient magnetic field. This, in turn, regulates its stability against fragmentation and/or gravitational collapse~\citep{Mestel1956, Price2008} and 
influences its
subsequent star-forming activities.
The propagation of charged CRs is governed by the local magnetic field, which can develop a very complicated structure~\citep{Padovani2013} with elevated field strengths arising as it co-evolves with its host cloud~\citep{Crutcher2012ARAA}.
In these strengthened complex magnetic fields, CRs may be focused by the field morphology to cause a convergence in their diffusive propagation. On the other hand, CRs may also be reflected/deflected as their pitch angles increase due to the stronger magnetic field. 
These antagonistic processes generally act concurrently~\citep{Cesarsky1978, Ko1992, Chandran2000, Desch2004ApJ, Padoan2005ApJ}, and
it has recently been argued that the mirroring/deflection effect always dominates over focusing 
    such that
    the CR flux reaching the densest magnetized core regions is slightly reduced overall by a factor of 2 or 3~\citep{Padovani2011, Padovani2013, Silsbee2018ApJ} compared to the exterior cloud environment (if other effects, e.g. energy losses and absorption interactions, are ignored; at MeV energies, for example, energy losses would dominate and could prevent CR propagation into the densest core regions entirely -- see~\citealt{Chernyshov2018NPPP}).

This study investigates the ionization and heating 
    of MCs in the presence of CR irradiation.
We consider a model for the propagation of CRs 
  in specified MC magnetic field and density structures, 
  also accounting for secondary electrons 
  arising through hadronic interactions of primary CR protons.
{We determine gas temperatures and molecular abundances 
 through calculating the chemical balances of species 
 including HCO$^{+}$, H$_3^{+}$, OH$^{+}$ and H$_2$O$^{+}$, and determine temperature profiles and chemical balances in model MCs 
 when irradiated by CR fluxes typical of the
 Galactic interstellar environment. 
We also predict the corresponding outcome if the irradiating CR intensity changes. 
We then apply our model to the interstellar cloud IC 5146, a complex known to host a broad variety of environments for which a plethora of polarization measurements are available~\citep{Wang2017ApJ, Wang2020ApJ}.}


The paper is organized as follows. 
  Section~\ref{sec:molecular_clouds} 
  outlines the characteristics of MC environments based on studies of the Milky Way and nearby galaxies. A model is introduced to describe the density and magnetic structures of MCs, identifying the dominant components engaging with CRs.  
Sections~\ref{sec:effects},~\ref{sec:cr_interactions} and~\ref{sec:cr_propagation} present the relevant CR physics 
 - CR propagation in MC environments and the interaction channels 
  for both hadronic and leptonic CRs. 
Section~\ref{sec:methods} details the analysis of magnetic field structures 
 in MC environments 
 and demonstrates how the propagation parameters of CRs can be determined from this.
Section~\ref{sec:results} presents our results, showing CR heating and ionization rates in a model MC and in the dense filamentary structures of IC 5146. 
We provide a brief summary and conclusions in section~\ref{sec:summary}.

\section{Cosmic rays in molecular clouds}
\label{sec:cr_physics}

\subsection{Molecular clouds}
\label{sec:molecular_clouds}

The interstellar medium is multi-phase, 
  with cold, dense neutral MCs inter-mingled
  with hot tenuous gases in pressure equilibrium. 
The clouds condense from the hot interstellar medium (ISM) gas as it cools and collapses under gravity, reaching typical densities of around $10^2~\text{cm}^{-3}$, temperatures of around $10~\text{K}$, and sizes of a few to a few tens of pc.
These are permeated by magnetic fields of a few $\mu$G, presumably having been swept in from the ISM by the collapsing material to create a `pinching' effect in the field orientation.
The resulting structure of the magnetic field vectors has been observed to resemble an `hourglass'~\citep{Girart2006, Rao2009, Tang2009}, where field lines are drawn closer together in regions 
of higher cloud density.
This produces a much stronger magnetic field within the MC compared to the surrounding ISM
~\citep[e.g.][]{Crutcher1999, Basu2000, Basu2009}. 
Further gravitational collapse within MCs to form clumps{/filaments} and cores is mediated by pressure support -- the source of this being either the magnetic fields~\citep{Mouschovias1991, Mouschovias1999} or turbulence~\citep{Padoan1999, MacLow2004, Zhang2017, Coude2019}, with recent studies  suggesting that both could have important roles~\citep{Kudoh2008, Vazquez2011, Seifried2015, Federrath2016, Planck2016_mag}. 
This leads to a hierarchical substructure,  
 wherein clouds host clumps of size 0.3 -- 3 pc {(or filaments of comparable width -- e.g. see~\citealt{Arzoumanian2011})} and density $10^3-10^4~\text{cm}^{-3}$. 
Moreover, cores on scales of 0.03 -- 0.2 pc 
  may develop within these {clumps/filaments}. 
Cores have higher densities of around $10^4-10^5~\text{cm}^{-3}$, 
  with some cases even reaching $10^6-10^7~\text{cm}^{-3}$ 
   \citep[see][for a review]{Bergin2007}.
Note that different distinctions between the components
are also proposed in literature and are equally valid~\citep[e.g.][]{Myers1995}. This is because an exact description of the continuous substructure of MCs cannot fully be captured by a simple hierarchy of just a few distinct elements~\citep{Rodriguez2005}.

Polarization observations 
  have indicated complex
  structures in the magnetic fields towards dense cores 
 (see e.g. ~\citealt{Hull2017}). 
This indicates that turbulence may be more important than magnetic fields in governing their internal dynamics.  
However, there are some massive cores 
  with ordered magnetic fields on 0.1 -- 0.01 pc scales which are mostly parallel or perpendicular to ordered fields on larger scales~\citep{Zhang2014}.  
These are generally also found to be either parallel or perpendicular to the observed orientation of their host cloud~\citep{Li2015Natur}, suggesting that they are of dynamical importance in regulating the initial cloud collapse/fragmentation processes~\citep[see also][]{Koch2014}.
\cite{Li2009} arrived 
at a similar conclusion when comparing magnetic fields on 1 pc and 100 pc scales, but \citet{Zhang2019} found 
that this correlation does not extend to sub-pc ($0.1-0.01$\,pc) scales where turbulence may have a more important role, possibly concentrated by gravitational collapse to create slightly super-Alfv\'{e}nic cores in otherwise sub-Alfv\'{e}nic host clouds~\citep[see][]{Ching2017, Tang2019}.
Being of higher density and smaller size, the cores themselves are observed to sustain typical magnetic field strengths of around $10-15~\mu$G
~\citep{Crutcher1999},\footnote{Note that observations show substantial spread in this value, which seems to be sensitive to the type of core observed~\citep{Crutcher2010}.} while the mean line-of-sight and total magnetic field strengths in the inter-core regions of molecular clouds have been observed to be $7.4~\mu$G and $14.8~\mu$G, respectively, via the Zeeman effect
\citep{Thompson2019ApJ}.

\begin{figure}
    \centering 
    \includegraphics[width=0.8\columnwidth]{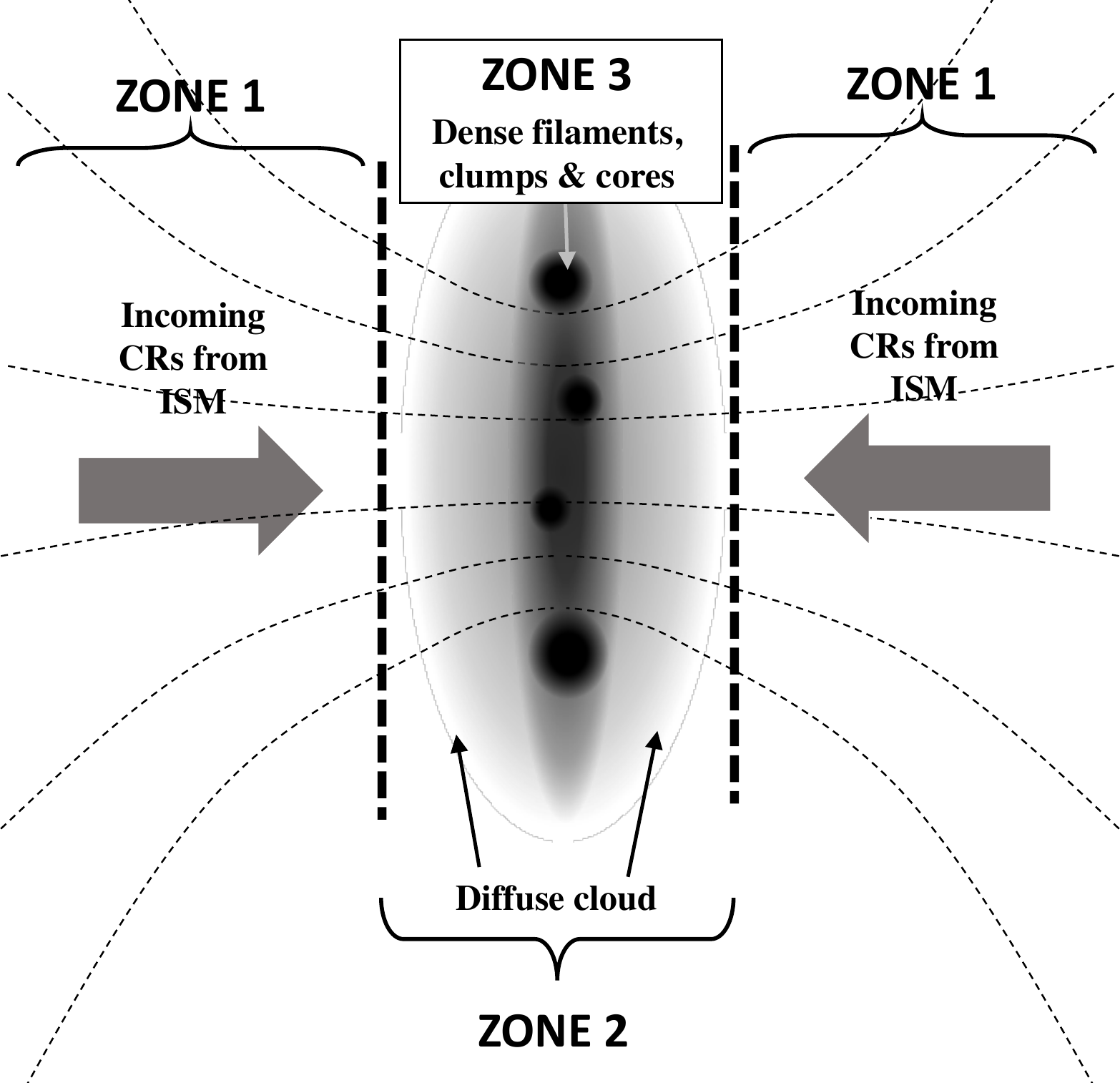}
    \caption{Schematic 
    of the different CR propagation regions within the density and magnetic field profiles of an idealized molecular cloud, characterising observed features (e.g. ~\citealt{Rao2009}). Zone 1 
    is the ISM region ($x_i\approx 1$).
    In Zone 2, the magnetic field pinching effect and increased density/neutral fraction of the cloud begins to affect the propagation of CRs, which becomes decreasingly diffusive as the higher neutral fraction damps the Alfv\'{e}n waves~\citep{Kulsrud1969, Zweibel1982}. 
     Zone 3 is the dense {filament/clump (containing dense cores)} of the MC where magnetic fields are the strongest and densities the highest. The gas is almost completely neutral ($x_i\approx 10^{-7}$ or less -- e.g. see~\citealt{Draine2011Book}) except for the ionization caused by CRs. Magnetic focusing/mirroring effects become important in Zone 3.}
    \label{fig:gmc_schematic}
\end{figure}

We consider a 3-zone model for
the structure of a MC and core in which we compute the interactions and propagation of energetic particles. 
The model, as illustrated in Fig.~\ref{fig:gmc_schematic}, 
has an external ISM region outside of the cloud, Zone 1,
where CRs propagate diffusively, and the medium is 
almost fully ionized $x_i\approx 1$.
From Zone 2 to 3, the magnetic field pinching effect resulting from the cloud's evolution would begin to influence the CR propagation, while the increased density/neutral fraction would lead it to become decreasingly diffusive (compared to the ISM outside the cloud).
As long as the magnetic field orientations and density structure are known at any given location, our approach can be adapted to account for any likely MC structure -- not just the `hourglass' shape indicated by the figure. In this simplified model, we consider only a single core (Zone 3) within the MC, but this can be extended to any number of cores as required.
Unless otherwise specified, $n_{\rm H}$ refers
to the number density of all gas within the cloud, whether atomic or molecular Hydrogen. 

\subsection{Effects of CRs}
\label{sec:effects}

\subsubsection{Ionization and chemical tracers}
\label{sec:cr_ionization}

CR ionization rates are studied observationally through the impact they have on the astrochemistry of MCs.
{Along a line of sight, the column density $N$ of certain species is considered to probe their relative abundances, assuming co-spatiality,\footnote{{This is usually necessary due to limited information about the line-of-sight structure.}} and $\zeta^{\rm H}$ (the total CR ionization rate)
can be inferred from these.}
Of particular interest are species which may be readily observed and are sensitive to CR ionization: CO$^{+}$, OH$^{+}$ and C$^{+}$ being three such examples which we consider here.
Analyzing chemical formation and destruction channels and 
their
rates allows abundances to be modeled in a steady-state chemical system (see Appendix~\ref{sec:appendixd}). 
Observationally, these would typically be probed using abundance ratios, integrated over a line of sight, between two chemically related species.
We consider the ratios $N({\rm CO}^+)/N({\rm H}_2)$, $N({\rm OH}^+)/N({\rm H}_2)$ and $N({\rm C}^+)/N({\rm C})$, being closely related species (via reactions P6, P4 of Table~\ref{tab:astrochemistry_rates} and carbon ionization, respectively), by invoking a simplified astro-chemical network (see Table~\ref{tab:astrochemistry_rates}) for key reaction chains involving 
C, O and molecular/atomic Hydrogen as initiated by CR ionizations, 
{in which a constant temperature is imposed. Our simplified network is strictly valid only in cold dense regions on scales of less than 1 pc, where the gas temperature is unlikely to exceed 10 K. However, for this proof-of-concept study, we adopt a binary ionization approximation where CRs either ionize or do not ionize a species in an interaction event. This allows us to neglect excited atomic/molecular states and associated modifications required for the rate coefficients of reactions, therefore relaxing the limit on temperature. We note that recent theoretical work~\cite[e.g.][]{Gaches2019} has shown that adopting such simplified chemical networks, where abundance ratios are determined from time-integrated chemical rate equations,
can lead to an under-prediction of ion column densities by as much as an order of magnitude. We therefore caution that the abundance ratio results presented in this work should be regarded as approximate, and could vary by up to an order of magnitude. }

While H$_3^+$ and HCO$^+$ are conventional probes of CR ionization rates in certain systems~\citep[e.g.][]{Tak2000}, the former is hampered by foreground contamination issues in infrared (IR) bands {and constrained by the need for a bright IR background source, while} the latter is strongly affected by photo-ionization. As such, we consider alternative species.
OH$^+$ is recommended by some studies as a more direct measure of CR ionization, relying on fewer interaction steps in its formation~\citep{Hollenbach2012}. The strongest OH$^+$ transition is at 972 GHz, with previous detections using \textit{Herschel}~\citep[e.g.][]{Neufeld2010}.
The ionization potential for the formation of OH$^+$ is around 13.0 eV, so any CR ionization signal could be easily contaminated by photo-ionization in diffuse, non-shielded cloud regions, thus is only appropriate as a tracer of CR ionization in dense, well-shielded clumps and cores. We derive the abundance ratio
\begin{equation}
    \frac{N({\rm OH}^+)}{N({\rm H}_2)} \approx \frac{\big[\;\!({4}/{f_{{\rm H}_2})} + 2\;\!\big] \zeta^{\rm H}}
    {n_{\rm H}(k_{\rm P4}\;\!f_{{\rm H}_2} + 2 x_i k_{\rm T1})}
    \label{eq:ohplus_ratio}
\end{equation}
from Appendix~\ref{sec:appendixd}, using equation~\ref{eq:ohplus_reduced} for the OH$^+$ balance,
assuming that 
the rate of process P2 and the ${\rm H}_3^+$ dissociative recombination rate are negligible compared to the ionization rate of H and H$_2$ (which would be appropriate to probe dense cores/clumps). Here, $f_{{\rm H}_2}$ is the molecular Hydrogen fraction (defined as the number of H nuclei in H$_2$ molecules as a fraction of the total number of H nuclei per unit volume), and the rate constants $k_{\rm T1}$ and $k_{\rm P4}$ relate to the processes T1 and P4 (see Appendix~\ref{sec:appendixd}). 

CO$^+$ is also a plausible tracer of CR ionization. It has accessible observable lines at 236.1 GHz (strongest), 235.8 GHz, 353.7 GHz and 354.0 GHz, which are all within the detection range of the {Atacama Large Millimeter/Submillimeter Array (ALMA)}. 
The
ionization potential in this case is greater than 13.6 eV, so CO$^+$ would not be susceptible to substantial ionization by interstellar radiation fields, making it a more robust probe of CR ionization.
Even though the strongest 236 GHz line is easily confused with emission from complex molecular species (in particular $^{13}$CH$_3$OH), several detections have now been reported~\citep{Latter1993, Fuente1997, Ceccarelli1998}. We derive the ratio
\begin{equation}
    \frac{N({\rm CO}^+)}{N({\rm H}_2)} = \frac{x_{\rm C}\zeta^{\rm H}}
    {n_{\rm H}\left(k_{\rm P5} 
    + \big( {k_{\rm P6} f_{{\rm H}_2}}/{2} \big)
    + x_i k_{\rm T5}\right)}
    \label{eq:coplus_ratio}
\end{equation}
using the chemical balance for CO$^+$ (eq.~\ref{eq:coplus_reduced}), where $x_{\rm C}$ is the Carbon abundance fraction.
We may further consider the ratio
\begin{equation}
    \frac{N({\rm C}^+)}{N({\rm C})} = \frac{2 \zeta^{\rm H}}{f_{{\rm H}_2}n_{\rm H} x_{\rm OH} k_{\rm I3}} \ ,
    \label{eq:cplus_ratio}
\end{equation}
from the chemical balance of C$^{+}$ from C ionization in equation~\ref{eq:cplus}. Here, the rate constants $k_{\rm T5}$, $k_{\rm P5}$, $k_{\rm P6}$ and $k_{\rm I3}$ are given in Table~\ref{tab:astrochemistry_rates}, and $x_{\rm OH}$ is the OH abundance fraction.
Although ${N({\rm C}^+)}/{N({\rm C})}$ would be contaminated by photo-ionization in photon-dominated regions and environments that are not well-shielded from ionizing interstellar radiation fields, it is 
a direct measure of CR ionization rate in denser clumps and 
cores,
which may be complementary to the other line ratios above.

We note that all of these abundance ratios require an estimate for the ambient volume density of the region being probed. Typically, estimates for density $n_{\rm H}$ can be found via rotational excitation analysis of observed ${\rm C}_2$ lines~\citep{Sonnentrucker2007}, analysis of ${\rm H}$ and the $J = 4$ level of ${\rm H}_2$~\citep{Jura1975}, or from thermal pressure analysis of \ion{C}{1}~\citep{Jenkins1983} -- see also~\citet{Indriolo2012}. Moreover, estimates for the fractional abundance of C and OH are required. For this proof-of-concept case, we adopt fiducial values of $f_{\rm H2} = 0.053$~\citep{Indriolo2015}\footnote{This is a mean value from a Milky Way sample -- see~\cite{Indriolo2015} for details.}, $x_{\rm C} = 1.6\times 10^{-4}$~\citep{Sofia2004} and $x_{\rm OH} = 1.0\times 10^{-9}$~\citep{Hollenbach2012}\footnote{Typically, variation of the $x_{\rm OH}$ fraction would be expected throughout a MC due to its formation by the recombination of ${\rm H}_3{\rm O}^+$ with electrons or destruction by reactions with C or ${\rm C}^+$. Studies indicate that this generally would lead to a build-up of OH in some parts of the cloud compared to the adopted 
fiducial limit~\citep{Hollenbach2012}. As such, the estimated  $N({\rm C}^+)/N({\rm C})$ ratio in equation~\ref{eq:cplus_ratio} should be regarded as an upper limit to constrain $\zeta^{\rm H}$.
Indeed, care must be taken if using this ratio alone as strong contamination by photo-ionization in non-shielded regions would be expected to boost the ${\rm C}^{+}$ abundance far above the levels driven by CR ionization processes.}. 
We discuss our model for the fractional ionization, $x_i$,  in section~\ref{sec:ion_frac_profile}.

\subsubsection{Heating}
\label{sec:cr_heating}

CR heating has been argued to be important in the ISM of galaxies~\citep{Field1969, Wiener2013_b, Walker2016, Owen2018, Owen2019AA}, in their circum-galactic environments~\citep[e.g.][]{Salem2016, Owen2019MNRAS, Owen2019AA},
and even in the intra-cluster medium between galaxies~\citep[e.g.][]{Loewenstein1991, Wiener2013, Ruszkowski2017}.
Its power is mediated by the thermalisation mechanism(s) at work,
as governed by the local conditions (e.g. density, ionization fraction, magnetic field). Here, we outline the dominant heating processes likely to arise from the presence of CRs within molecular cloud environments.

Traditionally, the excitation and subsequent non-linear Landau damping, ion-neutral collisional damping and/or turbulent damping of Alfv\'{e}n waves in magnetized environments, e.g. the intra-cluster medium~\citep{Loewenstein1991, Farmer2004, Fujita2011, Wiener2013, Fujita2013, Jacob2017, Ruszkowski2017}, has been regarded as the main process by which CR thermalise, and
it is likely that this would be dominant in a MC. 
CR heating by this channel
would arise at a rate of $n_0\langle \sigma v \rangle$ s$^{-1}$~\citep{Kulsrud1969, Zweibel1982} for $\langle \sigma v \rangle$ as the rate per particle of momentum exchange between ions and neutral particles, averaged over a thermal distribution. This indicates a damping timescale of around $3 \times 10^3$ yr, which would suggest that Alfv\'{e}n waves could not propagate far (less than 0.01 pc) into a MC \citep[cf.][]{Martin1997}, 
even if considering artificially-favourable conditions for wave propagation, i.e. low-density clouds ($n\approx 10^2~\text{cm}^{-3}$) appropriate for peripheral regions
yet a relatively strong $0.1~\text{mG}$ magnetic field more suitable for a dense core~\citep[e.g.][]{Crutcher2012ARAA}. 
At higher temperatures, this damping length would be even shorter, so thermalisation via this mechanism deep within a MC would depend on Alfv\'{e}n amplification by those CRs able to reach the heavily shielded cores. 
Amplification could be either driven by streaming instabilities,
or CR anisotropies within the cloud \citep{Kulsrud1969, Wentzel1969, Zweibel1982}, 
or any mechanical processes (e.g. motions due to gravitational collapse/condensation within the cloud) 
or turbulence that may act as a wave source
\citep{Carlberg1990, McKee1995, Gammie1996, Martin1997, Falceta2003}.

The timescale of magnetohydrodynamical (MHD) wave excitation via the streaming instability  
  is $\sim 10^6\;\! {\rm s}$~\citep{Ginzburg1964, Kulsrud1971}, 
  which is shorter than the corresponding damping rate. 
Alfv\'{e}n waves are therefore expected to persist, build up and thermalise within a MC, 
  with a power  
\begin{equation}
    \mathcal{Q}_{\rm A} = | {\boldsymbol v}_{\rm A, i}\cdot\nabla P_{\rm c}| 
\end{equation} 
\citep{Wentzel1971, Wiener2013_b}, 
where $\nabla P_{\rm c}$ is the (local) CR pressure gradient 
and ${\boldsymbol v}_{\rm A, i}$ is the generalized modified Alfv\'en velocity{,
which is defined by
\begin{equation}
    {v}_{\rm A, i} = {v_{\rm A}}{\left(1+ \varepsilon_i \right)^{-1/2}}
\label{eq:alf_gen}
\end{equation}
\citep[e.g.][]{Gomez2018ApJ}, where $\varepsilon_i = n_{\rm H}/n_{\rm ions}$, and  
$v_{\rm A}$~($=|{\boldsymbol v}_{\rm A}|= B/\sqrt{4\pi \rho}$, with 
$\rho = \sum_{\rm i}\;\!n_{\rm i} \;\! m_{\rm i}$ as the mass density, for number density $n_{\rm i}$ of particles with mass $m_{\rm i}$). 
We may write $n_{\rm H} = n_{\rm ions} (1-x_i)/x_i$ as the 
number density 
of neutral molecules (regardless of whether they are H or H$_2$), where $x_i$ is the ionization fraction, and $n_{\rm ions}$ is the number density of ions (assuming their abundance is equivalent to that of thermal electrons). From this, it follows that ${v}_{\rm A, i} \approx v_{\rm A}\;\!\sqrt{x_i}$.}

Given that CRs propagate preferentially along the magnetic field vectors,
  $|\nabla P_{\rm c}|$ in this direction can be estimated 
  from the number density and spectrum 
  derived from the relevant transport equation
(accounting for both proton and electron contributions), 
  with ${v}_{\rm A, i}$
  also calculated according to the local conditions. 
The heating power at a location $s$ is then 
\begin{equation}
\label{eq:MHD_heating}
    \mathcal{Q}_{\rm A}(s) = (\gamma_{\rm A} - 1)\;\! \left\{ v_{\rm A, i} \;\!\int 
     {\rm d}E \ \left| \;\! \nabla_{\parallel} 
     n(E) \;\! \right| \;\! E\;\!\right\}\biggr\vert_s \ .
\end{equation} 
Here we set the adiabatic index 
$\gamma_{\rm A} = 4/3$
for the CRs. 
The integral over $ n(E)\;\!E\;\!{\rm d}E$ gives the CR energy density, 
  and we retain $n$ in its general form to denote the total contribution from both CR protons ($n_{\rm p}$) and electrons ($n_{\rm e}$).

A CR population can also heat a medium by collisional ionization 
  and subsequent thermalisation. In~\citealt{Spitzer1968ApJ} (see also, e.g.~\citealt{Goldsmith2001ApJ}), this was found to arise at a rate of roughly $6.3\times 10^{-27}\;\!n_{\rm H}\;\!\text{erg}\;\!\text{cm}^{-3}\;\!\text{s}^{-1}$, where the heating power may be fully calculated using
\begin{equation}
    \mathcal{Q}_{\rm I}(s) = \sum_i \;\! f_{\rm i}\;\! n_{\rm i}(s)\;\! \int {\rm d}E \ \mathcal{E}_{\rm h}(E) \;\! \zeta^{\rm H'}(E)
\end{equation}
for $f_{\rm i}$ as the abundance fraction of species $i$ with number density $n_{\rm i}$ (we use this to account for the molecular vs. atomic Hydrogen fraction which is of importance in this calculation), and {where $\zeta^{\rm H'}(E)$ is the total differential (per energy interval) rate of all CR ionization channels.}
The energy a `knock-on' electron contributes towards heating its ambient gas is $\mathcal{E}_{\rm h}$, 
    which is a function of the energy of the CR initiating the ionization event.
Below a threshold of the excitation energy of an atom, being 3/4 of its binding energy $E_{\rm B}$ (13.6 eV for Hydrogen), the full CR knock-on electron energy is available for heating: 
collisional excitations cannot act to absorb its energy (to re-radiate as photons), while further secondary ionizations are also not possible. At higher energies between $3 E_{\rm B} /4 <E<E_{\rm B}$,
  a fraction of an electron energy $3 E_{\rm B} /4$ in any collision may now be lost to an excitation event, with only the remainder ($E_{\rm h} = E - 3 E_{\rm B} /4$) available for heating. 
  At energies between $E_{\rm B}<E<3 E_{\rm B}/2$, a single ionization or excitation process can proceed, with the probability of each determined by the relative weighting of the respective cross sections:
  \begin{equation}
      \mathcal{E}_{\rm h}(E) = \frac{\sigma^{\rm ex}\left[E-g\;\!E_{\rm B}\right] + \sigma^{\rm ion}\left[E-E_{\rm B}\right]}{\sigma^{\rm ex} + \sigma^{\rm ion}}
      \label{eq:e_h_ion_ex} \ , 
  \end{equation}
  where $g = 3/4$~\citep{Spitzer1968ApJ},  
  $\sigma^{\rm ex}$ 
  is the collisional excitation cross section, proportional to the energy-specific collision strength~\citep{Osterbrock_book}
  and $\sigma^{\rm ion}$ is the ionization cross section for the process in question (\citealt{Padovani2009AA}; hereafter~\citetalias{Padovani2009AA}).
  Given the energies of interest in this work, 
  collisions would overwhelmingly lead to an ionization, meaning equation~\ref{eq:e_h_ion_ex} reduces to $\mathcal{E}_{\rm h}(E) = E-E_{\rm B}$.
  In the range $3 E_{\rm B}/2 < E < 7 E_{\rm B}/4$,
  only one ionization may arise. However, if instead an excitation occurs then so must another and equation~\ref{eq:e_h_ion_ex} still applies, but with $g = 3/2$. When $E\geq 7 E_{\rm B}/4$, the situation is governed by the exact energy spectrum of the CRs, 
  and $\mathcal{E}_{\rm h}(E)$ is 
  determined numerically by~\cite{Dalgarno1972}, 
  although in  this work, we adopt the numerical approximation in~\cite{Draine2011Book} for computational efficiency.
  Overall, an ionization efficiency reduction would also arise to account for energy losses to other channels, which is calculated by weighting according to the respective timescales of each process.

Other mechanisms have also been proposed, but are unlikely to be as important as MHD wave damping -- in particular, see~\cite{Colafrancesco2008, Ruszkowski2017, Owen2018, Owen2019AA} where, on 0.1 kpc scales, thermalisation by Coulomb interactions in an ionized ISM is considered. In this process, secondary CRs are injected by pp interactions of CR primary protons (primary electrons would cool too quickly to propagate far from their source) to provide a channel by which CR protons can thermalise. 
This process can be the most effective in ionized media~\citep[see][]{Owen2019AA} but, in a predominantly neutral cloud, Coulomb thermalisation would only be able to 
 attain levels of around $10^{-31}~\text{erg}\;\text{cm}^{-3}\;\text{s}^{-1}$ for $n_{\rm H} = 10^4~\text{cm}^{-3}$ and $x_{\rm i} = 10^{-8}$ (scaled from the result in~\citealt{Owen2019AA}).

\subsubsection{Equilibrium temperature and gas cooling}
\label{sec:gas_cooling_cloud}

Cooling processes would also operate within MC environments, allowing an equilibrium temperature to be reached at which cooling and heating rates are comparable. Under typical dense molecular cloud conditions, gas cooling is dominated by CO and dust, with the latter only becoming important above densities of around $3\times 10^4~\text{cm}^{-3}$ -- below this, only CO cooling would operate effectively~\citep[e.g.][]{Galli2002AA, Goldsmith2001ApJ}.

To calculate the CO cooling rates, we adopt the analytic approximation presented in~\citet{Whitworth2018AA}, which was based on data in~\cite{Goldsmith1978}. To approximate dust cooling rates, we assume a uniform dust temperature of 10 K through the cloud as a reasonable upper bound for these dense regions, 
    otherwise we follow the treatment in~\cite{Goldsmith2001ApJ}. This allows equilibrium temperature profiles of MCs to be determined when subjected to different intensities of CR heating.

\subsection{Cooling and absorption of CRs}
\label{sec:cr_interactions}

\subsubsection{Electrons}
\label{sec:cr_leptonic}

In each electron interaction event, 
  only a small fraction of the particle energy is transferred.
Thus, the cooling due to electron interactions 
  is practically a continuous process.  
The following processes apply both to primary and secondary electrons/positrons (without losing generality, hereafter both $e^{+}$ and $e^{-}$ are referred to as ``electrons").
In a low-density plasma,
  the rate of cooling via electron-Coulomb interactions 
for ionization fraction $x_i$ is 
\begin{equation}
b_{\rm C} \approx m_{\rm e} c^2 \; n_{\rm H}\;\!x_i\;\! \; c \; \sigma_{\rm T} \ln \Lambda 
\end{equation}
\citep[see][]{Dermer2009book}, 
where $m_{\rm e}$ is the rest mass of an electron, $c$ is the speed of light and $\sigma_{\rm T}$ is the Thomson cross section. 
The Coulomb logarithm is taken as
    $\ln \Lambda\simeq 30$.
The rate of cooling due to electron 
  bremsstrahlung (free-free) is
\begin{equation}
b_{\rm ff}(\gamma_{\rm e}) \approx \alpha_{\rm f} \; c \; \sigma_{\rm T} \; n_{\rm H}\;\!x_i\;\! \; \gamma_{\rm e} m_{\rm e} c^2 \ 
\end{equation}
\citep[see][]{Dermer2009book}, 
where $\alpha_{\rm f}$ is the fine-structure constant
and 
$\gamma_{\rm e}$ is the Lorentz factor of the electrons.
Average ionization losses due to interactions with Hydrogen per CR electron are given by
\begin{equation}
b_{\rm ion}(\gamma_{\rm e}) = \left[n_{\rm H}/n_{\rm e}\right] \;\!(1-x_i)\;\! \zeta_{\rm e}^{\rm ion}(\gamma_{\rm e})  \;\!\gamma_{\rm e} m_{\rm e} {c}^2 
\end{equation}
\citep{Spitzer1968ApJ}. 
Here, the differential (per energy interval) direct ionization rate by electrons\footnote{This does not account for so-called `knock-on' ionizations caused by energetic electrons ejected by a primary ionization process -- these are only relevant for the efficiency of thermalisation/total ionization rate experienced by the cloud.} is defined as
\begin{equation}
    \zeta_{\rm e}^{\rm ion}(\gamma_{\rm e}) = n_{\rm e}(\gamma_{\rm e})\;\!c\;\!\sum_{\rm x}\sigma_{\rm e, x}^{\rm ion}(\gamma_{\rm e}) \ , 
    \label{eq:ion_rate_electrons}
\end{equation}
 where $\sigma_{\rm e, x}^{\rm ion}$ 
 is the cross section associated with
 each relevant ionization process ${\rm x}$ between electrons and the neutral medium.
 Strictly, the composition of molecular clouds is dominated by molecular Hydrogen, not atomic Hydrogen, and this changes the available channels through which ionization may proceed.
 Together with direct ionization ($e_{\rm CR}$ + H$_{2} \rightarrow e_{\rm CR}$ + H$_2^{+}$ + $e$), dissociative ionization ($e_{\rm CR}$ + H$_{2} \rightarrow e_{\rm CR}$ + H + H$^{+}$ + $e$) and double ionization ($e_{\rm CR}$ + H$_{2} \rightarrow e_{\rm CR}$ + 2H$^{+}$ + 2$e$) may also occur.
 We follow the approach by~\citetalias{Padovani2009AA} to model the ionization cross sections in this work.\footnote{\citetalias{Padovani2009AA} adopts the semi-empirical expression by \citet{Rudd1991} for the direct ionization process, the polynomial fit to data from~\cite{Straub1996} (see also~\citealt{Liu2004}) for the dissociative ionization process, and the fit to the data from ~\cite{Kossmann1990} for the double ionization process.} {Although this does not include the relativistic corrections introduced in later works (in particular, see~\citealt{Krause2015ICRC}), we find the impact of including these in our calculations is negligible.} 
 The cooling rate due to MHD wave excitation arises at a rate of 
 \begin{equation}
    b_{\rm MHD}(\gamma_{\rm e}) = (\gamma_{\rm A} - 1)\;\! v_{\rm A, i}  \ \left| \;\! \nabla_{\parallel} 
     n_{\rm e}(\gamma_{\rm e}) \;\! \right| \;\! \gamma_{\rm e} m_{\rm e} c^2 \ ,
     \label{eq:electron_mhd_cooling}
\end{equation} 
(see also equation~\ref{eq:MHD_heating}, where symbols retain the same definitions).
The rates for 
Compton and synchrotron cooling are
\begin{equation}
\label{eq:synch_ic_cooling}
b_{\rm rad}(\gamma_{\rm e}) = \frac{4}{3} \; \sigma_{\rm T} c \; {\gamma_{\rm e}}^2 \; \epsilon_{i}
\end{equation}
\citep[see e.g.][]{Blumenthal1970PRD}, 
  where $\epsilon_{i}$ is the energy density of the radiation field $\epsilon_{\rm ph}$ (Compton cooling)
  or magnetic field $\epsilon_{\rm B}$  
  ($=B^2/ 8\pi$, synchrotron cooling),
  respectively. 
In MC environments, $\epsilon_{\rm ph} \ll \epsilon_{\rm B}$, 
so Compton cooling is insignificant.\footnote{Other effects, e.g. triplet pair-production processes can arise at higher energies, but are not important in the weak radiation fields inside MCs~\citep{Schlickeiser2002_book}.}
The total cooling rate (at some position $s$) is the sum of 
 all contributing processes, 
where we note that ionization and MHD wave excitation losses are most important (in-line with the dominance of these processes in driving CR heating in the cloud -- see section~\ref{sec:cr_hadronic} for details).

\subsubsection{Protons}
\label{sec:cr_hadronic}

Protons predominantly lose energy in MC environments by MHD wave excitation. The cooling rate follows that for electrons, given in equation~\ref{eq:electron_mhd_cooling}, where the CR proton energy $\gamma_{\rm p} m_{\rm p} c^2$ and density gradient $\nabla_{\parallel} 
     n_{\rm p}(\gamma_{\rm p})$ along the magnetic field vector are used.
Additionally, protons interact with their environment either by ionization or, at kinetic energies above a threshold of $E_{\rm p}^{\rm th} = 0.28~{\rm GeV}$ 
   \citep{Kafexhiu2014}, by pion-producing hadronic interactions.
This hadronic threshold is the minimum energy required for the production of a pair of neutral pions, being the lowest energy particle forming in the resulting cascade, 
    where $E_{\rm p}^{\rm th} = 0.28 ~{\rm GeV} = 2m_{\pi^0} + {m_{\pi^0}}^2/2m_{\rm p}$, 
    for $m_{\pi^0}$ as the neutral pion rest mass and $m_{\rm p}$ as the proton rest mass.

The average proton cooling rate (per CR proton) due to collision-induced ionizations follows that for electrons, and is modeled as a cooling process arising at a rate given by
\begin{equation}
b_{\rm ion}(\gamma_{\rm p}) = \left[n_{\rm H}/n_{\rm p}\right] \;\!(1-x_i)\;\! \zeta_{\rm p}^{\rm ion}(\gamma_{\rm p}) \;\! \gamma_{\rm p} m_{\rm p} c^2  
\end{equation}
\citep{Spitzer1968ApJ}, 
 where 
  $n_{\rm p}$ 
  is the number density of protons and $\gamma_{\rm p}$ 
  is their Lorentz factor.  
The total differential 
ionization rate experienced by the cloud is defined as
\begin{equation}
    \zeta_{\rm p}^{\rm ion}(\gamma_{\rm p}) = n_{\rm p}(\gamma_{\rm p})\;\!c\;\!\sum_{\rm x}\sigma_{\rm p, x}^{\rm ion}(\gamma_{\rm p})
    \label{eq:ion_rate_protons}
\end{equation}
 for $\sigma_{\rm p, x}^{\rm ion}$ as the cross section associated with
 each relevant ionization process ${\rm x}$ between protons and neutral H$_2$. 
 The possible channels are
 direct ionization $p_{\rm CR}$ + H$_2 \rightarrow p_{\rm CR}$ + H$_2^{+}$ + $e$,
 electron capture ionization 
 $p_{\rm CR}$ + H$_2 \rightarrow$ H + H$_2^{+}$ (effectively a charge exchange process),
dissociative ionization
$p_{\rm CR}$ + H$_2 \rightarrow p_{\rm CR}$ + H + H$^{+}$ + $e$,
and 
double ionization
$p_{\rm CR}$ + H$_2 \rightarrow p_{\rm CR}$ + 2H$^{+}$ + $2e$, which we model by adopting the cross sections in \citetalias{Padovani2009AA}.\footnote{\citetalias{Padovani2009AA} 
used the empirical fit
to the cross section for direct ionization from \citet{Rudd1985}, and the fit by ~\citet{Rudd1983} for the electron capture (charge exchange) ionization cross section. The cross sections for dissociative ionization and double ionization are taken to be equivalent to those for the corresponding electron interactions.}

The pion-producing events are modeled as an absorption process
because a large fraction of the CR energy is transferred in a single interaction. These proceed via the major channels
\begin{align}
\label{eq:pp_interaction}%
\rm{p} + \rm{p} \rightarrow %
	\begin{cases}%
		&\rm{p}  \Delta^{+~\;} \rightarrow\begin{cases}%
				\rm{p} \rm{p} \pi^{0}  \xi_{0}(\pi^{0}) \xi_{\pm}(\pi^{+} \pi^{-}) \\[0.5ex]%
				\rm{p} \rm{p}  \pi^{+}  \pi^{-}  \xi_{0}(\pi^{0}) \xi_{\pm}(\pi^{+} \pi^{-}) \\[0.5ex]%
				\rm{p} \rm{n}  \pi^{+}  \xi_{0}(\pi^{0}) \xi_{\pm}(\pi^{+} \pi^{-})\\[0.5ex]%
			\end{cases} \\%
		&\rm{n} \Delta^{++} \rightarrow\begin{cases}%
				\rm{n} \rm{p} \pi^{+} \xi_{0}(\pi^{0}) \xi_{\pm}(\pi^{+} \pi^{-}) \\[0.5ex]%
				\rm{n} \rm{n} 2\pi^{+} \xi_{0}(\pi^{0}) \xi_{\pm}(\pi^{+} \pi^{-})\\[0.5ex]%
			\end{cases} \\%
	\end{cases} \ ,%
\end{align}%
    where $\Delta^{+}$ and $\Delta^{++}$ baryons are the resonances~\citep{Almeida1968PR, Skorodko2008EPJA}, and $\xi_{0}$ and $\xi_{\pm}$ are the multiplicities of the neutral and charged pions respectively, which are increasingly formed at higher energies. 
The hadronic products continue their interaction processes 
   until their energies fall below the interaction threshold $E_{\rm p}^{\rm th}$, 
    occurring
    within just a few interaction events \citep[see][]{Owen2018}.
    The neutral pions decay rapidly into two $\gamma$-rays, with a branching ratio of 98.8\%~\citep{Patrignani2016ChPh}, on timescales of $8.5 \times 10^{-17}\;\!{\rm s}$. 
    The charged pions undergo a weak interaction, either via
    $\pi^+	\rightarrow \mu^+ \nu_{\rm \mu} \rightarrow \rm{e}^+ \nu_{\rm e} \bar{\nu}_{\rm \mu} \nu_{\rm \mu}$
	or $\pi^- \rightarrow \mu^- \bar{\nu}_{\rm \mu} \rightarrow \rm{e}^- \bar{\nu}_{\rm e} \nu_{\rm \mu} \bar{\nu}_{\rm \mu}$, with a branching ratio of 99.9\%~\citep{Patrignani2016ChPh}  
 on a timescale of $2.6\times 10^{-8}\;\!{\rm s}$.  
    The rate at which protons are absorbed by the pp process
    is
    \begin{equation}
        S_{\rm p}(\gamma_{\rm p}, s) = n_{\rm H}(s)\;\! n_{\rm p}(\gamma_{\rm p}) \;\!c\;\!\sigma_{\rm p\pi}(\gamma_{\rm p}) \ , 
    \end{equation}
where ${\sigma}_{\rm p\pi}$ is the total inelastic pp interaction cross section,
being well parametrised by
\begin{equation}%
\label{eq:pp_cs}%
   {\sigma}_{\rm p\pi} = \left( 30.7 - 0.96\ln(\chi_E) + 0.18(\ln\chi_E)^{2} \right)\left( 1 - \chi_E^{-1.9}   \right)^{3}~\rm{mb} 
\end{equation}%
   \citep{Kafexhiu2014},
   where $\chi_E =  E/E^{\rm{th}}_{\rm p} =  (\gamma_{\rm p}-1)/(\gamma_{\rm p}^{\rm{th}}-1)$. Here, $E$ is the proton kinetic energy, $E^{\rm{th}}_{\rm p}$ is the threshold kinetic energy for the pp interaction and $\gamma_{\rm p}^{\rm th}$ is the Lorentz factor of a proton at this threshold energy.

\subsection{CR propagation}
\label{sec:cr_propagation}

CRs gyrate around magnetic field lines
  with a gyro-radius
\begin{equation}
    r_{\rm L} =  \frac{1.07\times 10^{-5}}{|Z|} \left(\frac{E}{100\;\text{MeV}}\right)\;\!\left(\frac{B}{\mu \text{G}}\right)^{-1} ~\text{pc} \ ,
    \label{eq:larmour_radius}
\end{equation}
   where $E$ is the particle energy, 
   $|Z|$ is the magnitude of the particle charge 
   and $B$ is the (uniform) magnetic field strength. 
 {The gyro-frequency is related to this by $\omega_{\rm L} = \beta_{\perp} c/r_{\rm L}$, where $\beta_{\perp} c$ is the particle velocity perpendicular to the magnetic field vector.} 
In general, CRs propagating through the tangled magnetic fields of interstellar space do not experience a uniform deflection -- instead, their propagation is better described as a series of random scatterings in the magnetic field domains. 
In a phenomenological perspective, this may be regarded as a diffusion process with a characteristic length scale set by $r_{\rm L}$.
In a MC, the magnetic field exhibits an ordered structure on the length scale of the cloud itself (cf. section~\ref{sec:molecular_clouds}).
Moreover, the size of the gyro-radius for particles in typical MC environments is many orders of magnitude smaller than this structure: the
gyro-radius of a 100 MeV CR in a 100 $\mu$G magnetic field
is around $10^{-7}$ pc, being even smaller at lower energies. This compares with a length scale of a few 10s of pc for a MC, a few
pc for clumps (with $n_{\rm H}\sim 10^3~\text{cm}^{-3}$) or a few tenths of a 
pc for the dense cores (with $n_{\rm H}\sim 10^4-10^5~\text{cm}^{-3}$) i.e. many orders of magnitude larger than $r_{\rm L}$ in all cases.
As such, propagating CRs would experience an effective locally-uniform magnetic field vector within a cloud, and would be strongly guided along them (following a helical path gyrating around the field vector).
This leads to strongly directed, anisotropic CR diffusion in the direction of the field vector, which facilitates magnetic focusing and CR entrapment in MCs. 

  \subsubsection{Magnetic mirroring and focusing}
\label{sec:mirroring_and_focusing}


Since CRs are constrained to gyrate and propagate along the magnetic field lines, their flux (per unit area) and any change in their number density (per unit volume) must be proportional to the density of the magnetic field lines per unit area.
  We quantify this using a magnetic concentration parameter, 
  $\chi (\equiv {B}/{B_0})$, 
  as in \citet{Desch2004ApJ}.  
This parameter is the ratio of 
  the magnetic field strength $B$ measured at some location 
  within the MC (including the core region) 
  compared to the mean ISM value, $B_0$. 
  This follows from the magnetic field strength being defined as the magnetic flux through a point, which is proportional to the concentration of magnetic field lines through that point.
As a CR propagates along a curved magnetic field vector into a MC, kinetic energy and magnetic moment must both be conserved. If it enters the cloud with a pitch angle
 (the angle between the incoming particle's velocity and the orientation of the magnetic field vector)
$\theta_{\rm in}$ and speed $v_{\rm in}$, the component of its velocity along the field vector would be 
$v_{||} = v_{\rm in} \tilde{\mu}_{\rm in}$, where $\tilde{\mu}_{\rm in}$ retains its earlier definition ($\cos\theta_{\rm in}$).
The CR gyration velocity around the magnetic field would then follow as
$v_{\rm \perp} = v_{\rm in}(1-\tilde{\mu}^2_{\rm in})^{1/2}$. When a CR has propagated into a cloud where magnetic field strength has attained 
$B = \chi B_0$, the cosine of its corresponding pitch angle must be
\begin{equation}
    \tilde{\mu}^2 = 1 - \chi + \chi \tilde{\mu}^2_{\rm in} \ ,
    \label{eq:pitch_relation}
\end{equation}
 in order to ensure that kinetic energy $\propto v_{\rm \perp}^2 + v_{\rm ||}^2$ and magnetic moment $\propto v_{\rm \perp}^2/B$ are conserved. 
 It therefore follows that $\tilde{\mu}^2 <0$ (i.e. the CR will be deflected and unable to propagate into the core) unless 
 $\tilde{\mu}^2_{\rm in} > 1-1/\chi$. This effect reduces the overall flux of CRs 
penetrating into the core of a MC, with CRs generally being reflected out of regions where 
$B \geq \chi B_0$ unless the pitch angle at which they enter the cloud is small (so-called magnetic mirroring).
Detailed treatments accounting the effect of mirroring and focusing in evolving magnetic fields are considered in the literature -- in particular, see \citealt{Kulsrud1969} and \citealt{Felice2001ApJ}.
In our approach we consider a simple, idealized system with non-evolving magnetic fields as a demonstrative study. 
More sophisticated prescriptions of CR propagation fall beyond the scope of this first model, and we leave these to future work. We instead adopt the treatment outlined in~\citealt{Desch2004ApJ} to characterise the approximate combined impact of magnetic mirroring and focusing by averaging over pitch angles of CRs within and outside the cloud, and apply this as an adjustment factor to the CR distributions we later calculate (see section~\ref{sec:the_transport_equation}). In this approach, 
the ratio of the internal and external fluxes
demonstrates the degree to which magnetic mirroring/focusing has modified the CR flux at a given point where the magnetic field strength is known.
Within the ISM, the angle-averaged CR flux (for $n$ as CR particle density) may be expressed as
\begin{align}
    \left\langle \frac{\partial n}{\partial t}\right\rangle \Bigg\vert_{\rm ISM} 
    & =  \int_{\Omega}\frac{{\rm d}\Omega}{4\pi}
    \frac{\partial n}{\partial t}  
     =  \frac{\partial n}{\partial t}  
     \int_{-\pi}^{\pi}\;\!\frac{{\rm d}\phi}{4\pi} 
     \int_0^{1}  \;\! {\rm d}\tilde{\mu} 
     = \frac{1}{2} \frac{\partial n}{\partial t} \ ,
    \label{eq:outside_cr_flux}
\end{align}
assuming that the external ISM CR flux is isotropic, and only the fraction of CRs directed within a solid angle of $\Omega_{\rm in}$ towards the cloud are able to enter it.
The same analysis through some boundary within the MC (where the propagation is now anisotropic) yields
\begin{align}
    \left\langle \frac{\partial n}{\partial t}\right\rangle \Bigg\vert_{\rm MC} 
    & = \int_{\Omega} {\rm d}\Omega \ 
    \xi(\tilde{\mu}_{\rm in}, \tilde{\mu})\;\!\frac{\partial n}{\partial t}   \nonumber \\
    & = \frac{\partial n}{\partial t} 
    \int_{-\pi}^{\pi}{\rm d}\phi  
    \int_0^{1}  {\rm d}\tilde{\mu} \   \xi(\tilde{\mu}_{\rm in}, \tilde{\mu})  \ ,
\label{eq:av_mc_flux}
\end{align}
where the term $\xi(\tilde{\mu}_{\rm in}, \tilde{\mu})$ accounts for the combined focusing and mirroring effects, with the effective CR flux being amplified by focusing through a factor of $B_0 \chi \tilde{\mu}$ 
and reduced by mirroring through a factor of $B_0 \chi \tilde{\mu}_{\rm in}$. 
By substituting equation~\ref{eq:pitch_relation}, $\xi$ may be expressed as
\begin{equation}
    \xi(\tilde{\mu}_{\rm in}, \tilde{\mu}) = \frac{\chi \tilde{\mu}}{\sqrt{\chi \tilde{\mu}^2 - \chi {+} \chi^2}}
\end{equation}
such that equation~\ref{eq:av_mc_flux} can be written as
\begin{align}
    \left\langle \frac{\partial n}{\partial t}\right\rangle \Bigg\vert_{\rm MC} 
    & =  2 \pi\;\! \frac{\partial n}{\partial t} 
    \int_0^{1}  \frac{\chi \tilde{\mu} \ {\rm d}\tilde{\mu} }{\sqrt{\chi \tilde{\mu}^2 - \chi {+} \chi^2}}  \nonumber \\
    & =  2 \pi \left[\;\! \chi - \sqrt{\chi^2-\chi}\;\!\right] 
    \frac{\partial n}{\partial t} \ .
    \label{eq:av_mc_flux_fin}
\end{align}
The magnetic scaling factor applicable to the CR flux and/or number density is then the ratio of the angle averaged result inside the cloud (equation~\ref{eq:av_mc_flux_fin}) compared to that outside the cloud (equation~\ref{eq:outside_cr_flux}), defined as
\begin{align}
\eta(\chi) & = 4\pi \left[\;\! \chi - \sqrt{\chi^2-\chi}\;\!\right] \ , 
\label{eq:correction_factor}
\end{align}
which is adopted in our subsequent calculations.

\subsubsection{The transport equation}
\label{sec:the_transport_equation}

If temporarily ignoring magnetic field mirroring/focusing effects, 
the propagation of CRs can be described using the transport equation,
\begin{align}
      \frac{\partial n}{\partial t}  & -
      \nabla \cdot \left[D(E, {\boldsymbol{s}})\nabla n \right] 
      + \nabla \cdot \left[{\boldsymbol{v}} n\right]  
         \nonumber \\
     &  \hspace*{2cm}+ \frac{\partial}{\partial E} \left[ \;\! b(E, {\boldsymbol{s}}) n \;\! \right]   =  Q(E, {\boldsymbol{s}}) - S(E, {\boldsymbol{s}}) \ ,  
\label{eq:transport_equation}     
\end{align}
\citep[e.g.][]{Schlickeiser2002_book}, where $n = n(E, {\boldsymbol{s}})$ is the differential number density of CRs (number of CR particles per unit volume per energy interval between $E$ and $E+{\rm d}E$) 
  at a location 
  ${\boldsymbol{s}}$. 
  The diffusive term
  $\nabla \cdot \left[D(E, {\boldsymbol{s}})\nabla n \right]$ is governed by the coefficient
  $D(E, {\boldsymbol{s}})$
  which depends on the gyro-scattering radius (or frequency)
  of the CRs of energy $E$ in their local magnetic field (cf. equation~\ref{eq:larmour_radius}), 
  as well as
  turbulence and 
  magnetohydrodynamical (MHD) perturbations along the local magnetic field vectors. 
  We deal with this empirically in section~\ref{sec:methods}.
  The second propagation term $\nabla \cdot \left[{\boldsymbol{v}} n\right]$ is usually an advection term which describes the propagation of CRs in the bulk flow of a magnetized medium (e.g. inflow/outflow).
  However, in this work, it 
 describes the propagation of CRs through a magnetized ISM (Zone 1 in Fig.~\ref{fig:gmc_schematic}) which, due to the CR streaming instability~\citep{Wentzel1974ARAA, Kulsrud2005book}, typically
  corresponds to the Alfv\'{e}n 
  speed~\citep[e.g.][]{Commercon2019AA}, 
  $v_{\rm A}$.\footnote{Strictly, this is only true for particles of tens of MeV, with high energy particles of energies above 10 GeV undergoing free-streaming without experiencing any substantial scattering at all~\citep{Chernyshov2018NPPP}. However, we argue that at GeV energies and below, where the bulk of the energy density lies in our adopted CR spectrum (and where the effects of the CRs are correspondingly strongest) this treatment is sufficiently informative for our first model. We leave a more detailed model accounting for the transition from streaming to advection to future work.} 
{As the cloud becomes more neutral, the generalized modified Alfv\'{e}n speed is adopted, which accounts for ionization fraction, $x_i$ (see equation~\ref{eq:alf_gen}).}
  
The mechanical and radiative cooling term $b(E, {\boldsymbol{s}})$ can also be considered as an advection of the CR ensemble in energy space due to cooling processes arising along their propagations.\footnote{Cooling effects are most severe for CR electrons. The CR protons, 
being of larger mass, 
have a substantially 
smaller 
Thomson cross section compared to the electrons. As such, proton cooling in this work -- apart from ionization losses -- is neglected.}
The injection of CRs by the source term is given by $Q(E,{\boldsymbol s})$, 
  while CR absorption/attenuation is encoded in the sink term $S(E,{\boldsymbol s})$. 
  As with the cooling term, the exact form of these source/sink terms depends on the CR species in question.
 The transport equation takes a different form for both CR protons (denoted as $n_{\rm p}$) and electrons (denoted $n_{\rm e}$ -- although, where necessary, we differentiate between primary and secondary electrons as $n_{\rm e, 1}$ and $n_{\rm e, 2}$, respectively, 
with the total electron number density  $n_{\rm e} = n_{\rm e, 1} + n_{\rm e, 2}$).\footnote{It is argued that both primary and secondary CRs contribute to the CR electron component of Galactic CRs and also in ISM environments of nearby galaxies. Of these, as much as 60-80\% could be secondary CRs~\citep[e.g.][]{Torres2004, Thompson2007, Lacki2010, Lacki2013}. In this work, we regard primary CRs as those which enter the MC through the boundary, so both contributions are taken into account in our definition of the primary flux. In this paper, secondary CRs are considered as those which are produced \textit{within} the MC.}

  Following \citet{Owen2019AA}, we consider 
  the transport equation for protons and electrons separately, where the absorption 
  of the protons is dominated by pp losses~\citep[see][]{Owen2018}. 
  This process injects some of the secondary CRs 
  into the source term of the transport equation for
  electrons (others being provided by ionizations).
  We adopt a Cartesian geometry, with a coordinate $s$ for the distance into a MC from a boundary set as the edge of the region of influence of the cloud
  (similar to,
  e.g.~\citealt{Morlino2015} and~\citealt{Phan2018MNRAS}).
  To account for the deflective effects of magnetic mirroring and focusing on the CR distribution (cf. section~\ref{sec:mirroring_and_focusing}), we multiply the solution of equation~\ref{eq:transport_equation} by the adjustment factor $\eta(\chi)$ (equation~\ref{eq:correction_factor}).

\subsubsection{Primary protons}

For protons in typical cloud environments,  
  the time scale of radiative loss (of order Gyr) is generally longer than 
  that of advection ($\sim$10s of kyr) and diffusion ($\sim$ a few kyr).  
The cooling is therefore due to their ionizations of the cloud medium only. 
We consider no additional sources of CR protons within or in the vicinity of the MC environment,   
  and ignore CR acceleration in the system.
The absorption is dominated by the hadronic (pp) interaction, 
 implying that the transport equation is simply
\begin{align}
      \frac{\partial n_{\rm p}}{\partial t}   \  - & \frac{\partial}{\partial s} \left\{D(E_{\rm p}, s)\;\! \frac{\partial n_{\rm p}}{\partial s} \right\} + v_{\rm A, i}\frac{\partial n_{\rm p}}{\partial s} + \frac{\partial}{\partial E_{\rm p}} \left[ \;\! b_{\rm p}(E_{\rm p}, s) \;\! n_{\rm p} \;\! \right] \nonumber \\
      & \hspace{1.5cm}  = Q_{\rm p}(E_{\rm p}, s) 
       - n_{\rm H}(s)\;\!{\sigma}_{\rm p\pi}(E_{\rm p})\;\! n_{\rm p}\;\!c \ ,  
\label{eq:transport_equation_protons}     
\end{align}
where ${\sigma}_{\rm p\pi}$ is the hadronic (pp) interaction cross section responsible for pion-production (see section~\ref{sec:cr_hadronic}), 
   and the cooling term is given by ionization loss (see equation~\ref{eq:ion_rate_protons}). We apply the mirroring/focusing term $\eta(\chi)$ as an adjustment to the resulting solution, so a mirroring term does not appear in the transport equation. 
In the steady-state,  
equation~\ref{eq:transport_equation_protons} becomes 
\begin{align}
       -\frac{\partial}{\partial s} \left\{D \frac{\partial n_{\rm p}}{\partial s} \right\} + v_{\rm A, i}\frac{\partial n_{\rm p}}{\partial s} 
        = -\frac{\partial}{\partial E_{\rm p}} \left[ \;\! b_{\rm p} \;\! \;\!n_{\rm p} \;\! \right] 
      - n_{\rm H}\;\!{\sigma}_{\rm p\pi}\;\!n_{\rm p}\;\!c \ ,   
\label{eq:reduced_transport_equation_protons}     
\end{align}  
  implying a balance between cooling, absorption,  
  diffusion and advection of the energy carried by the CR particles. 
The source term is treated as a boundary condition at $s=0$.  
This may be solved numerically subject to appropriate boundary conditions, as outlined in Appendix~\ref{sec:appendixa}, after which the magnetic mirroring/focusing factor $\eta(\chi)$ is applied to yield the final particle distribution. 
Note that, in solving equation~\ref{eq:reduced_transport_equation_protons}, the same boundary condition applies symmetrically 
  at $s = s_{\rm c}$, where $s_{\rm c}$ is the size of the region of influence of the cloud. Strictly, this boundary should be taken at $\pm \infty$ as the presence of the cloud affects the cosmic ray intensity in the vicinity of the cloud~\citep{Cesarsky1978, Morfill1982MNRAS}. However, since such a boundary condition could not yield a self-consistent determination of the CR flux flowing through some boundary, \citet{Morfill1982ApJ} argues that practically each cloud may be considered to have its own sphere of influence from which the boundary condition may be taken. In this case, we take $s_{\rm c}=5~\text{pc}$ to reflect the approximate separation of clouds and/or filaments found in the types of astrophysical setting in which we would expect our model to apply (see, e.g.~\citealt{Arzoumanian2011, Wang2020ApJ} from which distances between filaments in the IC 5146 molecular cloud complex can be estimated to be just a few pc apart). 
  A further condition is required 
on $D\;\!\partial n_{\rm p} /\partial s$, which is taken to be the CR flux through the boundary as estimated by $j(E)$ (see section~\ref{sec:cr_spectrum}). We provide the resulting CR spectrum at different locations in the cloud in Appendix~\ref{sec:spectral_evolution}.

\subsubsection{Primary electrons}
\label{sec:prim_elecs_info}

We refer to primary electrons as those which enter the cloud through some model boundary -- these could be injected into the ISM as secondaries beyond the limits of the model, but we do not require these to be distinguished as such.
The form of the electron transport equation in this case is similar to that for the protons, with the exception that there is now no absorption term (electrons would cool much more rapidly than protons and are not subject to catastrophic processes like the pp interaction). In the steady state, 
this gives
\begin{align}
      -\frac{\partial}{\partial s} \left\{D \;\! \frac{\partial n_{\rm e, 1}}{\partial s}  \right\} + v_{\rm A, i}\frac{\partial n_{\rm e, 1}}{\partial s}   = \  & - \frac{\partial}{\partial E_{\rm e}} \left[ \;\! b_{\rm e} \;\! n_{\rm e, 1} \;\! \right] \ , 
\label{eq:transport_equation_electron_primary}     
\end{align}
where $D$ and $v_{\rm A}$ are the same as for protons (since they are relativistic -- see, e.g.~\citealt{Kulsrud2005book}), and 
the electron cooling term is the sum of all relevant contributions, 
given in section~\ref{sec:cr_leptonic}.
Equation~\ref{eq:transport_equation_electron_primary} may be solved numerically 
(see Appendix~\ref{sec:appendixa}) 
subject to boundary conditions at $s=0$
and $s = s_{\rm c}$. Again, the resulting spectrum at different locations in the cloud is shown in Appendix~\ref{sec:spectral_evolution}, which includes the adjustment to account for  
the magnetic mirroring/focusing.


\subsubsection{Secondary electrons}
\label{subsec:secondary_elecs_theory}

We consider secondary electrons as those which are injected within the MC environment, at a rate encoded by the source term $Q_{\rm e}(E_{\rm e}, s)$. In this case, the transport equation reduces to
\begin{align}
      -\frac{\partial}{\partial s} \left\{D \frac{\partial  n_{\rm e, 2}}{\partial s}  \right\} +& v_{\rm A, i}\frac{\partial n_{\rm e, 2}}{\partial s}   
      = -\frac{\partial}{\partial E_{\rm e}} \left[ \;\! b_{\rm e} \;\! n_{\rm e, 2} \;\! \right] + Q_{\rm e} \ .
\label{eq:transport_equation_electron_secondary}     
\end{align}
The cooling terms retain their definitions from equation~\ref{eq:transport_equation_electron_primary}, while the injection term is mediated by the solution to the proton transport equation. 
We relate the injection of CR electron secondaries $Q_{\rm e}^{\rm had}$ to the local number density and interaction rates of the CR protons, and we refer the reader to
\citet{Owen2019AA}
for 
discussions on multiplicities, by-products and energy transfer efficiencies from primary to secondary species. We note that we convert the pp injection term into differential units of electron energy\footnote{This assumes that multiple secondaries produced in a given interaction would  
have a roughly equal share of energy. This follows from the presence of a strong peak in 
the differential production cross section in the pp interaction~\citep{Murphy1987ApJS, Berrington2003ApJ}, indicating that many of the electrons are produced at similar energies. This is particularly the case at the dominating energy range of our calculation, which typically leads to the injection of electrons at a few tens to hundreds of MeV, given that the secondary electrons characteristically inherit a few percent of the energy of the primary proton~\citep{Owen2018}.}, 
instead of in terms of the energy of the 
initiating 
proton flux.
The inclusive 
pion formation cross sections are adopted from \cite{Blattnig2000}. 
Moreover, to simplify the computation (since we do not require detailed particle spectra), we assume the pion and muon decay processes yield secondaries of equal energies, instead of calculating their energy distribution in full with the secondary electron and primary proton Lorentz factors related by $\gamma_{\rm e} = \gamma_{\rm p} m_{\rm p} \bar{\kappa}_{\rm \pi}/4 m_{\rm e}$~\citep{Sikora1987ApJ}, 
    where the average pion-production inelasticity is $\bar{\kappa}_{\rm \pi} = 4 m_{\rm \pi}/m_{\rm p}$;
    for $m_{\rm e}$, $m_{\rm \pi}$ and $m_{\rm p}$ as the electron, (charged) pion and proton rest masses, respectively.
Electrons may also be injected by so-called `knock-on' production, where the ionization of the ambient MC gas leads to the emission of an electron of sufficiently high energy to cause further ionizations.
We represent this with the knock-on injection term
\begin{equation}
    Q_{\rm e}^{\rm K}(E_{\rm e}) = 1.75\ n_{\rm H}(s) \int_{E_{\rm 1}} {\rm d}E_{\rm 1} \ \sigma^{\rm ion}_{\rm H}(E_{\rm e}; E_{\rm 1}) \;\! c \;\! n_{\rm 1}(E_{\rm 1})
\end{equation}
\citep{Brunstein1965,  Brown1977ApJ}, where $n_{\rm 1}$ is the differential number density (i.e. per energy interval) of CRs of energy $E_{\rm 1}$ which initiate the first ionization -- either protons or electrons (including those provided by the pp process for completeness, although from comparison of cross sections, this would be substantially less important in a MC environment except at very high energies when CR fluxes would be relatively small) -- and $\sigma^{\rm ion}_{\rm H}(E_{\rm e}; E_{\rm 1})$ is the cross section for the production of knock-on electrons of energy $E_{\rm e}$ due to an initial CR energy $E_{\rm 1}$~\citep[see][although we use the energy-integrated form here]{Abraham1966}. This gives the contribution of knock-on secondary electrons per unit volume per energy interval between $E_{\rm e}$ and $E_{\rm e} + {\rm d}E_{\rm e}$, with the total secondary CR electron injection term $Q_{\rm e} = Q_{\rm e}^{\rm had} + Q_{\rm e}^{\rm K}$. Equation~\ref{eq:transport_equation_electron_secondary} can then be solved (see Appendix~\ref{sec:appendixa}) subject to the boundary conditions that both $n_{\rm e}$ and $D\;\!\partial n_{\rm e}/\partial s = 0$ at $s = 0$ and $s = s_{\rm c}$ for all energies (as no secondary electrons would be expected to be flowing through or be present at the boundaries), with the adjustment factor $\eta(\chi)$ to account for magnetic mirroring/focusing being applied to the solution of equation~\ref{eq:transport_equation_electron_primary} to account for this effect in the final particle distribution.

\subsection{Cosmic ray spectrum}
\label{sec:cr_spectrum}

{The irradiating incident CR spectrum is used as a boundary condition for equation~\ref{eq:reduced_transport_equation_protons}, and is split into two components based on their spectral shape (e.g. see~\citetalias{Padovani2009AA}).}
We note that the high-energy component is largely responsible for hadronic interactions, while that below a GeV is more important in directly driving ionization processes.
The differential spectrum of CRs above a GeV observed in the Milky Way follows a distinctive power-law
\begin{equation}
\label{eq:ref_cr_eqn}
   n_i(E) = \frac{{\rm d}n_i(E)}{{\rm d}E} =  \tilde{n}_{\rm HE, i} \left(\frac{E}{E_0}\right)^{-\Gamma_{\rm HE, i}} \ ,
\end{equation}
for species $i$ being electrons or protons, and $n_{\rm i}$ being their volume density. The spectral index in the 1 GeV to 1 PeV regime (largely attributed to internal Galactic CR sources, and where the vast majority of the CR energy density lies) may be characterized by $\Gamma_{\rm HE,p} = 2.7$,\footnote{Note that the index would be less steep in regions of `fresh' CR acceleration, e.g. in the Galactic ridge or a starburst galaxy~\citep[e.g.][]{Aharonian2006, Gaggero2017, HESS2018b, HESS2018a}.} being appropriate for primary protons~\citep[e.g.][]{Kotera2011ARAA}, or the steeper index of $\Gamma_{\rm HE,e} = 3.3$ for primary electrons~\citep[e.g.][]{Hillas2006astroph}. The spectral index of secondary electrons is determined by the solution of the injection/transport equation rather than being adopted as an intrinsic boundary condition.
We specify
$\tilde{n}_{\rm HE, i}$ as the normalization of the high-energy component
\begin{equation}
\label{eq:cr_norm}
     \tilde{n}_{\rm HE, i}
       = \frac{\epsilon_{\rm CR, i}\;\! f_{\rm U}\;\!(2-\Gamma_{\rm HE, i})E_0^{-\Gamma_{\rm HE, i}}}{E_{\rm max}^{2-\Gamma_{\rm HE, i}} - E_0^{2-\Gamma_{\rm HE, i}}}  
          \ ,
\end{equation}
with $E_{\rm max} = 1~\text{PeV}$, and 
   the reference energy taken as $E_0 = 1~\text{GeV}$.
    $\epsilon_{\rm CR, i}$ is the CR energy density across all components, which takes a value of around 1.8 eV cm$^{-3}$ for the Galactic ISM, of which (roughly) around 0.4 eV cm$^{-3}$ may be attributed to the component below 1 GeV and the rest to higher energy CRs~\citep[e.g.][]{Webber1998ApJ, Ferriere2001RvMP}. 
    The parameter
    $f_{\rm U} = 1.4/1.8$ 
    is the fraction of CR energy density attributed to the high-energy component of the spectrum, with the remainder at lower energies being described by
\begin{equation}
\label{eq:ref_cr_eqn_low}
    n_i(E) = \tilde{n}_{\rm LE, i}\;\!\left(\frac{E}{E_0}\right)^{-\Gamma_{\rm LE, i}} \ .
\end{equation}
The normalisation $\tilde{n}_{\rm LE, i}$ is specified by the high-energy spectrum at 1 GeV (i.e at $E_0$) 
to ensure continuity across the spectral break. 
The spectral continuity is a more physically meaningful condition than 
maintaining an exact energy density ratio between the two CR components, which is subjected to substantial uncertainties 
as a result of
e.g. modulation effects 
in the solar neighbourhood,
and spatial inhomogeneities~\citep[e.g.][]{Webber1998ApJ, Cummings2016ApJ}.
The ratio $n_{\rm e}/n_{\rm p}$ at 1 GeV is set to be 1\%, in line with observations~\citep[e.g., see][]{Hillas2006astroph}. We set two values for each of $\Gamma_{\rm LE, e}$ and $\Gamma_{\rm LE, p}$ to account for  the broad range of 
spectral indices considered in literature.
The `minimum' spectra use $\Gamma_{\rm LE, p} = -0.95$ (\citealt{Webber1998ApJ}, 
hereafter \citetalias{Webber1998ApJ})
and $\Gamma_{\rm LE, e} = -0.08$ (the `conventional' model C of~\citealt{Strong2000ApJ} -- hereafter C00, following the notation of~\citetalias{Padovani2009AA}).
The `maximum' spectra use $\Gamma_{\rm LE, p} = 1$ (\citealt{Moskalenko2002ApJ}, hereafter \citetalias{Moskalenko2002ApJ}) 
and $\Gamma_{\rm LE, e} = 1$ (model SE in ~\citealt{Strong2000ApJ}, hereafter E00 by the~\citetalias{Padovani2009AA} convention).
We also adopt a CR flux model $j(E)$ based on the above treatment, normalized to that of~\citetalias{Padovani2009AA} for Milky Way conditions to provide an appropriate flux boundary condition when later solving the transport equations~\ref{eq:reduced_transport_equation_protons} and~\ref{eq:transport_equation_electron_primary}.

\section{Magnetic field structure and cosmic ray propagation}
\label{sec:methods}

\subsection{Empirical characterisation of CR propagation}
\label{sec:3_1}


We adopt the same approach as~\citealt{SA1993_1, SA1993_2} (see also~\citealt{Schlickeiser2002_book, Kulsrud2005book}) 
  to characterise CR propagation in the MCs. In this approach,  
     the background cloud-scale magnetic field structure 
    is taken to vary on length scales 
    which are substantially larger 
    than both the magnetic-field fluctuations
   and gyrating (scattering) radii of CRs.
Using a quasi-linear approximation~\citep{Jokipii1966ApJ, Schlickeiser2002_book}, the Fokker-Planck (FP) equation~\citep{Kirk1988} can be greatly simplified, 
    assuming that (i) the turbulence driving the field fluctuations is purely magnetic\footnote{Essentially, this means neglecting density perturbations as well as the electric field component of associated
    Alfv\'{e}n waves.}
and of low-frequency, 
(ii) the turbulence components on different scales are uncorrelated and non-interacting i.e. $v_{\rm A} \ll c$, 
and
(iii) the CR pitch angle is small.
When the flow of CRs 
and the orientation of the fluctuations are largely parallel to the orientation of the background cloud-scale magnetic field vector and independent of that in the perpendicular direction, the only non-vanishing FP coefficient is then
\begin{equation}
    P_{\mu \mu} \approx \frac{\mathcal{J}(\lambda_1)}{\;\!v_{\rm A}\;\! \lambda_1}\left(\frac{\omega_{\rm L}\;\!B_0}{B}\right)^2\;\!\mathcal{I}_{\perp} \ .  
    \label{eq:pmumu_original}
\end{equation}
Here $\lambda_1 =\lambda_{\rm td}\;\!(|\omega_{\rm L}|/\omega_{\rm p,0})$ 
  is the CR resonant scattering length scale 
  parallel to the background magnetic field line.  
The turbulent decay length scale is
  $\lambda_{\rm td} \approx v_{\rm A}\;\!\tau_{\rm td}$, 
  where we adopt 
  a turbulent decay timescale $\tau_{\rm td} =2$\,Myr in our calculations
  \citep[see][]{Gao2015ApJ, Larson2015ApJ}.
The CR gyro-frequency is
    $\omega_{\rm L}$ 
with a sign convention set by the charge (in units of proton charge). {In the relativistic limit, this is given by}
${\omega}_{\rm L} = (10^{1}/\gamma_{\rm p})\;\!(B/1\,{\rm m}\text{G})~\text{s}^{-1}$ for protons, and 
${\omega}_{\rm L} = (1.8\times 10^4/\gamma_{\rm e})\;\!(\;\!B/1\,{\rm m}\text{G})~\text{s}^{-1}$ for electrons
~\citep[see e.g.][]{Kulsrud2005book}.
The normalisation
 $\omega_{\rm p,0}$ is taken to be
 the gyro-frequency of a CR proton at a reference energy of 100 MeV.
We adopt a magnetic field strength normalisation
$B_0 = 1\,{\rm m}\text{G}$ 
to be comparable to the magnetic field strength outside the densest parts of clumps/cores -- see e.g.~\citealt{Crutcher2010, Li2015Natur}.
Hence what remains to be evaluated are the two variables 
  $\mathcal{J}(\lambda_1)$ and  $\mathcal{I}_{\perp}$ 
  in the above equation.


The dimensionless variable $\mathcal{J}(\lambda_1)$ 
  characterises the magnetic field fluctuations 
  along the direction of the background large-scale magnetic field vector 
  and is 
  defined as 
\begin{equation}
    \mathcal{J}(\lambda_1) \equiv
    \int_0^{\lambda_1} {\rm d}\lambda \ \frac{ \lambda}{\lambda_1} 
     \hat{P}_\parallel(k_{\rm c}\lambda) 
    + \int_{\lambda_1}^{\infty} {\rm d}\lambda\ \frac{\lambda_1}{\lambda}\;\! 
    \hat{P}_\parallel(k_{\rm c}\lambda) \ ,  
\end{equation} 
   where $\hat{P}_\parallel$ is the power spectrum of the fluctuations 
  along the large-scale magnetic field vector.
The wavenumber normalisation is defined as
  $k_{\rm c} = \omega_{\rm p,0}/v_{\rm A}$.
In terms of the dimensionless variable $\kappa_{\parallel} = \lambda k_{\rm c}$,
\begin{align}
    \mathcal{J}(\lambda_1) =  k_{\rm c}^{-1}\;\!\int_0^{\lambda_1 k_{\rm c}} {\rm d}\kappa_{\parallel} \;\! & \frac{\kappa_{\parallel}}{k_{\rm c} \lambda_1}\;\! \hat{P}_\parallel(\kappa_{\parallel}) \nonumber \\
    & {\hspace*{-0.5cm}}+ k_{\rm c}^{-1}\;\!\int_{\lambda_1 k_{\rm c}}^{\infty} {\rm d}\kappa_{\parallel}\;\! \frac{k_{\rm c} \lambda_1}{\kappa_{\parallel}}\;\! \hat{P}_\parallel(\kappa_{\parallel}) \ . 
    \label{eq:j_k1_resonant} 
\end{align}  
The variable $\mathcal{I}_{\perp}$ specifies the contribution 
 from the orthogonal components of the magnetic field fluctuations. 
Note that $\mathcal{I}_{\perp}$ is not dimensionless. 
Without losing generality, we denote the two orthogonal components 
  of the perpendicular wave vector 
  by $y$ and $z$, 
  corresponding to the wave vectors ${\boldsymbol k}_y$ and ${\boldsymbol k}_z$, 
  respectively. 
We have 
${\boldsymbol k}_\perp   =  
  {\boldsymbol k}_y + {\boldsymbol k}_z$, 
  with ${k_\perp}\!^2  = {k_y}\!^2 + {k_z}\!^2$. 
In terms of these wave vectors and the normalisation $k_c$, 
\begin{align}
\mathcal{I}_{\perp} &\equiv 
  \int_{-\infty}^{\infty} \frac{ {\rm d}{k}_{y}}{k_{\rm c}} 
  \;\!\int_{-\infty}^{\infty} \frac{ {\rm d}{k}_{z}}{k_{\rm c}} 
  \; \hat{P}_\perp(k_y,k_{z};k_{\rm c}) \nonumber \\ 
  &=  \int_{\perp} 
  \frac{{\rm d}^2{\boldsymbol k}_{\perp}}{k_{\rm c}\!^2} \; 
   \hat{P}_\perp({\boldsymbol k}_\perp; k_{\rm c}) \ ,  
\label{eq:pspec_int}
\end{align}   
where $\hat{P}_\perp$ is the power spectrum 
 of the perpendicular component of the magnetic field fluctuations, 
 which effectively takes the form of a scalar delta function, i.e.   
$\hat{P}_\perp({\boldsymbol k}_\perp; k_{\rm c}) 
  = \delta(k_\perp\!^2 - k_{\rm c}\!^2)$ 
  (cf. the `slab' approximation, \citealt{Hasselmann1968}).    
For isotropic fluctuations in the $y$-$z$ plane,  
  we have 
\begin{align}
\mathcal{I}_{\perp} & = 
  2\pi \int_{0}^\infty  
  \frac{{\rm d} {k}_{\perp} k_\perp} {k_{\rm c}\!^2} \; 
  \delta(k_\perp\!^2 - k_{\rm c}\!^2) \nonumber \\ 
 & =  \pi \int_{0}^\infty  
  \frac{{\rm d} {k}_{\perp} k_\perp} {k_{\rm c}\!^3} \; 
 \Big[\;\! \delta(k_\perp + k_{\rm c}) 
  + \delta(k_\perp - k_{\rm c}) \;\! \Big] \ . 
\label{eq:pspec_int_a}
\end{align}     
As $k_{\rm c} \neq 0$, $k_\perp + k_{\rm c}>0$, and hence,  
\begin{align} 
\mathcal{I}_{\perp} & =   \pi \int_{0}^\infty  
  \frac{{\rm d} {k}_{\perp} k_\perp} {k_{\rm c}\!^3} \; 
   \delta(k_\perp - k_{\rm c}) = \frac{\pi}{k_{\rm c}\!^2} \ . 
\label{eq:pspec_int_more}
\end{align} 
Finally we may relate the spatial diffusion coefficient $D$ to $P_{\mu\mu}$, 
  with 
\begin{equation}
    D \approx \frac{c^2}{8}\left(1-\frac{1}{\gamma^2}\right)\int_{-1}^{1} {\rm d}\mu \;\! \frac{(1-\mu^2)^2}{P_{\mu \mu}} \ .
    \label{eq:diff_coeff_full}
\end{equation}
Since the pitch angles are small, 
  $P_{\mu \mu}$ 
  is not strongly dependent {on} $\mu$ \citep[see][]{SA1993_2}, 
  and we can take it outside the integral. 
Thus, we obtain 
 $D \approx {2\;\!c^2}/{15\;\!P_{\mu \mu}}$\footnote{Note that this formulation would be invalid if the MC is not in a steady-state, i.e. if there are large-scale flows of the cloud medium, resulting from e.g. ongoing gravitational collapse, inflows or outflows.},  
 for $\gamma\gg 1$. 
{Although this approximation is adopted in our calculations, 
  our results would not differ much 
  if fully accounting for 
  the presence of non-relativistic particles.}

%


\subsection{Diffusion estimation}
\label{sec:3_2}

\subsubsection{Angular dispersion function}

Dust polarization can be used to probe the magnetic field structure in dense MC environments over a range of scales. 
The field structure is often characterized by
the angular dispersion function, defined as 
\begin{equation}
    \mathcal{S}_d (\ell) =
    \frac{1}{N_{\rm pair}} 
    \sum_{i = 1}^{N_{\rm pair}}
    [\varphi_{i}(s+\ell)-\varphi_{i}(s)]^d \ ,
    \label{eq:adf_def}
\end{equation}
\citep[e.g][]{Redaelli2019} which is sometimes also referred to as the `structure function'. 
Here $d$ is the order number (in this case, we adopt $d=2$ to constrain the power spectrum),
    while the normalisation  $N_{\rm pair} = N_{\rm P}(N_{\rm P}-1)/2$ is the number of unique pairs in a data-set of $N_{\rm P}$ individual points. 

While similar to the correlation function, the structure function can be computed to higher accuracy with less data~\citep{SchulzDubois1981ApPhy}, making it more appropriate for smaller data-sets with large uncertainties, such as those being used in this study. 
The angular dispersion function has often been used to study astrophysical magnetic fields, using e.g. rotation measure, RM 
~\citep[e.g.][]{Minter1996ApJ, Lazarian2016ApJ, Xu2016ApJ}, as well as polarization angle (PA) measurements, including in the analysis of MC environments~\citep[e.g.][]{Hildebrand2009ApJ, Houde2009ApJ, Planck2016_mag, Wang2019ApJ, Redaelli2019}.

In this work, differences between pairs of measured dust polarization angles $\varphi$ over separations $\ell$ are computed 
to quantify the similarity in the orientation of magnetic field fluctuations on different scales, with the intention of encoding the deviation of local perturbations from the background mean field vector.\footnote{The strength of the local field is estimated separately -- see section~\ref{sec:observing_b_fields}.}
    In practice, to calculate $\mathcal{S}_n(\ell)$ from a set of $N_{\rm P}$ polarization angles, every unique pair in that set must be identified and binned according to their 
    angular
    separation distance $\ell$. This would yield $N_{\rm pair}$ unique pairs, 
    indexed sequentially as
    \begin{equation}
      N_{\rm pair} = j(N_{\rm P}-1) - \frac{j^2-j}{2} - (N_{\rm P}-i) \ ,
    \end{equation}
    for $i$ and $j$ as the indices of contributing $N_{\rm P}$ data points to that pair in the original data-set (where $i\neq j$).
    Equation~\ref{eq:adf_def} would then be applied to all points within that bin to give an estimate for the angular dispersion function at that scale.

\subsubsection{Diffusion parameter estimation}
\label{sec:d_ang_sep_method}

    The application of equation~\ref{eq:diff_coeff_full} to empirical data requires the computation of the parallel fluctuation term $\mathcal{J}(\lambda_1)$. This depends on the power spectrum $\hat{P}(k)$ of fluctuations along the large-scale magnetic field vector which, we argue, is well characterized by the fluctuations 
    in the measured PAs.
The power spectrum $\hat{P}(k)$ and angular dispersion function $\mathcal{S}_2(\ell)$ are related (via the Wiener-Khinchin theorem; see~\citealt{Wiener1930_book, Percival1995_book}) by
    \begin{equation}
    \hat{P}(k) = \frac{1}{2}\mathcal{F}\left[\mathcal{S}_2(\ell)\right] \ , 
    \label{eq:relation_s2_sk}
    \end{equation}
(see Appendix~\ref{sec:appendix_sk_s2} for details) 
where ${\mathcal F}[...]$ denotes a Fourier transform (FT).
    When applied to a discrete data-set separated into $N_{\rm bins}$ bins according to scale (e.g. measured PA difference between a pair of points, binned according to the angular separation of each pair -- cf. section~\ref{sec:application}), the discrete FT\footnote{We use the discrete Fourier transform implementation in the \texttt{SciPy} Python package \citep{SciPy2019}, and
    \texttt{NumPy} \citep{harris2020} which uses the algorithm set out in \cite{NRbook3_2007} \citep[see also][]{Cooley1965_book}.} 
    of $S_2(\ell_n)$ can be taken for each scale-bin $\ell_n$ to find $P(\kappa_n)$. The fluctuation statistic then follows 
    a discretized form of equation~\ref{eq:j_k1_resonant}:
\begin{equation}
    \mathcal{J}(\lambda_1) \approx
    k_c^{-1}\;\!\sum_{n=1}^{i_{\rm b}} \frac{\kappa_{n}}{k_c \lambda_1}\;\! P(\kappa_n) + 
    k_c^{-1}\;\!\sum_{n=i_{\rm b}}^{N_{\rm bins}} \frac{k_c \lambda_1}{\kappa_{n}}\;\! P(\kappa_n) \ ,
    \label{eq:j_n1_discrete}
\end{equation}
where $i_{\rm b}$ is
the bin index corresponding to the (normalized) resonant length scale $\lambda_1\;\!k_c$,
and $P(\kappa_n)$ is the discrete FT of $S_2(\ell_n)$ for $\ell_n$ as the characteristic separation length between data pairs in the bin (taken simply as the bin center-point in $\ell$). 
$\kappa_n$ is the normalized wavenumber associated with $\ell_n$. 
An empirical estimation for the diffusion parameter $D$ within the observed system 
is obtained by substituting equation~\ref{eq:j_n1_discrete} into~\ref{eq:pmumu_original}. 

This allows us to calculate the diffusion coefficient from observations. For instance, we obtain a value of around $10^{29}-10^{30}~\text{cm}^2\;\!\text{s}^{-1}$ (depending on waveband -- see Appendix~\ref{sec:appendixc} for details) for a 1 GeV CR in a magnetic field of reference strength $B_{\rm ref} = 2.46 \mu\text{G}$ (see section~\ref{sec:filamentary_structures}) when we apply this to observations of the IC 5146 molecular cloud complex (section~\ref{sec:application}). This compares with estimates of around $10^{26}~\text{cm}^2\;\!\text{s}^{-1}$ for molecular clouds in the W28 region~\citep{Gabici2011crpa}\footnote{We note that this value is likely attributed to CRs in the 10 TeV range, and would correspond to a diffusion coefficient value of around $10^{28}~\text{cm}^2\;\!\text{s}^{-1}$ if scaled to GeV energies.}, $10^{29}~\text{cm}^2\;\!\text{s}^{-1}$ in the Galactic Center ridge~\citep{Gabici2011crpa}, or values of between $10^{25} - 10^{27}~\text{cm}^2\;\!\text{s}^{-1}$ in the Sgr B2 giant molecular cloud~\citep{Protheroe2008MNRAS, Dogiel2015ApJ}.
To account for the energy-dependence of the diffusion coefficient, we adopt the parameterised scaling 
 \begin{equation}
    D(E, s) = D_0 \left[ \frac{r_{\rm L}\left(E, \langle | B | \rangle|_s \right) }{r_{{\rm L},0}} \right]^{\delta} \ ,
    \label{eq:general_diff_coeff}
\end{equation}
based on the local gyro-radius $r_{\rm L}$ compared to $r_{{\rm L},0}$, i.e. that of a 1 GeV CR in a reference magnetic field of characteristic mean strength $B_{\rm ref}$.
Here, $\langle | B | \rangle|_s = |B(s)|$ is the characteristic mean magnetic field strength at some position $s$,  
   and $D_0$ is the value of the characteristic diffusion coefficient calculated as outlined above.
   The index $\delta=1/2$~\citep[see also][]{Berezinskii1990, Strong2007} accounts for 
   the cloud turbulence spectrum 
   which is set here to be the same as the broader ISM. The exact choice of this parameter does not strongly influence our results, and we leave a more careful assessment of its value in specific regions to future work.

\subsection{Observing magnetic fields in molecular clouds}
\label{sec:observing_b_fields}

Polarization of starlight by dust (as well as polarized sub-millimeter emission from the dust itself) is often used to probe the orientation and structure of interstellar magnetic fields on the plane of the sky~\citep[see][for a review]{Crutcher2012ARAA}. 
This polarization arises from selective absorption (or re-emission) by dust grains, whose magnetic moments tend to align themselves with the local interstellar magnetic field due to radiative torque alignment, leading to a preferred perpendicular orientation of the grains to ambient magnetic fields in low-extinction regions~\citep{Dolginov1976ApSS, Draine1996ApJ, Lazarian1997ApJ, Draine1997ApJ}. 
As a result,
a larger grain extinction cross section occurs perpendicular to the background magnetic field vector, 
causing linear polarization. 
While there is some debate over whether this alignment mechanism could operate effectively in denser high-extinction regions where radiation fields are lacking -- e.g. in MCs and cores therein -- results are so far largely consistent with the radiative torque model, but with substantial variation in alignment efficiency between individual sources~\citep[see][]{Whittet2008ApJ, Cashman2014ApJ}.\footnote{Other mechanisms, e.g. Zeeman or Goldreich-Kylafis (GK) effects, can also be used to probe MC magnetic fields, with the Zeeman effect being the only 
direct method 
to measure the magnetic field strength~\citep{Crutcher1993ApJ}. 
However in MCs these mechanisms are more difficult to be observed than dust polarization~\citep[see e.g.][]{Crutcher2012ARAA}.}
The strength of magnetic fields in MCs can be estimated by the
Davis-Chandrasekhar-Fermi~\citep{Davis1951PhRv, Chandrasekhar1953ApJ} method (hereafter DCF), being the `standard' approach to infer magnetic field information from dust polarization data. This assumes equipartition between turbulent magnetic energy and turbulent kinetic energy 
such that
the observed PA dispersion and velocity dispersion 
can be related to 
the magnetic field strength on the sky plane.


\section{Model and Results}
\label{sec:results}

In the first instance (section~\ref{sec:mol_cloud_idealized}), we consider an idealized MC model 
to assess the propagation and interactions of CRs. 
The model
has a simple Plummer-like density profile 
with a frozen-in magnetic field
and a simplified CR diffusion coefficient. Our intention is to demonstrate the magnitude of the ionization and heating effects CRs may impart in such an idealized scenario. However, we acknowledge that such a treatment is not necessarily representative of all ISM molecular clouds and star-forming regions, where CR diffusion can be different to that in the broader ISM (e.g. see~\citealt{Dogiel2015ApJ}, which finds a value of $3\times 10^{27}~\text{cm}^2~\text{s}^{-1}$ for the diffusion coefficient in the molecular cloud Sgr B2 near the Galactic Center, around an order of magnitude lower than the typical value quoted for the ISM -- see, e.g.~\citealt{Berezinskii1990, Aharonian2012, Gaggero2012}). As such, we later compute an empirically derived diffusion coefficient (using the method detailed in section~\ref{sec:3_2}) specifically for the IC 5146 molecular cloud complex in section~\ref{sec:application}. This uses the inferred structure of the local magnetic field estimated from dust polarization to determine a diffusion coefficient appropriate for this region. We use this to apply our model to identified filamentary structures in IC 5146 to assess a more realistic level of CR ionization and heating for a typical Milky Way star-forming environment.

\subsection{Molecular cloud model}
\label{sec:mol_cloud_idealized}

\subsubsection{Fractional ionization and density profile}
\label{sec:ion_frac_profile}


The ionization of molecular gas in different cloud components may largely be considered as a balance between the CR ionization rate and recombination processes. The resulting ionization 
fraction per H$_2$ molecule may be approximated by 
\begin{equation}
    x_i(s) = x_{i,0} \left( \frac{Y}{Y_{\odot}} \right)^{1/2} \left( \frac{\zeta^{\rm H}}{10^{-17}\;\!\text{s}^{-1}} \right)^{1/2} \left(\frac{n(s)}{10^5\;\!\text{cm}^{-3}}\right)^{-1/2} 
\end{equation}
\citep{Elmegreen1979ApJ} where $x_{i,0} = 8.7 \times 10^{-8}$ 
for small dust grains
and $Y_{\odot}$ 
is the solar metallicity. 
Strictly, the ionization fraction should be calculated dynamically in our model, 
but to reduce computational time we consider it sufficient to conservatively adopt a value of the `standard' CR ionization rate of 
$\zeta^{\rm H} = 10^{-16}\;\!\text{s}^{-1}$,  typically appropriate for the diffuse inter-clump medium~\citep{Black1978ApJ, vanDishoeck1986ApJS, Federman1996ApJ}.
The density profile $n(s)$ comprises a clump and a core, described by two superimposed Plummer profiles~\citep{Whitworth2001ApJ, Lee2003ApJ, Dib2010MNRAS}, 
such that $n(s) = n_0[1+(s/s_0)^2]^{a}$
where 
$-2.5 \le a \le -1.5$ 
for different clouds~\citep{Federrath2013ApJ},
including IC 5146~\citep[see][]{Arzoumanian2011}.
For our calculations, we adopt 
$a = -2$, 
assuming that gravitational pressure and thermal pressure 
are in equilibrium, with $n_0 = 10^3~\text{cm}^{-3}$ and $s_0 = 2~\text{pc}$ for the clump region, and $n_0 = 10^5~\text{cm}^{-3}$ and $s_0 = 0.2~\text{pc}$ for the core~\citep[e.g.][]{Bergin2007}.


\subsubsection{Magnetic field and diffusion coefficient}
\label{sec:b_field_mc_idealized}

We consider 
the magnetic field strength to simply scale with 
the cloud density: 
\begin{equation}
  B(n) = \begin{cases}
B_{\rm ICM} \ ,  \hspace*{2cm} n\leq n_1 \ ;  \\
B_{\rm ICM}\left({{n}/{n_1}}\right)^{\bar{q}} \ , \hspace*{0.77cm} \ \ n> n_1  \ , 
\end{cases}    
\label{eq:crutcher_model}
\end{equation}
where
we set 
$n_1 = 300~\text{cm}^{-3}$, 
$B_{\rm ICM} = 30~\mu\text{G}$ as a background reference magnetic field strength appropriate for the inter-clump medium,\footnote{This is estimated from line-of-sight field strengths in~\cite{Crutcher2010} by omitting the diffuse cloud components; see also~\cite{Thompson2019ApJ} which specifically considers inter-core regions of MCs (with $30~\mu\text{G}$ as a line-of-sight magnetic field strength falling well within the suggested range, albeit stronger than average for their sample).}
and 
$\bar{q} \approx 2/3$, 
    assuming spherical collapse and magnetic flux-freezing
    \citep[see][]{Mestel1966, Crutcher2010}.
 In this first approach (which is not specific to any particular molecular cloud complex), we do not account for the orientation or structure of the magnetic field, and simply assume isotropic diffusion with a simplified coefficient of the form given by equation~\ref{eq:general_diff_coeff}. Moreover, we adopt the empirical normalisation reference for diffusion coefficient to be that based on the broader Milky Way ISM 
   as $D_0 = 3.0\times 10^{28}$ cm$^2$ s$^{-1}$,~\citep{Berezinskii1990, Aharonian2012, Gaggero2012}
   and $B_{\rm ref} = 5~\mu\text{G}$.

\subsubsection{Results}
\label{sec:results_idealized}

\begin{figure}
    \centering
    \vspace*{0.2cm}
    \includegraphics[width=\columnwidth]{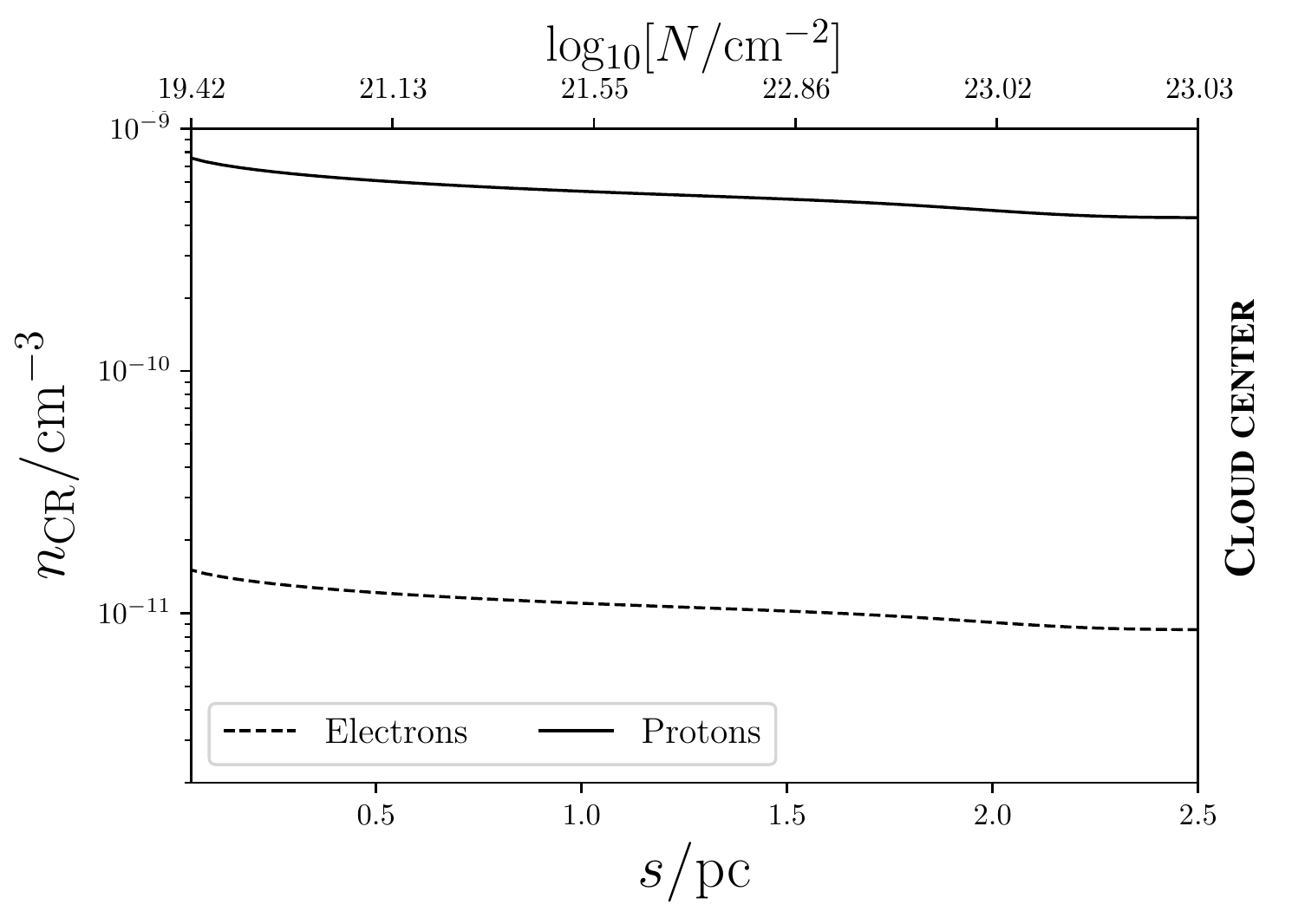}
    \vspace*{-0.5cm}
    \caption{Distribution of energetic primary CR protons and electrons above a GeV propagating  into the MC model prescribed in sections~\ref{sec:ion_frac_profile}--~\ref{sec:b_field_mc_idealized}{, with distance into the cloud $s$ shown on the lower $x$-axis, and the corresponding gas column density $N$ on the upper axis}. 
    The small decrease in the CR number density towards the core is due to increased magnetic mirroring effects.}
    \label{fig:primary_crs_number_density}
\end{figure}
\begin{figure}
    \centering 
    \vspace*{0.2cm}
    \includegraphics[width=\columnwidth]{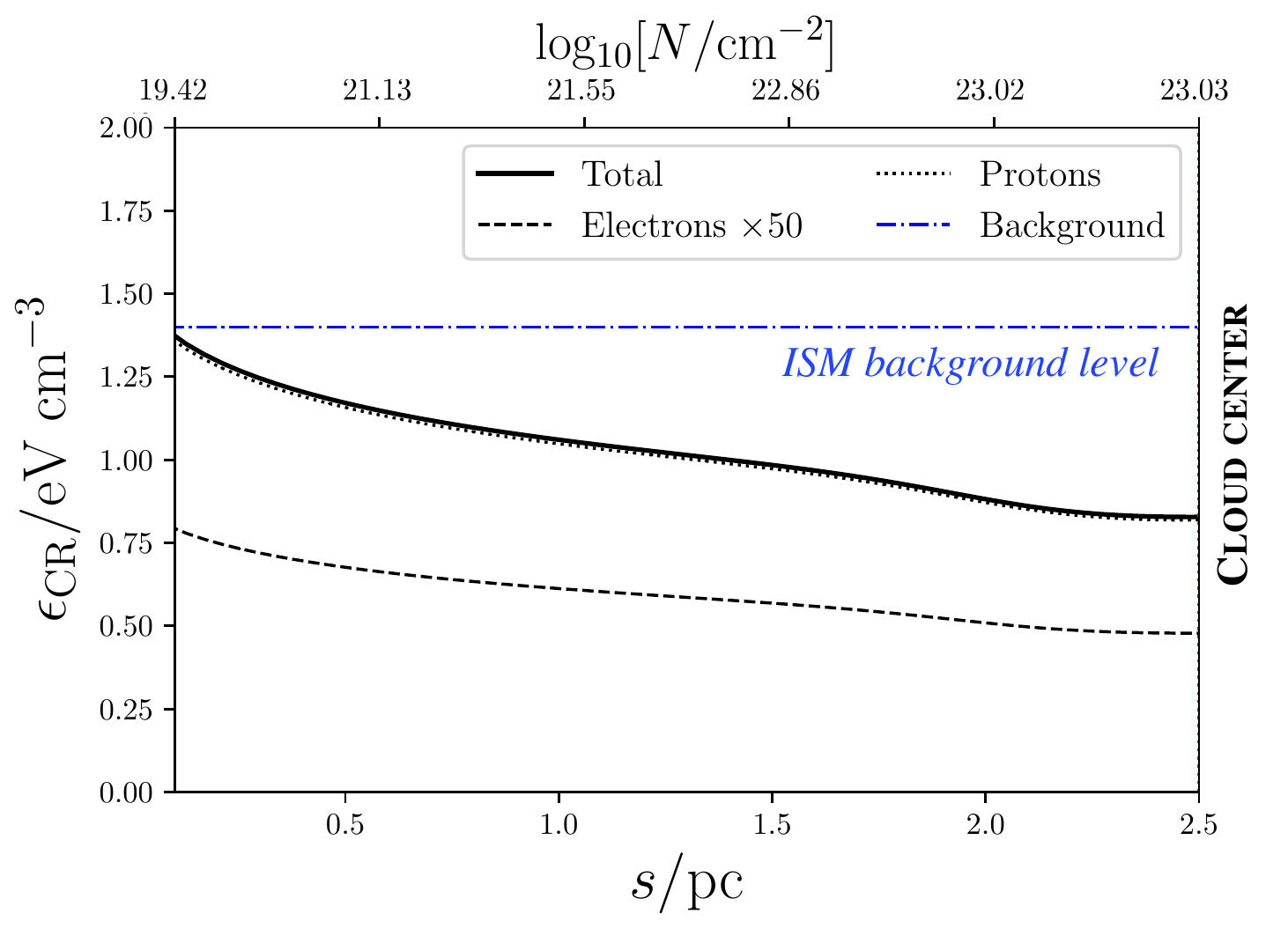}
    \vspace*{-0.5cm}
    \caption{Energy densities of 
     primary CRs
    throughout the cloud under Galactic conditions. The background interstellar CR energy density is indicated by the blue line, while the contribution from the secondary CRs is negligible (not shown). The primary electron component is multiplied by 50 in this plot for clarity.
    }
    \label{fig:energy_densities_mc}
\end{figure}

The transport equations
\ref{eq:reduced_transport_equation_protons},~\ref{eq:transport_equation_electron_primary} and~\ref{eq:transport_equation_electron_secondary} are solved according to the scheme in Appendix~\ref{sec:appendixa} for the cloud model 
specified in sections~\ref{sec:ion_frac_profile} -- \ref{sec:b_field_mc_idealized} to 
calculate the distribution of CRs 
in an idealized environment.
Fig.~\ref{fig:primary_crs_number_density} 
shows the distribution of primary CRs,
gradually decreasing in number density 
 towards the cloud center, which 
 is also reflected in the CR energy density distribution in Fig.~\ref{fig:energy_densities_mc}.
 The latter arises from mirroring effects, coupled with a stronger magnetic field hampering CR propagation into the central region. 
 The decrease in CR abundance is relatively moderate -- roughly half of ISM levels -- suggesting that the combined effects of CR containment, mirroring and focusing 
 by MC environments are not particularly severe. 

\subsubsection{Comment on secondary electrons}
\label{sec:secondary_elecs}

The impact of the secondary CRs in this study 
is found to be negligible:
their abundance is several orders of magnitude
lower than the primary CRs
and their impacts are correspondingly small.
A low secondary abundance would also be consistent with the view that $\gamma$-rays from neutral pion decays would presumably dominate the high-energy emission from molecular clouds~\citep{Brown1977ApJ, Gabici2009MNRAS, Casanova2010PASJ, Dogiel2018ApJ}\footnote{Leptonic $\gamma$-rays and X-rays may also arise as a result of bremsstrahlung by the high-energy tail of the energetic electron distribution~\citep{YusefZadeh2002ApJ, YusefZadeh2013ApJ} -- but these would be attributed to primary electrons rather than locally-injected secondaries.}, and any synchrotron emission would come from the primary electron component of the CR flux
~\citep{YusefZadeh2013ApJ, Strong2014arXiv, Padovani2018Lett}.\footnote{Secondaries could begin to dominate at very high densities on much smaller scales, e.g. in circumstellar discs~\citep{Padovani2018Lett}. Very compact non-thermal sources in MCs may be a signature of synchrotron emission from secondary electrons~\citep{Jones2014ApJL}, although these may depend on in-situ re-acceleration in, e.g. proto-stellar jets~\citep{Padovani2015AA, Padovani2016AA, Cecere2016ApJ}. Note that this is in tension with earlier studies, which argued that synchrotron radiation from secondary electrons in dark clouds could actually dominate Galactic radio emission~\citep{Brown1977ApJ}; synchrotron emission from secondaries was also considered in \citet{Dogel1990AA} and \citet{Jones2014ApJL}, but these studies did not compare with the emission from primary electrons.} This is in line with expectations: knock-on production and pion-production processes injecting secondary electrons operate fairly competitively with one another (with knock-on processes slightly dominating at lower energies). This occurs at a rate of $\dot{n}_{\rm e} \sim n_{\rm H}\;\!c\;\!\sigma_{\rm [\pi^{\pm}]}\;\!n_{\rm p}$, (as protons are more abundant than electrons in the primary flux), where
$\sigma_{\rm [\pi^{\pm}]} \approx 10^{-29}~\text{cm}^2$ at a GeV~\citep{Blattnig2000} is the 
effective pion-production cross section (note that this is different from the total pp inelastic cross section ${\sigma}_{\rm p\pi}$ defined in equation~\ref{eq:pp_cs}). This can be balanced against the `loss' of secondary electrons via cooling over a timescale of $t_{\rm e} \approx (\sigma^{\rm ion}\;\! c\;\! n_{\rm H})^{-1}$. Radiative processes would be more important at higher energies, but the power-law nature of the spectrum means that the number of these higher energy electrons makes a negligible contribution to their total number density. So, lower-energy ionization losses can be regarded as the dominant secondary electron cooling process. Typically, $\sigma^{\rm ion} \approx 10^{-20}~\text{cm}^2$ is the electron-ionization cross section~\citep{Padovani2009AA} for a secondary electron generated by a 1 GeV hadronic primary (retaining a few percent of the primary's energy -- see \citealt{Owen2018}).
It then follows that $n_{\rm e} \approx \dot{n}_{\rm e}\;\!t_{\rm e}$, which may be re-arranged to give 
the ratio
${n_{\rm e}}/{n_{\rm p}} \approx \sigma_{\rm [\pi^{\pm}]}/\sigma^{\rm ion}\sim 10^{-9}$,  
thus confirming the negligible level of secondaries.

\subsubsection{Observable and astrophysical impacts}

\begin{figure}
    \centering
    \vspace*{0.2cm}
    \includegraphics[width=\columnwidth]{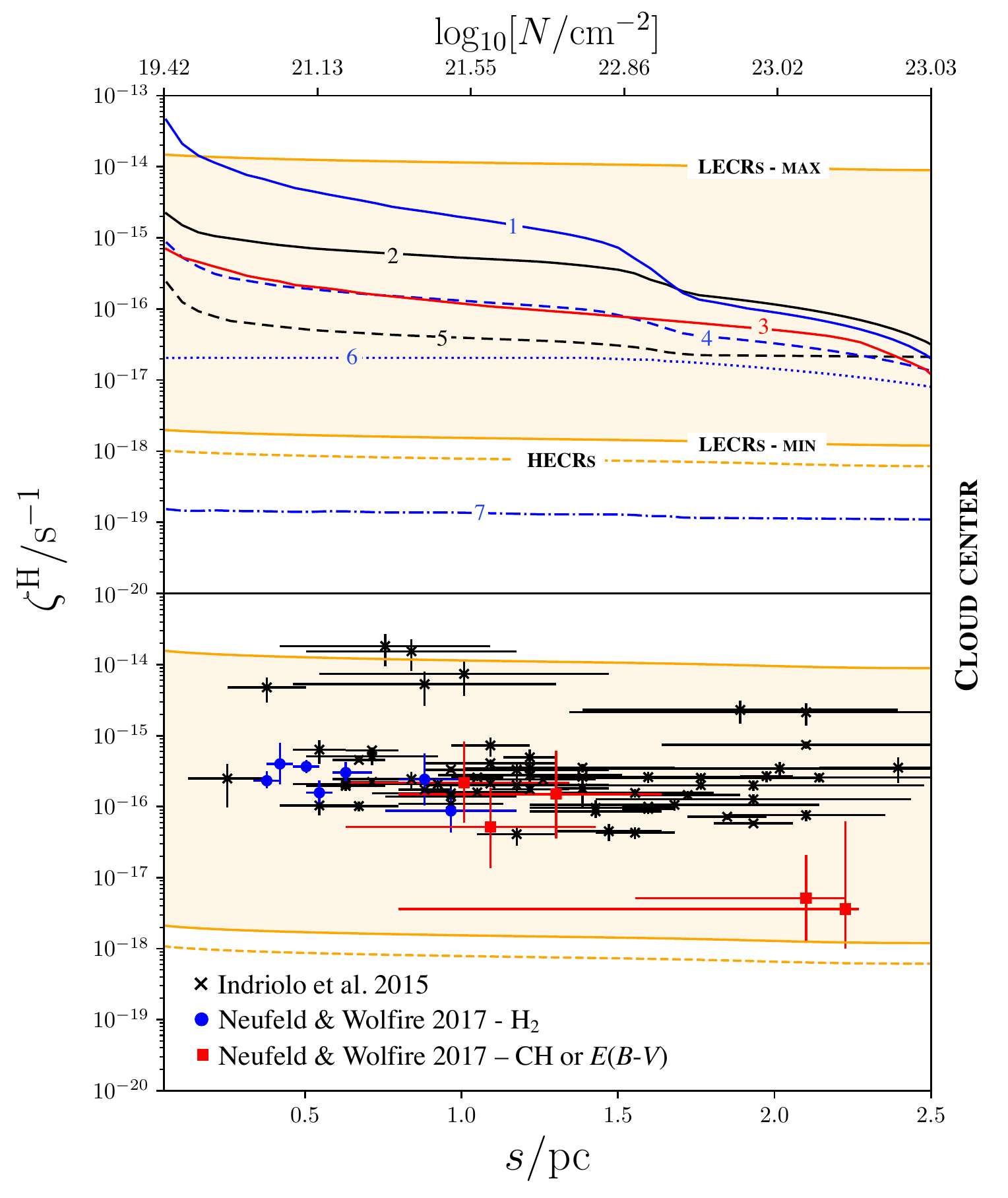}
    \vspace*{-0.5cm}
    \caption{{Ionization profile due to the HECR component (above 1 GeV) under Galactic conditions and the LECR component (the possible range is shaded, considering the minimum W98 and maximum M02 LECR spectra), as labeled in the top panel -- but shown in both panels. 
    The {\bf top} panel compares our result with other models: 
    (1) \citetalias{Padovani2009AA}, with their E00 CR spectrum; 
    (2) \cite{Padovani2018}, interstellar CR model (upper bound); 
    (3) \cite{Silsbee2019ApJ} pure diffusion model, using same CR spectral model as \cite{Padovani2018} (lower bound); 
    (4) \citetalias{Padovani2009AA}, with their M02 spectral model; 
    (5) \cite{Padovani2018} interstellar CR model (lower bound); 
    (6) \citetalias{Padovani2009AA}, with their W98 spectral model; 
    (7) \citetalias{Padovani2009AA}, with their C00 spectral model.
    The {\bf bottom} panel compares our result with observed CR ionization rates in molecular clouds, with data from \cite{Indriolo2015} and \cite{Neufeld2017ApJ} as labeled. The \cite{Neufeld2017ApJ} points are separated into measurements where column densities were obtained either from direct observations of H$_2$ or were inferred indirectly from observations of CH, or $E$($B$-$V$).}}
    \label{fig:ionization_profiles}
\end{figure}

CRs can influence their host environment by ionization, heating, 
and
modifying the chemical balance
of certain species~\citep[see, e.g.][]{Ivlev2018ApJ}.
The ionization profiles
due to the CR distributions 
in section~\ref{sec:results_idealized} are shown in Fig.~\ref{fig:ionization_profiles}.
{The impact of the high-energy CR component (HECR), 
above a GeV, is shown separately (as labeled)}
with rates 
of $\zeta^{\rm H} = 0.6 - 1.1 \times 10^{-18}~\text{s}^{-1}$, 
which are somewhat lower than the widely-adopted values of $\zeta^{\rm H} = 10^{-17} - 10^{-15}~\text{s}^{-1}$ in dense cores~\citep{Caselli1998, Tak2000, Doty2002},
and
$\zeta^{\rm H} = 10^{-16}~\text{s}^{-1}$ in the diffuse inter-clump medium~\citep{Black1978ApJ, vanDishoeck1986ApJS, Federman1996ApJ}. 
This difference can be attributed to the low-energy CR (LECR) component, {below a GeV, as indicated by the shaded region in} Fig.~\ref{fig:ionization_profiles}. Although their energy density is less than the HECRs, LECRs more strongly engage in ionization processes and so can elevate the rate to $\zeta^{\rm H} = 10^{-17}~\text{s}^{-1}$
when adopting the `minimum' LECR species spectra
in
the W98 \citep{Webber1998ApJ} and C00 \citep{Strong2000ApJ} models,
or $\zeta^{\rm H} = 10^{-14}~\text{s}^{-1}$ if adopting `maximum' spectra
in the 
M02 \citep{Moskalenko2002ApJ} and E00 \citep{Strong2000ApJ} models. {In the upper panel of Fig.~\ref{fig:ionization_profiles}, we show a comparison between our model and ionization rates from other models in the literature. We note that the results of~\citetalias{Padovani2009AA}, shown by the lines labeled 1, 4, 6 and 7, adopt the same CR spectra as considered in this work, so differences can be attributed to the treatment of CR transport. The lower panel of Fig.~\ref{fig:ionization_profiles} additionally shows comparisons with recently observed CR ionization rates in molecular clouds, which are broadly consistent with the results of this work.}

\begin{figure}
    \centering
    \includegraphics[width=\columnwidth]{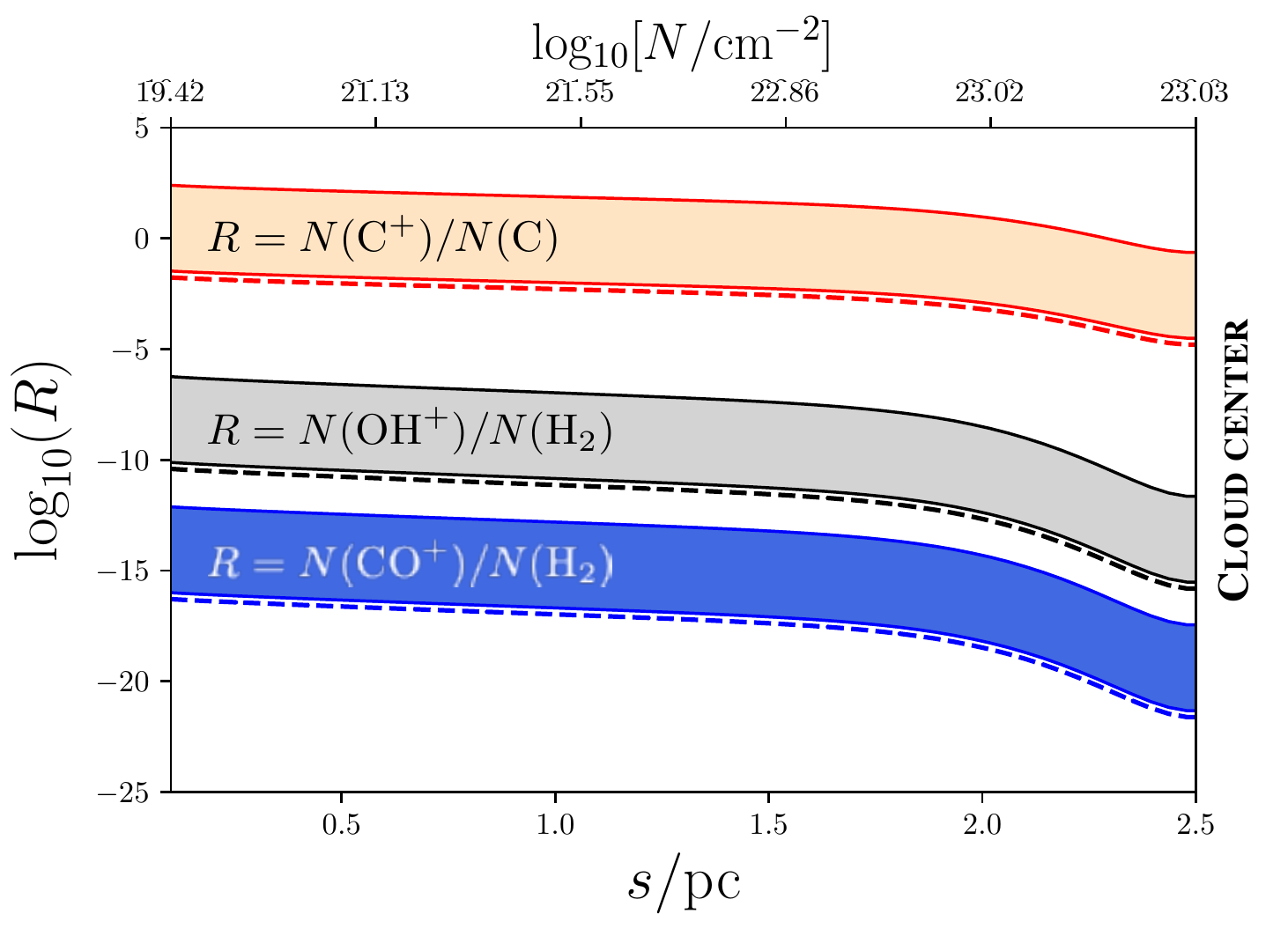}
    \caption{Abundance ratios due to CR ionization within the MC model, showing $N({\rm C}^+)/N({\rm C})$ (red), $N({\rm CO}^+)/N({\rm H}_2)$ (blue) and $N({\rm OH}^+)/N({\rm H}_2)$ (black). The shaded band represents the range of possible values from LECR ionizations only, 
    while the single dashed 
    line below each band represents the signature expected if only HECRs are present. Typically, any signal would therefore be dominated by the low-energy CR ionizations.}
    \label{fig:chem_profiles}
\end{figure}

These ionization profiles (again, accounting for the variation between the maximum and minimum LECR contribution to the ionization rates) can also be used to estimate abundance ratios for the species considered in section~\ref{sec:cr_ionization}, i.e. $N({\rm CO}^+)/N({\rm H}_2)$, $N({\rm OH}^+)/N({\rm H}_2)$ and $N({\rm C}^+)/N({\rm C})$. 
We present the resulting profiles for the three ratios in Fig.~\ref{fig:chem_profiles},\footnote{For reference, we find our results for $N({\rm OH}^+)/N({\rm H}_2)$ abundance ratio is broadly consistent with those presented in~\cite{Hollenbach2012}.} where 
the shaded ranges are due to LECR ionization
and the dashed lines due to HECR ionization.
The propagation of higher energy CRs may differ substantially from their lower energy counterparts due to, e.g. energy-dependent diffusion coefficients and the greater loss channels open to the higher energy particles. 
It is therefore worthwhile to consider the signature of both components, even though the LECRs would usually dominate the signal, with the higher-energy ratios typically around an order of magnitude lower.
While a single ratio should not be used independently to infer CR ionization rates, we note that the ratio $N({\rm C}^+)/N({\rm C})$ (in red) is much less sensitive to the underlying MC model than both $N({\rm CO}^+)/N({\rm H}_2)$ (blue) and $N({\rm OH}^+)/N({\rm H}_2)$ (black). This would make it a less stringent quantity to constrain CR ionization rates alone, and less powerful to diagnose the presence of CRs if alternative ratios are available. Indeed, this is in addition to possible contamination issues the $N({\rm C}^+)/N({\rm C})$ ratio would be susceptible to from any intervening non-shielded regions in the vicinity of the cloud which would boost the ${\rm C}^{+}$ abundance far above the levels driven by CR ionization processes.

\begin{figure}
    \centering
    \includegraphics[width=\columnwidth]{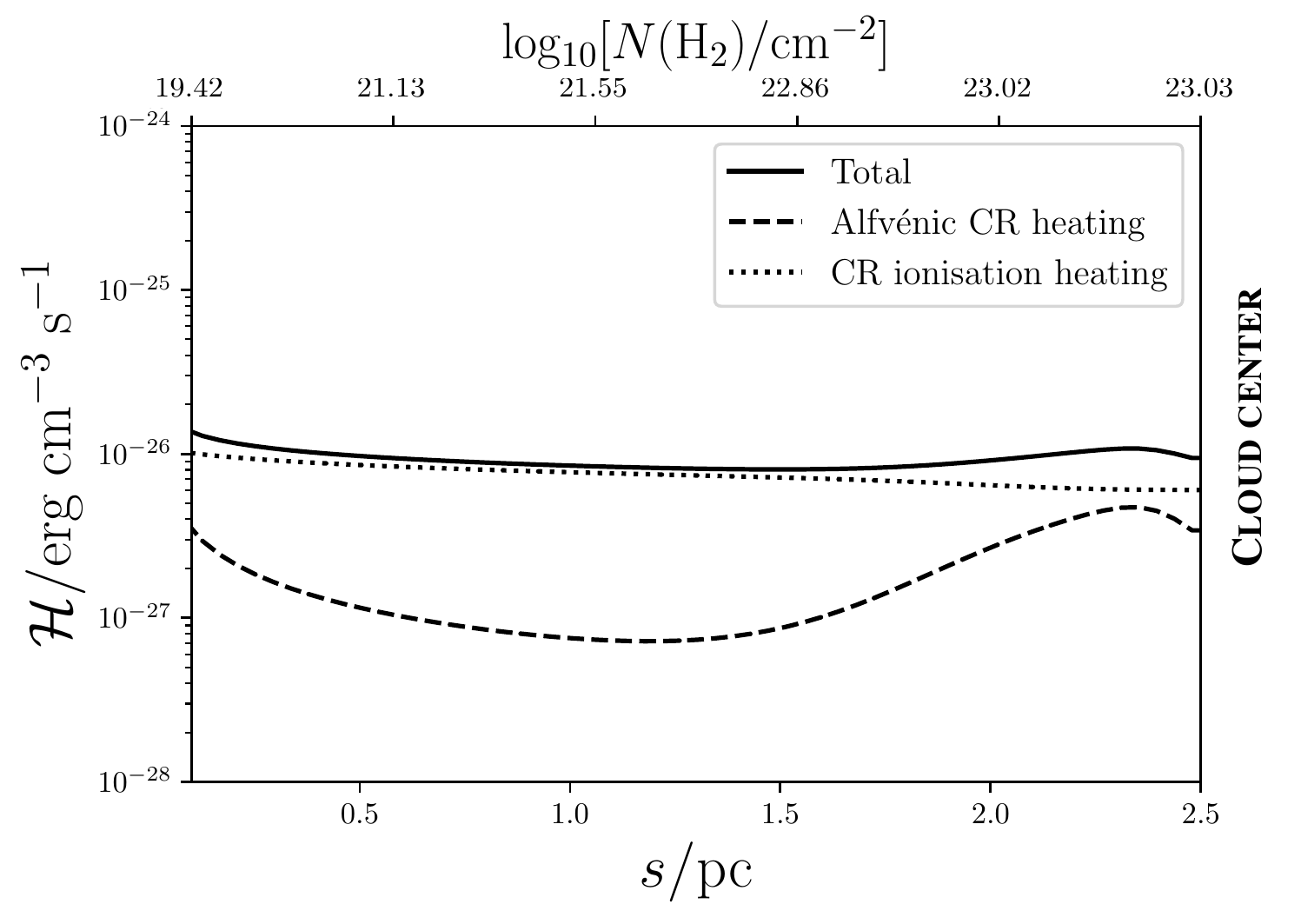}
    \caption{Total specific CR heating rates with contributions from Alfv\'{e}nic and ionization heating in the idealized molecular cloud model under Galactic conditions. $\mathcal{H}$ is the sum of all CR heating contributions outlined in section~\ref{sec:cr_heating}.}
    \label{fig:cr_heating_rates_mw}
\end{figure}

\begin{figure}
    \centering
    \includegraphics[width=\columnwidth]{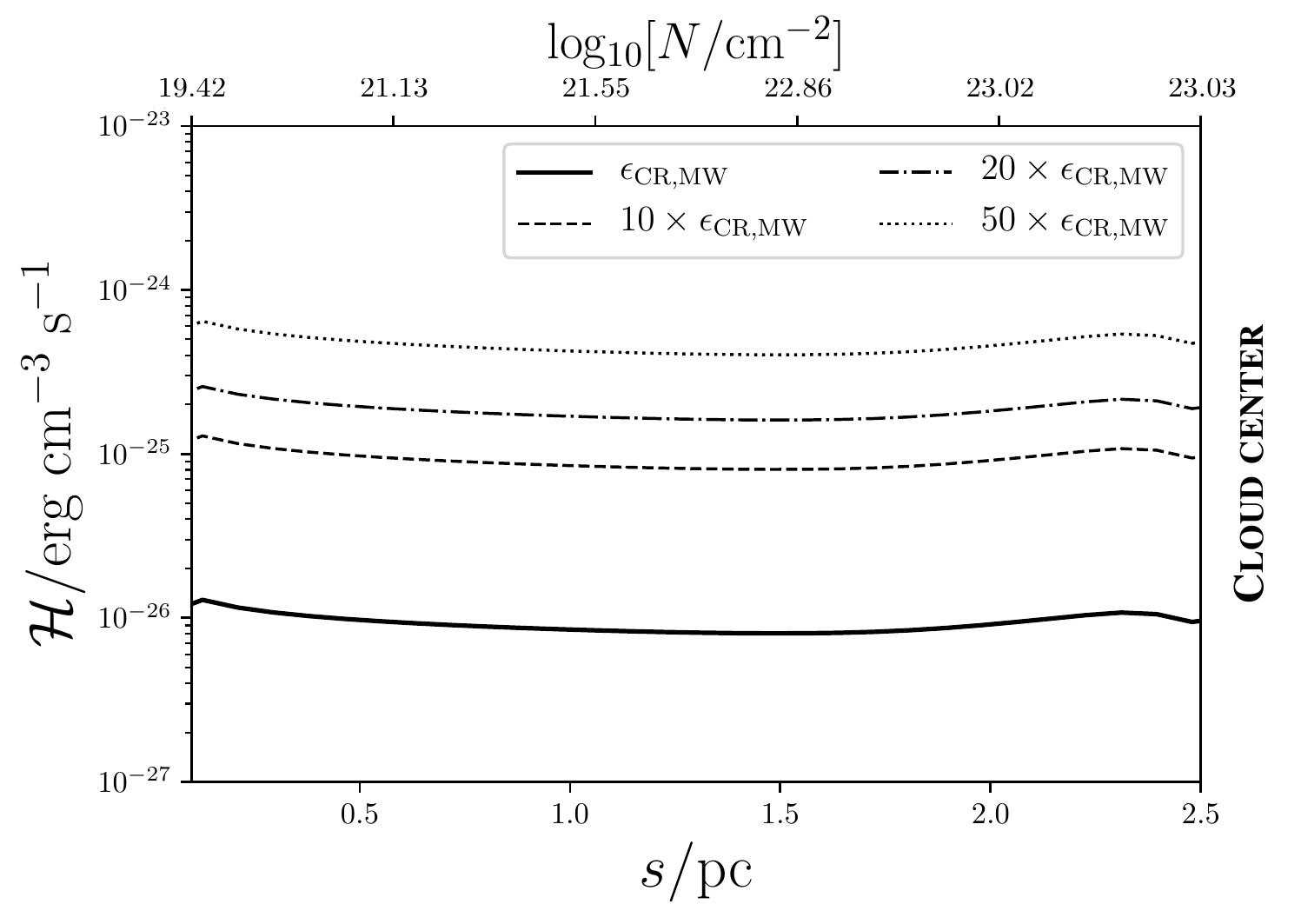}
    \caption{Total specific CR heating rates in the cloud under increasing ambient CR energy densities (as labeled), demonstrating the impact of different environments on the cloud's thermal evolution, even when shielded from the interstellar radiation field. The baseline case adopts the Milky Way CR energy density ($\epsilon_{\rm CR, MW}$), which is then scaled by the factors indicated. $\mathcal{H}$ is the sum of all CR heating contributions outlined in section~\ref{sec:cr_heating}.}
    \label{fig:cr_heating_rates_edense}
\end{figure}

Despite being more effective ionizers, the amount of energy available in LECRs to subsequently drive ionization-mediated heating effects is lower than in their higher-energy counterparts. Moreover, HECRs can heat a magnetized medium via Alfv\'{e}nic mechanisms (see section~\ref{sec:cr_heating} for details), making them more effective than LECRs in this capacity. 
The specific heating rate
due to HECRs in the MC is shown in Fig.~\ref{fig:cr_heating_rates_mw}, where $\mathcal{H}$ is the sum of all heating contributions outlined in section~\ref{sec:cr_heating}.
Generally, ionization-driven heating dominates -- however, in central regions the Alfv\'{e}nic specific heating rate
is boosted by the stronger magnetic field to become more competitive. 
The peak specific heating rate
reaches around $10^{-26}\;\!\text{erg}\;\!\text{cm}^{-3}\;\!\text{s}^{-1}$ in the core, which is largely unchanged from the level experienced in the outer regions of the cloud. This value is slightly lower
than less sophisticated estimates of $6.3\times 10^{-27}\;\!(n_{\rm H}/0.83~\text{cm}^{-3})\;\!\text{erg}\;\!\text{cm}^{-3}\;\!\text{s}^{-1}$~\citep{Spitzer1968ApJ, Goldsmith2001ApJ}, 
but around one order of magnitude higher than the estimated specific CR heating rate
in the warm ionized medium outside the cloud~\citep{Wiener2013_b}.
If the CR energy density is increased, the corresponding specific heating rates are boosted proportionally (see Fig.~\ref{fig:cr_heating_rates_edense}). This indicates that the CR impact on the evolution of MCs could be strongly influenced by their surrounding conditions, e.g. due to the impact of a nearby SN event (see also~\citealt{Gabici2009MNRAS}), or if located in a star-forming or high-redshift galaxy (see section~\ref{sec:summary}).

\begin{figure}
    \centering
    \includegraphics[width=\columnwidth]{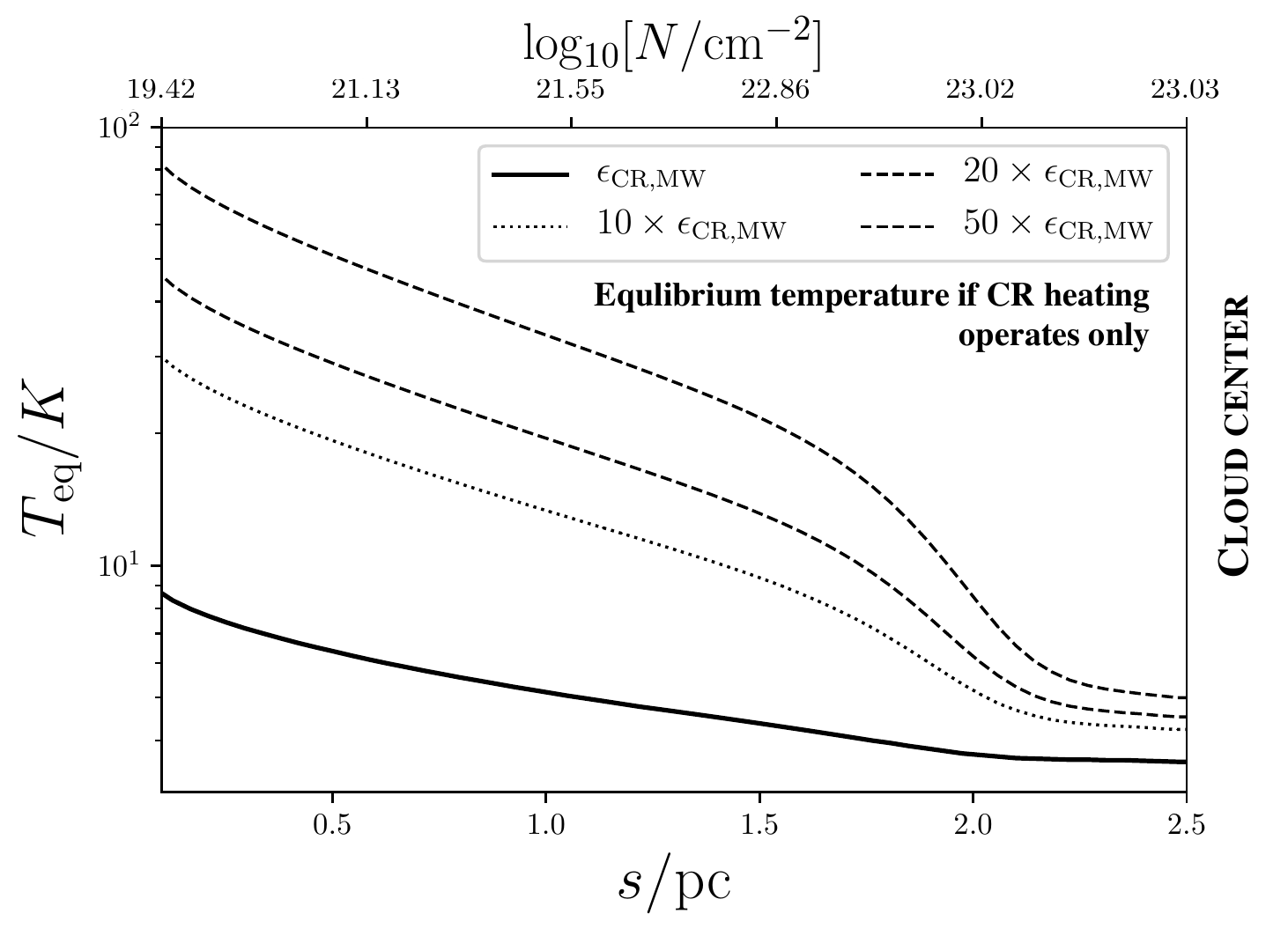}
    \caption{Equilibrium temperature profile arising from CR heating balanced against gas cooling 
    under Galactic conditions, and with CR energy densities boosted by factors of 10, 20
    and 50. Other processes may operate to maintain higher temperatures of around 10 K observed in most Galactic MCs.} 
    \label{fig:cr_temp_profile}
\end{figure}

By balancing CR heating rates against a simple cooling function (cf. section~\ref{sec:gas_cooling_cloud}), an equilibrium temperature can be estimated under different levels of CR irradiation. Fig.~\ref{fig:cr_temp_profile} suggests that, in a Galactic scenario, the effect of an enhanced CR energy density would be minimal in the core regions of a MC where the temperature is only boosted marginally compared to the surrounding clump.
Moreover, the core region 
 only experiences modest increases in its equilibrium temperature, even when faced with substantial enhancements of the CR flux. Instead, the surrounding clump and inter-clump medium are more strongly impacted.
The core temperatures estimated here are somewhat lower than expected, given that Galactic observations indicate core temperatures 
of around $8-12\ {\rm K}$~\citep{Bergin2007}. 
However, some starless cores are found to 
have temperatures of $T<7\ {\rm K}$
\citep{Pagani2007AA, Pagani2009AA, Pagani2015AA, lin2020AA}, 
while theoretical studies have also found temperatures comparable to those calculated here in idealized cases~\citep[e.g.][]{Juvella2011ApJ}.
%
%
Although our results are consistent with these certain extreme cases
(where 
core temperatures were argued to be as low as 6 K or even less, e.g.~\citealt{Harju2008}),  
they suggest that
either our adopted 
cooling function is 
overstated,
or 
other processes, 
e.g. turbulent heating~\citep{Pan2009ApJ},
residual unattenuated interstellar radiation fields and/or 
dust reprocessing/heating~\citep{Goldsmith2001ApJ},
may be operating alongside CRs to 
maintain higher core temperatures.
%
If other heating mechanisms operate in competition with CRs, 
CR heating would only 
dominate the regulation of MC core temperatures in environments where the CR energy density is substantially greater than that typically estimated in the Galaxy. 
A more thorough assessment of the 
competition between various heating effects throughout MC sub-regions merits a dedicated study and is left to future work.

\subsection{Illustrative case}
\label{sec:application}


\subsubsection{The IC 5146 cloud region and data-set}

The IC 5146 molecular cloud complex is located in Cygnus, 
exhibiting a converging filamentary system of dark clouds
and elongated sub-structures extending from a main filament
as seen
with \textit{Herschel} observations~\citep{Arzoumanian2011}. 
Throughout the cloud, structures are in various stages of their evolution: the main filament and nearby Cocoon Nebula appear to be at different stages of ongoing star-formation episodes~\citep{Harvey2008ApJ, Dunham2015ApJS}, while some other dark cloud regions remain quiescent~\citep{Arzoumanian2011}.
The formation of the system is believed to be driven by large-scale turbulence~\citep{Arzoumanian2011}, which 
would have
introduced perturbations into the otherwise well-ordered large-scale magnetic field morphology~\citep{Wang2017ApJ, Wang2019ApJ}, thus making it an ideal test-case for us to apply our model.

We note that there is some debate over the distance to the system
 \citep[see e.g.][]{Lada1999ApJ, Harvey2008ApJ}, 
 and whether it is one system or two clouds along the same line of sight at different distances~\citep{Wang2020ApJ}.
Such matters would impact the conversion between angular separations and physical distances in this work. A recent re-analysis (if assuming it to be a single cloud)
using \textit{Gaia} DR2 data~\citep{Gaia2018AA} indicates it to be located
$813\pm106~\text{pc}$ away~\citep{Dzib2018ApJ}. 
We adopt this distance in our calculations, as in  \citet{Wang2019ApJ}.  

For our analysis, we use optical and near infra-red stellar polarization observations towards IC 5146
\citep{Wang2017ApJ} to trace the magnetic fields and its underlying fluctuations.
The data-set was compiled by matching polarization data to the positions of 2022 independent background stars to within 0.5'' (corresponding to 0.002 pc at a distance of 813 pc) from the 2MASS all-sky survey~\citep{Skrutskie2006AJ}
in at least one of the Rc-, i', H- and K- bands. 
In total, only 3 stars were present in all four bands, with around 71\% of the stars being detected in the H-band, 24\% in the Rc-band, 10\% in the i'-band and 8\% in the K-band~\citep{Wang2017ApJ}.

\begin{figure}
    \centering
    \includegraphics[width=\columnwidth]{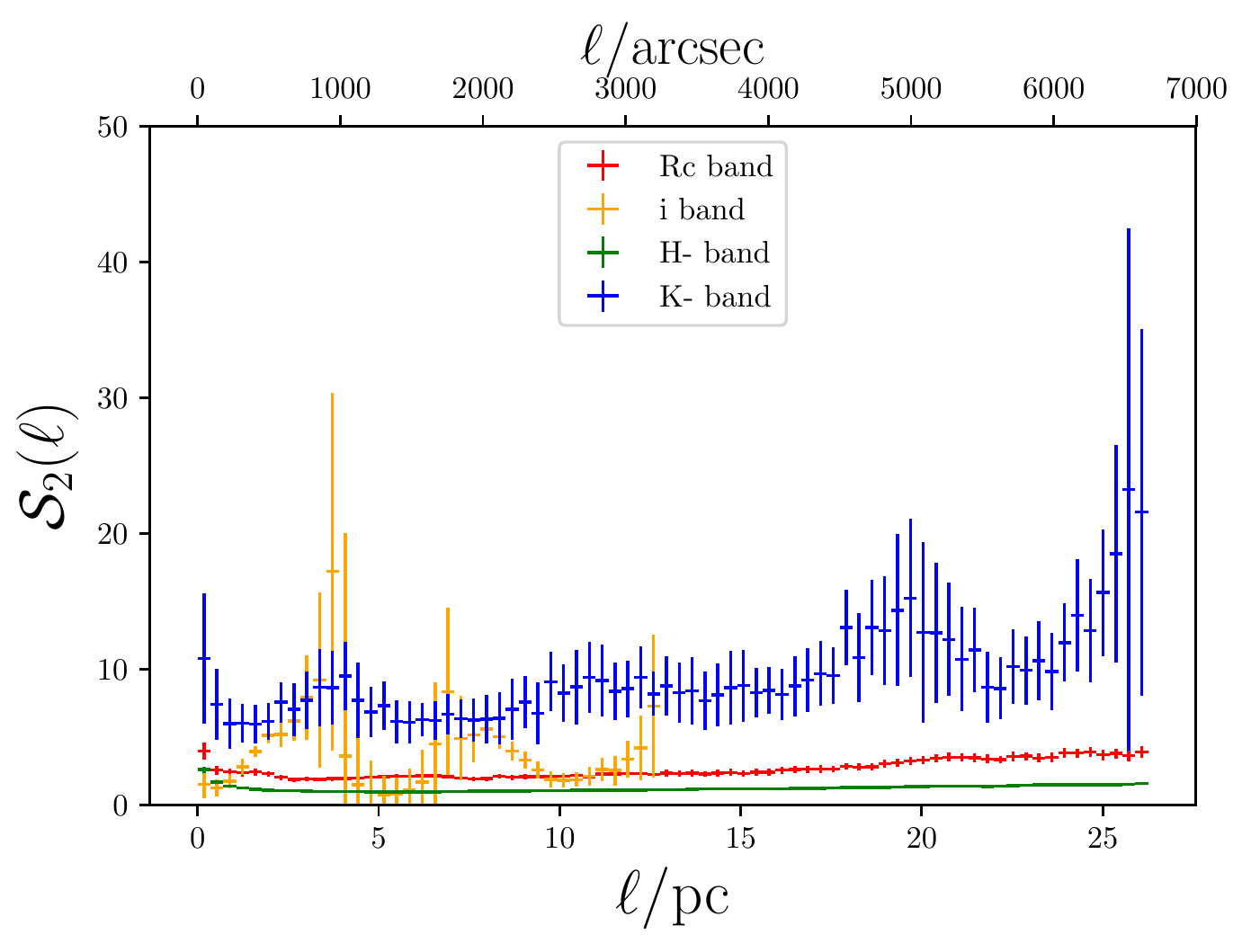}
    \caption{
    Dispersion functions calculated for the 4 bands. 
    The increasing power at larger scales due to the curved hour-glass magnetic field structure
    is not shown here as it is not relevant to our fluctuation analysis. 
    $x$ error bars indicate the bin size. $y$ error bars are 1$\sigma$ Gaussian errors estimated by a Monte Carlo approach with 10,000 perturbations. A distance of 813 pc to IC 5146 is assumed -- see main text for details.} 
    \label{fig:sf_profile}
\end{figure}

To estimate the diffusion coefficient of CRs through the region, 
we compute the angular dispersion function 
in every band
using a 
bin size
of 90'' (corresponding to a physical size of 0.35 pc) which gives reasonable signal-to-noise ratios over the length scales of interest. 
Uncertainties are estimated using 10,000 Gaussian Monte Carlo perturbations to the Stokes parameters $Q$ and $U$ (see Fig.~\ref{fig:sf_profile}, where 1$\sigma$ errors bars are shown). 
From our 
assessment of the distribution of 
separations of observed points, which do not reflect the features seen in Fig.~\ref{fig:sf_profile}, we argue that the structures evident in our dispersion analysis are likely to be physical in origin rather than due to instrumental or sampling effects.
The FT of the dispersion function is calculated, and
the diffusion coefficient for the region follows from  
equation~\ref{eq:diff_coeff_full}. 
We find that any spatial variation of the diffusion parameter $D$ 
arises from 
variations in the magnetic field~\textit{strength} alone, while spatial variations due to the field~\textit{structure} are not significant\footnote{There is also no compelling support for a significant variation in the empirical value of $D$ between the different wavebands.} --  see Appendix~\ref{sec:appendixc} for details.
These findings are applied to our subsequent analysis of IC 5146, where we adopt an average value for the structural contribution $\mathcal{J}$ to the diffusion parameter (see equation~\ref{eq:j_k1_resonant}) across the region, weighted by the number of background stars in each of the bands of the data-set (Rc-, i', H- and K-). {We find some variation of the diffusion coefficient, between $10^{29}-10^{30}~\text{cm}^2\;\!\text{s}^{-1}$ for a 1 GeV CR and a magnetic field of reference strength $B_{\rm ref} = 2.46\,\mu\text{G}$ (see section~\ref{sec:filamentary_structures}), depending on which band of the data-set is used in the analysis (see Appendix~\ref{sec:appendixc} for details). Our range of values is somewhat larger than the diffusion coefficients found in other studies -- e.g.   $10^{29}~\text{cm}^2\;\!\text{s}^{-1}$ in the Galactic Center ridge~\citep{Gabici2011crpa}, or between $10^{25} - 10^{27}~\text{cm}^2\;\!\text{s}^{-1}$ in the Sgr B2 giant molecular cloud~\citep{Protheroe2008MNRAS, Dogiel2015ApJ}. 
We believe this may result from some small-scale structure in the magnetic field being missed by our analysis, as would arise from limited/unavailable PA measurements on certain (smaller) separation scales, and/or projection effects that could mask magnetic field structures along the line of sight.}

\subsubsection{Representation of the filamentary structures}
\label{sec:filamentary_structures}

The IC 5146 region consists of a network of filamentary structures. \cite{Arzoumanian2011} identified 27 such filaments, to which they fit a cylindrical density profile
of the form
\begin{equation}
    n_{\rm H}(s) = {n_{\rm c}}{\left\{1+\left(\frac{s}{R_{\rm flat}}\right)^2\right\}^{-p/2}} \ ,
\end{equation}
with $n_{\rm c}$ being the density of the filament ridge and $R_{\rm flat}$ is the characteristic length scale of the flat innermost portion of the profile. Observationally, it is convenient to express this in terms of column density along a line of sight to a radial position $s$ within the projected filament
\begin{equation}
    N_{\rm H}(s) = \mathcal{P}{n_{\rm c}\;\!R_{\rm flat}} \; {\left\{1+\left(\frac{s}{R_{\rm flat}}\right)^2\right\}^{-\frac{p-1}{2}}} \ ,
\end{equation}
where $\mathcal{P}\propto 1/\cos \psi$ for $\psi$ is the angle of the filament to the sky plane.
The variation of this parameter due to the inclination of IC 5146 to the sky plane is small, within a factor of 2 if assuming random orientations, and less than many other sources of uncertainty in our model. 
As such,
we assume this system is oriented at $\psi = 0^{\circ}$ to the plane of the sky.
The magnetic field permeating this cloud is 
observed
to be arranged predominantly perpendicularly to the filament orientations on the sky plane 
\citep{Wang2017ApJ, Wang2020ApJ},
and the mean magnetic field strength across the filaments can be well-described by a simple power-law
\begin{equation}
    B_{\rm mean}(n) = B_0 \left(\frac{n}{150~\text{cm}^{-3}}\right)^{\bar{p}} \ , 
    \label{eq:bfield_ic_region}
\end{equation}
with best-fit values of $B_0 = 2.46^{+0.50}_{-0.51}~\mu\text{G}$ 
and $\bar{p} = 0.50^{+0.12}_{-0.13}$ \citep{Wang2020ApJ}.
    This prescription is specifically based on the analysis of IC 5146 and 
    is different
    from our idealized case following \citealt{Crutcher2010} 
    (see equation~\ref{eq:crutcher_model}).
    The filamentary structures are therefore ideal test-cases of our model, where the heating and ionization rate profiles can be calculated through each filament by adopting a natural boundary condition at the filament edge. In this case, we take this edge to be at $\pm$1.5~pc from the filament center, estimated from the proximity of the filaments in the region~\citep{Arzoumanian2011}.
    
\subsubsection{Results} 

Of the 27 distinct filamentary structures identified in~\cite{Arzoumanian2011}, there is sufficient parametric information (column density $N_{\rm H}$ and inner length scale $R_{\rm flat}$) available to allow our model to be applied to 15 cases (see Table~\ref{tab:filaments}).
For each case, we compute the ionization rate and specific heating rate, 
  and deduce the equilibrium temperature. 
The resulting heating and ionization rates in the filament ridge (being their densest point) are shown in Table~\ref{tab:filaments}, 
together with an estimate of the equilibrium temperature that could be sustained if only CR heating processes were operating. 

\begin{table*}
\centering
\hspace{-3.0cm}
\resizebox{0.90\textwidth}{!}{\begin{minipage}{\textwidth}
\begin{tabular}{|l|c|c|c|c|c|c|c|c|}
\hline 
\hline
  ID & $p$ & $R_{\rm flat}\;\!/\;\!\text{pc}$ & $n_{\rm c}\;\!/\;\!10^4\;\!\text{cm}^{-3}$ & $\mathcal{H}\;\!/10^{-26}\;\!\text{erg}\;\!\text{cm}^{-3}\;\!\text{s}^{-1}$ & $\zeta^{\rm H, min}_{\rm LECRs} \;\!/\;\!10^{-20}\;\!\text{s}^{-1}$ & $\zeta^{\rm H, max}_{\rm LECRs} \;\!/\;\!10^{-15}\;\!\text{s}^{-1}$ & $\zeta^{\rm H}_{\rm HECRs} \;\!/\;\!10^{-17}\;\!\text{s}^{-1}$ & $T_{\rm eq, CR}\;\!/\;\!~\text{K}$ \\
 \hline  
 \hline 
 \!\!
 1   & 2.1 & 0.09 & 0.3 & 0.59 & 2.1 & 4.4 & 1.1 & $0.8$ \\ 
 2   & 1.9 & 0.1 & 0.7 & 9.2 & 2.1 & 4.5 & 3.1 & $1.7$ \\ 
 4   & 1.4  & 0.04 & 0.7 & 4.3 & 2.1 & 4.5 & 1.9 & $1.3$ \\ 
 5   & 1.5  & 0.02 & 7 & 68 & 2.1 & 4.4 & 9.5 & $2.5$ \\ 
 6   & 1.7  & 0.07 & 4 & 290 & 2.1 & 4.5 & 14 & $4.0$ \\ 
 7   & 1.6  & 0.05 & 2 & 33 & 2.1 & 4.5 & 5.6 & $2.2$ \\ 
 8   & 1.5  & 0.09 & 0.4 & 6.8 & 2.1 & 4.6 & 2.0 & $1.8$ \\ 
 9   & 1.5  & 0.07 & 0.8 & 16 & 2.1 & 5.0 & 3.3 & $2.0$ \\ 
 10   & 2.1  & 0.1 & 0.5 & 2.5 & 2.1 & 4.5 & 2.1 & $1.2$ \\ 
 11   & 1.9  & 0.07 & 1 & 6.5 & 2.1 & 4.4 & 3.2 & $1.5$ \\ 
 12   & 1.5  & 0.05 & 4 & 240 & 2.1 & 4.5 & 12 & $3.7$ \\ 
 13   & 1.6  & 0.04 & 3 & 4.3 & 2.1 & 4.4 & 6.8 & $2.3$ \\ 
 20   &  1.5 & 0.05 & 0.2 & 0.34 & 2.1 & 4.5 & 0.66 & $0.7$ \\ 
 21   &  1.7 & 0.09 & 0.3 & 1.9 & 2.1 & 4.5 & 1.4 & $1.2$ \\ 
 25   & 1.5  & 0.05 & 0.7 & 4.9 & 2.1 & 4.5 & 2.2 & $1.4$ \\ 
\hline
\end{tabular}
\end{minipage}}
\caption{Specific rates of CR heating $\mathcal{H}$ and ionization $\zeta^{\rm H}$
  in 15 of the 27 filaments   
  in the IC 5146 region identified by \protect\cite{Arzoumanian2011}, where 
  filament ID numbers here correspond to those used in that paper. We have omitted filaments 16, 17, 18 and 19, which are located in a photon-dominated region, so temperature estimates and comparisons would not be useful or reliable for these cases. 
Filaments 22 and 27 have unresolved widths so are excluded. 
Filaments 3, 15, 16, 18, 22 and 24 are also omitted due to the lack of data on  
the filament width $R_{\rm flat}$, the $p$ index in the density profile
and/or
a single best-fit value. 
Filaments 14, 23 and 26 are excluded due to an asymmetric profile. 
Our CR propagation model 
uses
the derived diffusion coefficient specific to the IC 5146 region, where we adopt a mean value for the diffusion coefficients calculated in the Rc-, i'-, H- and K- bands, weighted according to the number of data points in each band. 
The filament ridge volume density, 
$n_{\rm c}$, 
is estimated from the measured column densities $N_{\rm H}$ using $n_{\rm c} \approx N_{\rm H}/(2\;\!R_{\rm flat})$ following~\protect\cite{Wang2020ApJ}. 
This assumes each filament has a cylindrical geometry, and we note that $R_{\rm flat}$ has been adjusted to account for the updated distance to IC 5146 of 813 pc (compared to a distance of 460 pc used in~\protect\citealt{Arzoumanian2011}). 
Ionization rates are due to the minimum W98 LECR spectrum ($\zeta^{\rm H, min}_{\rm LECRs}$), maximum M02 LECR spectrum ($\zeta^{\rm H, max}_{\rm LECRs}$) and HECRs above a GeV ($\zeta^{\rm H}_{\rm HECRs}$) 
-- see section~\ref{sec:cr_spectrum} for details. 
Note that the estimated equilibrium temperature is calculated assuming that only
CR heating operates (calculated as the sum of all CR heating contributions outlined in section~\ref{sec:cr_heating}), and is found by balancing against the cooling rate 
(see section~\ref{sec:gas_cooling_cloud}). 
We estimate uncertainties in calculated quantities to be within an order of magnitude of the stated values, being dominated by the uncertainty in the empirical CR diffusion parameter.
} 
\label{tab:filaments}
\end{table*}

We have found that stronger heating typically occurs 
  in the filaments that have larger volume densities. 
The effect is not linearly proportional to the density 
   (e.g. compare filaments 6 and 7).  
This is because an increase in density will also lead 
  to an increase in magnetic field strength inside a filament.  
This, in turn, increases the amount of deflection experienced by the CRs, but also the containment. 
The resulting impact of these antagonistic processes is determined by the extent of the cloud and the variation of the density profile -- e.g. smaller values of $p$ yield a less steep magnetic profile and a relatively lower degree of mirroring to increase the heating efficiency. 
As shown, the strongest heating occurs 
  in filaments 6 and 12.  
These have a high density   
  but not a particularly steep density profile,  
  thus reducing magnetic mirroring/deflection effects and allowing a substantial amount of CRs to penetrate inside. 
CR ionization rates due to both low and high energy CRs are much less susceptible to 
   density variations deep inside the cloud. 
This would imply relatively consistent CR ionization rates throughout Galactic molecular clouds regardless of their internal configuration or exact filament/clump properties.

\subsection{Remarks}
\label{sec:remarks}

A clear implication of our results in Table~\ref{tab:filaments} is that CR-driven processes alone, while able to sustain relatively high rates of ionization within the cloud and heating rates comparable to literature estimates (see, e.g.~\citealt{Goldsmith2001ApJ, Wiener2013_b} and section~\ref{sec:cr_heating}), cannot maintain substantial core/filament temperatures if operating \textit{alone}. 
Although caveats in our model would offer scope for larger values, this result taken 
by itself 
would suggest that other processes (e.g. heating from turbulence dissipation driven by gravitational collapse or other mechanical processes -- 
see~\citealt{Carlberg1990, McKee1995, Gammie1996, Martin1997, Falceta2003}) may be more important in maintaining clump/filament temperatures at the expected level of around $8-12~{\rm K}$~\citep{Bergin2007} under Galactic conditions. 
This would point towards clump destruction being reliant on the emergence of proto-stars heating and ionizing dense regions from within, or photo-evaporation/conduction from the external environment 
(i.e. outside-in destruction) rather than CR heating.

However, the models described in this work rely on a number of assumptions. Chief among these is the use of dust polarization of stellar radiation to trace the magnetic fields: 
the reliability of this indirect means of probing magnetic fields remains under debate -- particularly towards the dense higher-extinction regions of principal interest in this work, and variation in alignment efficiency between individual sources~\citep[see][]{Whittet2008ApJ, Cashman2014ApJ} is well known.
The observed PA dispersion is used to estimate the diffusion coefficient, however in high density regions with poor radiative alignment efficiencies, the correspondence between dust alignment and magnetic field vectors could be weak -- or timescales for radiative grain alignment to operate could become very long and unable to fully reflect rapidly-varying turbulent magnetic field structures or fluctuations on the very small scales. Moreover, CR heating of dust grains themselves~\citep{Kalvans2018ApJS} could also influence torque alignment efficiencies \citep[cf. thermal wobbling, see][]{Lazarian2007MNRAS}, 
and hence dust polarization. 
This would decrease the diffusion parameter and could substantially increase the CR heating rate
felt in the densest regions. Future observations probing a wide range of magnetic field structures in molecular clouds to higher resolution and avoiding the need to rely on dust alignment mechanisms may be possible with the upcoming Square Kilometer Array (SKA) 
\citep[see e.g.][]{Strong2014arXiv}.
Moreover, an assessment of the impact of various magnetic field configurations on the CR diffusion parameter and subsequent heating effect via simulations is worthy of a dedicated future study.

A further assumption is that the cooling is dominated by CO and dust~\citep{Galli2002AA, Goldsmith2001ApJ}, 
and we model this by invoking 
an analytical approximation~\citep{Whitworth2018AA, Goldsmith1978},
 which assumes
 dust temperatures of around $10~{\rm K}$ (a compromise value between hotter regions in the cloud peripheries and cooler regions within the denser parts). While suitable under typical galactic molecular cloud conditions, variations in dust fraction and/or composition may result in different cooling rates in alternative settings 
 -- e.g. in dusty galaxies or more/less chemically pristine or primordial environments~\citep{Jaacks2019MNRAS}. 
 Moreover, substantial variation in dust and gas temperatures throughout a cloud would likely yield different cooling rates: presumably the rates adopted in this work are somewhat overstated in the densest core/filament regions where gas temperatures and subsequent cooling rates would be lower and less able to balance CR heating.

Variations in the irradiating CR flux or spectrum are also plausible. 
While this work has considered the effect of scaling the CR energy density and maintaining the same spectral form 
  (see Figs.~\ref{fig:cr_heating_rates_edense} and~\ref{fig:cr_temp_profile}), 
there is wide-ranging evidence in literature for different CR spectral indices at high-energies in different environments 
  -- a harder spectrum in the vicinity of the galactic ridge, where more CRs are freshly accelerated \citep{Aharonian2006, Gaggero2017, HESS2018b, HESS2018a},
would likely increase the CR heating rate. 

\subsubsection{Implications}
\label{sec:implications}

\begin{figure}
    \centering   
    \vspace*{0.2cm}
    \includegraphics[width=\columnwidth]{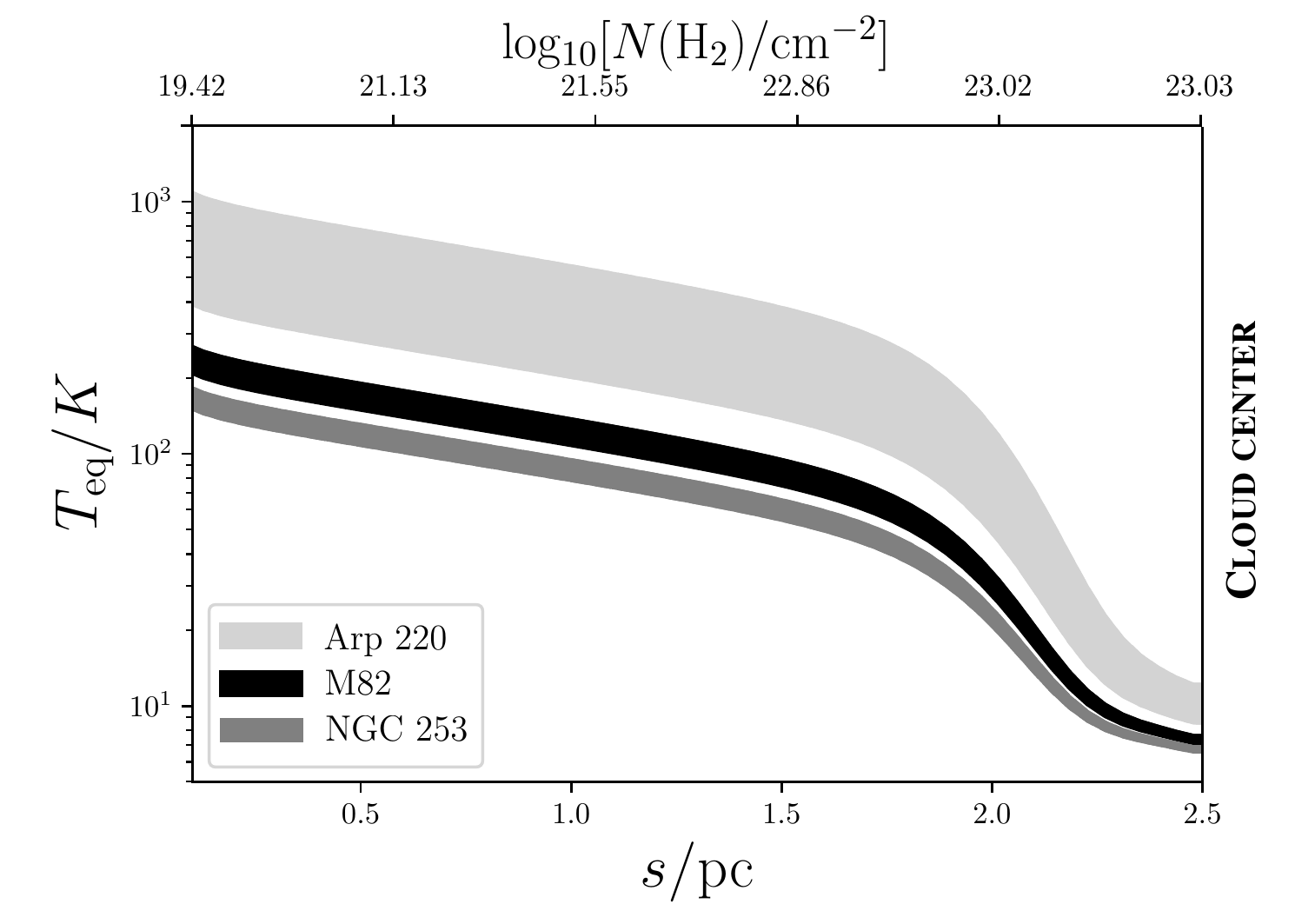}
    \vspace*{-0.5cm}
    \caption{Equilibrium temperature accounting only for CR heating 
    when considering a cloud in different starburst environments. 
    The CR energy density is estimated from \protect\cite{Yoast-Hull2016MNRAS}, where we adopt the range calculated by $\gamma$-ray and radio emission. For Arp 220, we use the values associated with the surrounding torus (ST) region -- see also \protect\cite{Yoast-Hull2015MNRAS}, and assume a Galactic CR spectral shape.}
\label{fig:gmc_temp_SBs}
\end{figure}

Although relatively well-shielded by the magnetic field configuration, 
 the modest level of CR heating experienced in a filamentary clump or core 
 would have substantial implications for subsequent star-formation. 
Fig.~\ref{fig:gmc_temp_SBs} explores the effect of different irradiating energy densities of CRs, scaled according to the estimated energy densities in nearby star-forming galaxies. In the exterior clump regions and inter-clump medium, a roughly linear relationship between CR energy density and resulting heating and equilibrium temperature is evident (see also Figs.~\ref{fig:cr_heating_rates_edense} and~\ref{fig:cr_temp_profile}), 
but the effect on the core temperature is much more modest. Aside from magnetic mirroring/deflection effects, 
CR heating is much more important in the inter-clump (diffuse cloud) medium in general where the ionization fraction is higher and CRs are deflected less. This would favour clump heating via conduction rather than direct internal heating by the CRs. 

Despite the core temperature increases remaining modest even when CR energy densities are enhanced substantially, the astrophysical implications can still be significant. The Jeans mass, for example, is dependent on temperature (as well as volume density of gas, $n_{\rm H}$) as this sets the level of thermal pressure against gravitational collapse:
\begin{equation}
    M_{\rm J} = 1.9 \times 10^2\;\!\left(\frac{T}{10~\text{K}}\right)^{3/2}\;\!\left(\frac{n_{\rm H}}{10^3\;\!\text{cm}^{-3}}\right)^{-1/2} ~\text{M}_{\odot} \ .
\end{equation}
%
While this is a crude measure of the maximum stable mass against collapse, 
    neglecting cloud/clump fragmentation,
    small-scale magnetic/turbulent support and 
    the micro-physics of the stellar initial mass function, 
it can still give an idea of the size of molecular clouds and the stellar clusters they develop into.
Without CR heating, Galactic clouds could reach around $200~\text{M}_{\odot}$ before they become unstable and gravitationally collapse. 
However, with Arp 220 levels of CR irradiation (see Fig.~\ref{fig:gmc_temp_SBs} for equilibrium temperatures resulting from CR heating in Arp 220, M82 and NGC 253 -- three nearby starburst galaxies), this could increase to around $4\times 10^3~\text{M}_{\odot}$.
The mass distribution of clouds/clumps in systems with higher CR energy densities (e.g. in star-forming galaxies) can therefore become distorted,
as higher star-formation rates would yield larger interstellar cloud sizes. 
This may favour more stochastic burst-like star-forming episodes. 
More importantly, the larger mass threshold required for gravitational collapse would likely lead to a period of quenching in a galaxy rich in CRs, as it would take longer for sufficiently large molecular clouds to gain enough mass to begin their collapse.

\section{Summary and Conclusions}
\label{sec:summary}

We investigate the heating effect of CRs  
  inside molecular clouds and the dense clumps within. 
We determine how CRs propagate and deposit their energy   
  inside the molecular clouds 
  using a CR transport formulation 
  for a model specified by density and magnetic field profiles.  
We find that for 
  an irradiating CR flux comparable 
  to in the Milky Way,  
  the specific heating rate can reach a level of
  $\sim 10^{-26}~{\rm erg~cm}^{-3}~{\rm s}^{-1}$ 
  in the densest regions of the cloud.
This heating is somewhat suppressed by magnetic deflection effects, 
 but comparable to the surrounding inter-clump medium. 
We also investigate the ionization caused by CRs
  and find observed rates 
  in the range 
  $\zeta^{\rm H} = 10^{-17} - 10^{-15}~\text{s}^{-1}$~\citep{Black1978ApJ,Tak2000, Doty2002}
  can be reproduced by our model, 
  with ionization being dominated by the contribution from low-energy CRs below a GeV. Higher energy CRs are more engaged with heating processes.

We further apply our model to the IC 5146 star-forming region in Cygnus, where we use observed dust polarization angles to estimate the CR diffusion coefficient through the region due to fluctuations in the local magnetic field (which we assume to be traced by the dust grain alignment). We calculate the CR heating and ionization rates in 15 of the filamentary structures identified in IC 5146, and find a broad variation of specific heating rates 
despite ionization rates remaining quite uniform between different filaments. 
While a specific CR heating rates 
  of 10$^{-27}-10^{-24}~{\rm erg~cm}^{-3}{\rm s}^{-1}$ 
  is not shown to lead to high equilibrium temperatures 
   of the filaments, 
   molecular clouds in environments 
   with a strong CR flux (e.g. in star-forming galaxies) 
   can be heated sufficiently, 
   resulting in an increase of the Jeans mass. 
 This would favour larger ISM clump sizes and presumably would yield a greater tendency for stochastic burst-like star-formation histories to emerge. Moreover, it could also arguably lead to periods of quenching.

\section*{Acknowledgements}

This work used high-performance computing facilities at the Center for Informatics and Computation in Astronomy (CICA), operated by the National Tsing Hua University (NTHU) Institute of Astronomy. This equipment was funded by the Taiwan Ministry of Education and the Taiwan Ministry of Science and Technology.
ERO and AYLO are supported by CICA at NTHU
  through a grant from the Ministry of Education of the Republic of China (Taiwan).
ERO's visits to NTHU 
  were supported by the NTHU International Exchange Scholarship, 
  hosted by SPL, 
  and by the Ministry of Science and Technology of the Republic of China (Taiwan) grants 105-2119-M-007-028-MY3 and 107-2628-M-007-003, 
  hosted by Prof Albert Kong. 
AYLO's visit to NTHU was supported 
 by the Ministry of Science and Technology of the Republic of China (Taiwan) 
 grant 105-2119-M-007-028-MY3, hosted by Prof Albert Kong. 
We thank Sheng-Jun Lin (NTHU) 
 for carefully reading through the manuscript. 
We also thank Dr Jia-Wei Wang (Academica Sinica Institute for Astronomy and Astrophysics), 
 Sheng-Jun Lin and Hao-Yuan Duan (NTHU) 
   and Dr Kate Pattle (NUI Galway)
   for discussions on molecular clouds and the IC 5146 polarization data, and Dr Ignacio Ferreras (Instituto de Astrof\'{i}sica de Canarias) 
  on the astrophysical implications of this work. 
ERO and AYLO also thank Dr Curtis Saxton (University of Leeds) 
  and Y. X. Jane Yap (NTHU) for 
  discussions and assistance with angular dispersion function (structure function) analyses.
  We are grateful to Prof Vladimir Dogiel (PN Lebedev Institute of Physics), Prof Chung-Ming Ko (National Central University) for their comments on this article, and the anonymous referees for their helpful and constructive feedback which substantially improved the manuscript.
This research has made use of NASA's Astrophysics Data System.
    
{\software{SciPy \citep{SciPy2019}, 
NumPy \citep{harris2020}}}

\newpage
\appendix

\section{Numerical scheme for solving the transport equation}
\label{sec:appendixa}

We introduce the variables  $X(E, s) = n(E, s)$ and $U(s) = D(E, s)\;\!\partial X/\partial s$, 
and the indices $q$ 
and $r$ for the energy $E_{\rm q}$ and spatial $s_{\rm r}$ grid points respectively when discretising the 
partial differential 
equations~\ref{eq:reduced_transport_equation_protons},~\ref{eq:transport_equation_electron_primary} and~\ref{eq:transport_equation_electron_secondary}. We thus seek a numerical solution for $X_{\rm q, r}$ in each case.
Due to the large differences in the variations of each of the terms (especially in sections~\ref{sec:primary_electrons_solver} and~\ref{sec:secondary_electrons_solver}), the greater numerical stability afforded by an implicit Runge-Kutta (RK) 4/5 scheme was required in order to respond to the inherent `stiffness' of the problem. 
The \verb RADAU5 \; implementation from~\cite{Hairer1993book} was adopted.\footnote{When parameter choices were selected to avoid stiffness issues, the explicit RK-Fehlberg implementation in~\cite{Press1992book} gave equivalent results.} 

\subsection{Primary protons}
\label{sec:primary_protons_solver}

Equation~\ref{eq:reduced_transport_equation_protons} is discretized as follows, when rewritten as a system of two 
difference 
equations: 
\begin{align}
    U_{\rm q, r+1} &= U_{\rm q, r} + \frac{\partial U}{\partial s}\Big\vert_{\rm q, r} \;\!\Delta s \label{eq:outerrk1} \ ; \\
    X_{\rm q, r+1} &= X_{\rm q, r} + \frac{U_{\rm q, r}}{D(E_{\rm q})} \;\!\Delta s  \ ,
    \label{eq:outerrk2}
\end{align} 
where $\Delta s$ is the step-size in the $s$ direction into the MC.
To solve them we consider a two-step scheme:
\begin{align}
    \frac{\partial U}{\partial s}&\Bigg\vert_{\rm q, r+1} = \frac{\partial U}{\partial s}\Bigg\vert_{\rm q, r} 
    + \Delta s \;\! \Bigg\{v_{\rm A}(s_{\rm r}) \frac{\partial X}{\partial s}\Bigg\vert_{\rm q, r} 
    \label{eq:inner_scheme_u}
    \nonumber \\
    & \hspace*{-0.3cm}
    +\;\!\frac{X_{\rm q+1, r}b_{\rm q+1, r} - X_{\rm q-1, r}b_{\rm q-1, r} }{2 \Delta E} + n_{\rm H}(s_{\rm r}) X_{\rm q, r} {\sigma}_{\rm p\pi}(E_{\rm q})\;\!{c} \;\! 
    \Bigg\} \ ;  \\
\frac{\partial X}{\partial s}&\Big\vert_{\rm q, r} = \frac{U_{\rm q, r}}{D(E_{\rm q}, s_{\rm r})} \ .
\end{align}
The boundary conditions are
\begin{align}
    X_{1,1} & = X_{\rm 1,{\rm r}_{\rm max}} = n_{\rm p}(E_1, s_1) 
     \ ; \\
    U_{1,1} & = U_{\rm 1,{\rm r}_{\rm max}} = j(E_1, s_1) \ ,
\end{align}
where ${\rm r}_{\rm max}$ is the maximum index on the spatial grid (i.e. the upper boundary of the system), and the CR influx, $j$, is estimated from~\citet{Padovani2009AA}.

\subsection{Primary electrons}
\label{sec:primary_electrons_solver}

In this case, we solve equation~\ref{eq:transport_equation_electron_primary} using the 
same iterative scheme as set out in equations~\ref{eq:outerrk1} and~\ref{eq:outerrk2}, but with equation~\ref{eq:inner_scheme_u} replaced with:
\begin{align}
    \frac{\partial U}{\partial s} \Bigg\vert_{\rm q, r+1} = \ &  \frac{\partial U}{\partial s}\Bigg\vert_{\rm q, r} 
     + \Delta s \;\! 
    \Bigg\{v_{\rm A}(s_{\rm r}) \frac{\partial X}{\partial s}\Bigg\vert_{\rm q, r}  
    \nonumber \\
    & \ \ +\;\!\frac{X_{\rm q+1, r}b_{\rm q+1, r} - X_{\rm q-1, r}b_{\rm q-1, r} }{2 \Delta E} \;\! 
    \Bigg\} \ ,
\end{align}
and an appropriate choice of cooling function $b$.
The boundary conditions are outlined in section~\ref{sec:prim_elecs_info}.

\subsection{Secondary electrons}
\label{sec:secondary_electrons_solver}

The scheme for secondary electrons follows from the discretisation of equation~\ref{eq:transport_equation_electron_secondary}. Again, the form is the same as in section~\ref{sec:primary_protons_solver}, but where the inner equation~\ref{eq:inner_scheme_u} is replaced by:
\begin{align}
    \frac{\partial U}{\partial s} \Bigg\vert_{\rm q, r+1} = \ & \frac{\partial U}{\partial s}\Bigg\vert_{\rm q, r} 
     + \Delta s \;\! \Bigg\{v_{\rm A}(s_{\rm r}) \frac{\partial X}{\partial s}\Bigg\vert_{\rm q, r}  \nonumber \\
      & \hspace*{-0.5cm}
       +\;\! \frac{X_{\rm q+1, r}b_{\rm q+1, r} - X_{\rm q-1, r}b_{\rm q-1, r} }{2 \Delta E}  - Q_{\rm e}|_{\rm q, r} \;\! \Bigg\} \ ,
\end{align}
with the appropriate choice of cooling function $b$.
This time, the boundary conditions on $X$ are $X_{1,1} = X_{\rm 1,{\rm r}_{\rm max}} = 0$, with those on $U$ following similarly as  $U_{1,1} = U_{\rm 1,{\rm r}_{\rm max}} = 0$. The source term depends on the primary proton/electron solutions above, with the $Q_{\rm e}|_{\rm q, r}$ injection term following as per the prescription in section~\ref{subsec:secondary_elecs_theory} at each point of the discretized grid.

\section{Spectral evolution}
\label{sec:spectral_evolution}

\begin{figure}
    \centering 
    \includegraphics[width=\columnwidth]{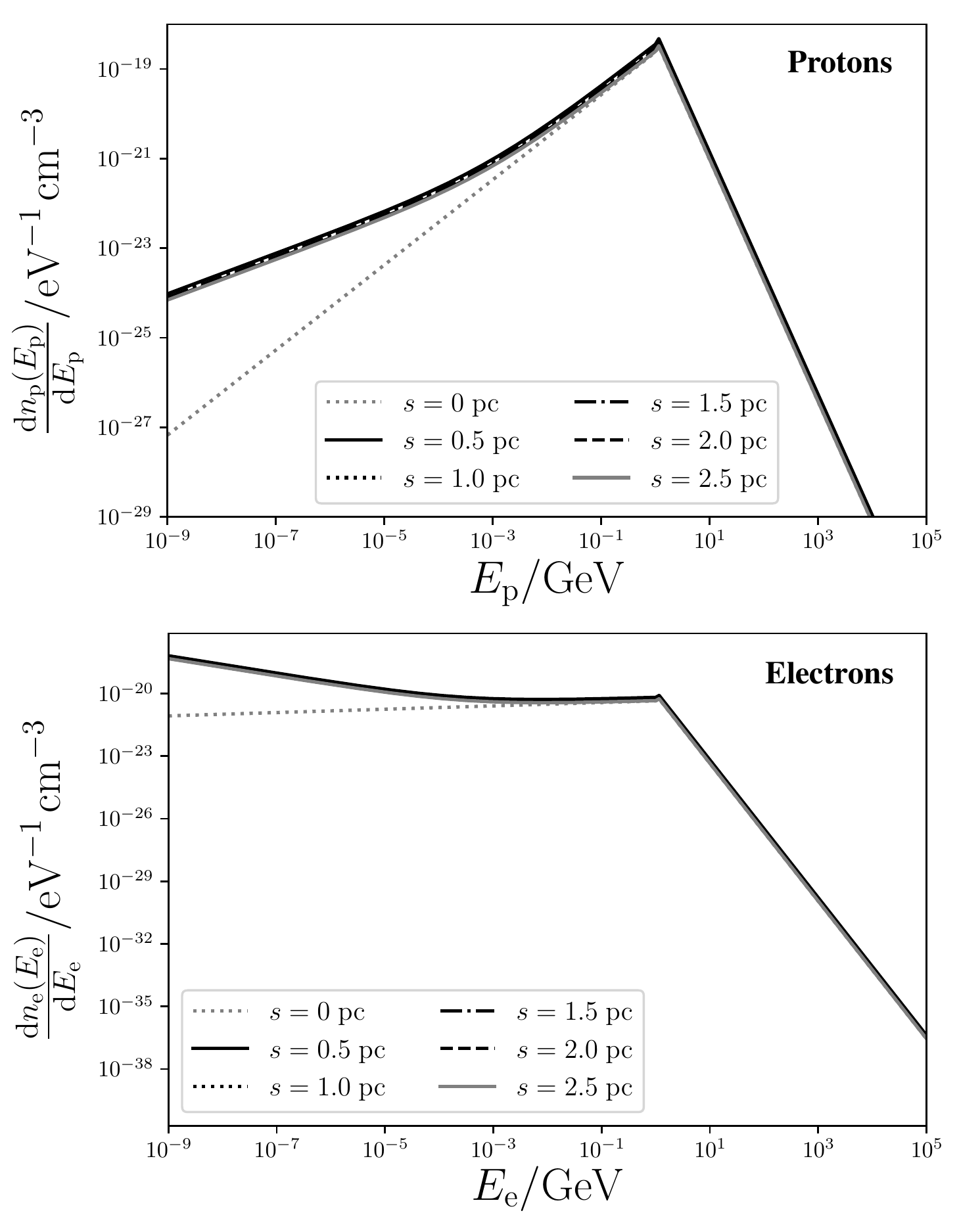}
    \caption{Proton and (primary) electron CR spectra at different positions within the cloud, with $s = 0\,{\rm pc}$ at the boundary {corresponding to a column density of $2.6\times 10^{19} {\rm cm}^{-2}$} (and shows the adopted boundary spectrum -- see section~\ref{sec:cr_spectrum}), {then at distances (column densities) of $s = 0.5 {\rm pc}$ ($1.3\times 10^{21} {\rm cm}^{-2}$), 1.0 pc ($3.5\times 10^{21} {\rm cm}^{-2}$), 1.5 pc ($7.2\times 10^{22} {\rm cm}^{-2}$), 2.0 pc ($1.0\times 10^{23} {\rm cm}^{-2}$) and $2.5$ pc ($1.1\times 10^{23} {\rm cm}^{-2}$).} The spectral evolution of the high-energy component of both spectra is insignificant, as seen by the effective preservation of the spectral shape deep into the cloud. Losses due to ionizations become severe at lower energies, and exhibit a similar energy-dependence for both protons and electrons. This leads to the deformation of both low energy spectra.}
    \label{fig:spectral_evo_full}
\end{figure}

\begin{figure}
    \centering 
    \includegraphics[width=\columnwidth]{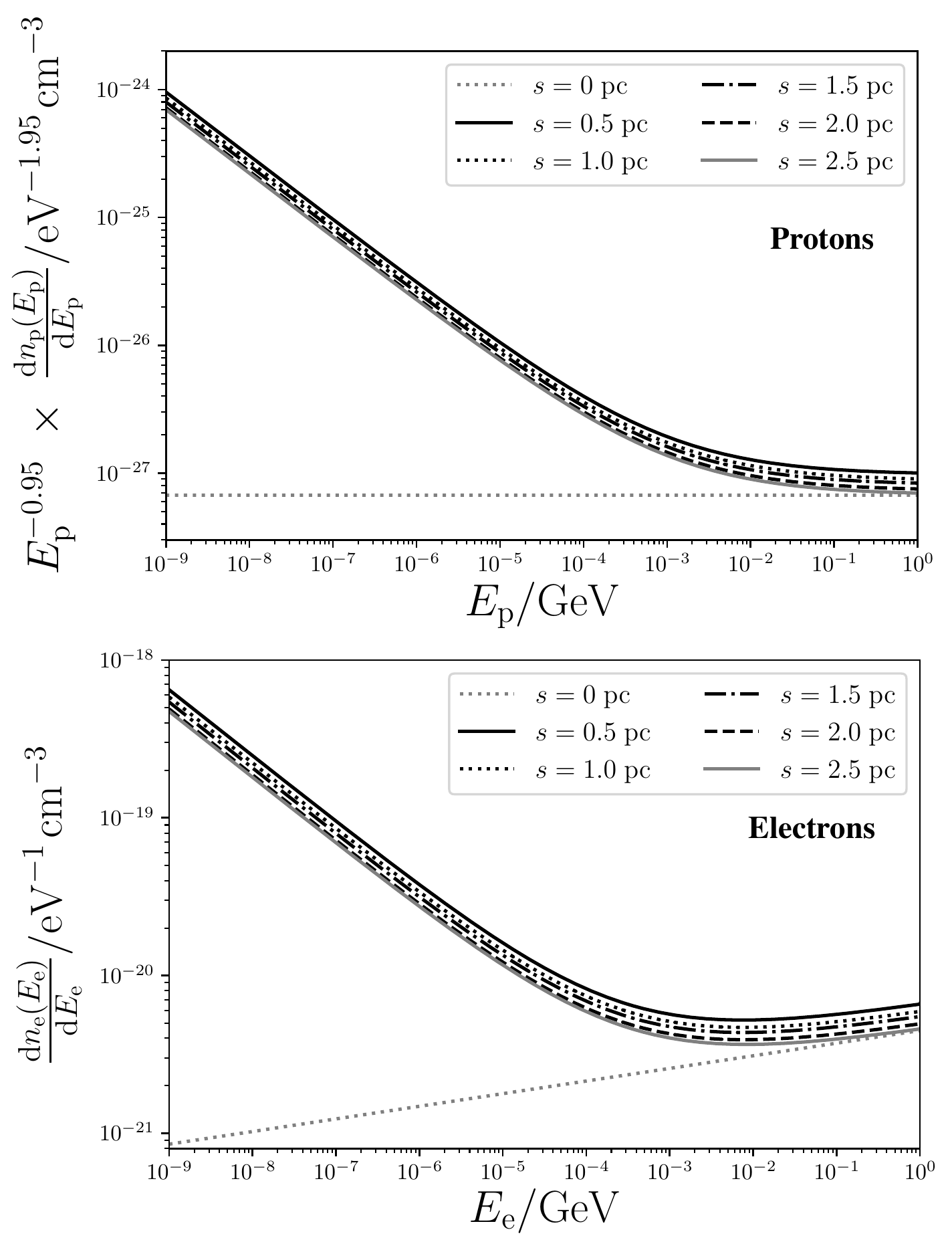}
    \caption{Same as Fig.~\ref{fig:spectral_evo_full}, but showing energies below 1 GeV only. Note that the proton spectrum is multiplied by $E_{\rm p}^{-0.95}$ to more clearly show the spectral evolution.}
    \label{fig:spectral_evo_full_LECR}
\end{figure}

Cooling and absorption losses as CR electrons and protons propagate through a cloud are relatively minimal for most energies of interest in this work. This is particularly the case at high energies, where MHD scattering losses operate for protons (presumably at a rate comparable to their heating power -- see \citealt{Wiener2013_b} for estimates of the heating power by this process), while additionally radiative, free-free and Coulomb losses also arise for electrons (see~\citealt{Owen2018} for associated timescales and length-scales for relevant processes and their energy dependence; also~\citealt{Owen2019AA, Dermer2009book}). However, at low energies, ionization losses are more severe~\citep{Padovani2009AA} and can lead to significant spectral evolution through the cloud. In Fig.~\ref{fig:spectral_evo_full}, we show the spectral evolution for the model MC described in section~\ref{sec:mol_cloud_idealized}. The spectrum shown at $s=0$ is the initial boundary spectrum entering the region of influence of the cloud (see section~\ref{sec:cr_spectrum}). This first shows a moderate increase at all energies, which is then followed by a moderate decrease deeper into the cloud. 
The initial increase near the cloud surface is largely driven by magnetic containment arising from slower CR diffusion in the strengthening magnetic field, while the antagonistic deflective effect of magnetic mirroring deeper into the cloud accounts for the subsequent decrease in CR density.
The spectral evolution at lower energies (see Fig.~\ref{fig:spectral_evo_full_LECR}, which re-plots the sub-GeV component of the spectra in Fig.~\ref{fig:spectral_evo_full} for clarity; note that the proton spectrum is multiplied by $E_{\rm p}^{-0.95}$ in the top panel to better show the spectral change) is driven by ionization losses. This exhibits a similar energy dependence for both protons and electrons (see the adopted ionization cross sections in~\citealt{Padovani2009AA}). 
This is countered, and ultimately overcome, by the mirroring effects that reduce the CR density -- but the decrease due to mirroring is noticeably moderated at these lower energies, which gain from the CR ionization losses.
Although not shown in Fig.~\ref{fig:spectral_evo_full_LECR}, a linear plot with mirroring effects removed
indicates that the low energy CR density is increased by a comparable amount in each spatial increment (of order $10^{-25}\;\text{eV}^{-1}\;\text{cm}^{-3}$ for protons, or $10^{-19}\;\text{eV}^{-1}\;\text{cm}^{-3}$ for electrons). This is in-line with estimations: the ionization cross section peaks at around $\sigma^{\rm ion} \sim 10^{-16}\;\text{cm}^2$ for both protons and electrons~\citep{Padovani2009AA}. In a medium of density $n_{\rm H} = 10~\text{cm}^{-3}$, this corresponds to an interaction rate of $R_{\rm ion} \approx v \sigma_{\rm ion} n_{\rm H}$, for $v$ as the CR effective macroscopic velocity. The time taken for CRs to propagate over a distance of $\ell_{\rm step} = 0.5$ pc (i.e. the distance increments in Figs.~\ref{fig:spectral_evo_full} and~\ref{fig:spectral_evo_full_LECR}) is then $t_{\rm step} \sim \ell_{\rm step}/v$. The number of ionizing interactions experienced by a CR beam in this interval then follows as approximately $t_{\rm step} R_{\rm ion}\sim \ell_{\rm step} \sigma^{\rm ion} n_{\rm H} \sim 10^3$. Thus for CR electrons (protons), of number density initially around $10^{-22}\;\text{eV}^{-1}\;\text{cm}^{-3}$ ($10^{-28}\;\text{eV}^{-1}\;\text{cm}^{-3}$) there would be around $10^3$  ionizing events in a 0.5 pc interval, moving this number of particles from higher energy bands to lower energies, where (without mirroring effects) they would increase the number density by around $10^{-19}\;\text{eV}^{-1}\;\text{cm}^{-3}$ for electrons ($10^{-25}\;\text{eV}^{-1}\;\text{cm}^{-3}$ for protons).

\section{Angular dispersion function and power spectrum}
\label{sec:appendix_sk_s2}
    
We compute the 
    empirical CR diffusion parameter 
    from the power spectrum $\hat{P}(k)$ of the
    magnetic fields permeating a MC 
    (cf. section~\ref{sec:d_ang_sep_method}).
     CR diffusion is determined by the properties and structures of the magnetic field. If the effects due to magnetic field dominate, we can compute the dispersion function $\mathcal{S}_2(\ell)$ from the magnetic field power spectrum $\hat{P}(k)$ under a slab approximation, i.e. where it only depends on the parallel component of ${\boldsymbol k}$~\citep{Hasselmann1968}. 
     Therefore,
     $\hat{P}(k)$
     can be related to the dispersion function $\mathcal{S}_2(\ell)$ (more generally referred to as the second order structure function) of observed polarization angles through the cloud via a Fourier transform, denoted here as $\mathcal{F}[...]$.
     From the Wiener-Khinchin theorem~\citep{Wiener1930_book, Percival1995_book},
    \begin{equation}
        \hat{P}(k) = \mathcal{F}[\mathcal{A}(\ell)] 
        =   \int_{-\infty}^{\infty}\;\!{\rm d}\ell 
        \;\!\mathcal{A}(\ell)  \;\!\exp(ik\ell)
        \ ,
    \end{equation}
    where 
    the autocorrelation function 
    $\mathcal{A}(\ell)$
    in the case of a statistically homogeneous and isotropic field
    can be expressed in terms of
    $\mathcal{A}(\ell) = \mathcal{A}(0) -\mathcal{S}_2(\ell)/2$ 
        \citep[e.g.][]{SchulzDubois1981ApPhy}. 
Therefore,   
    \begin{align}
        \hat{P}(k) &=   \int_{-\infty}^{\infty}\;\!{\rm d}\ell\;\!\left[\mathcal{A}(0) -\frac{1}{2}\mathcal{S}_2(\ell)\right]\;\!\exp(ik\ell) \nonumber \\
        &=
        \mathcal{A}(0)\delta(k)
        -\frac{1}{2}      
        \mathcal{F}\left[\;\! \mathcal{S}_2(\ell)\;\! \right] \ ,         
    \end{align}
    where
    the first term on the right hand side is unphysical when $k = 0$
    and 
    vanishes
    when $k \neq 0$.  
Hence,
      \begin{align}
    \hat{P}(k) &= \frac{1}{2}\mathcal{F}\left[\mathcal{S}_2(\ell)\right] \ , 
    \end{align}
since $\mathcal{S}_2(\ell)$ is real and 
the Fourier Transform of $\mathcal{S}_2(\ell)$ has a Hermitian symmetry.    

\section{Spatial dependence of the empirical diffusion parameter in IC 5146}
\label{sec:appendixc}


Our analysis in section~\ref{sec:3_2} assumes that the spatial dependence of the diffusion parameter 
is derived
only from the spatial variation in the magnetic field strength, and we argue that there is no clear empirical evidence for variation of the diffusion parameter within the IC 5146 region due to the magnetic fluctuations. 
Here we analyze four sub-regions of IC 5146 to demonstrate that there is no strong evidence to support large spatial variations in the diffusion parameter, if adopting a fixed magnetic field strength and gas density.
We select four circular regions (of radius 10', labeled A, B, C and D) around the hub/core filament structures of the region, shown in \cite{Arzoumanian2011} 
and \cite{Wang2017ApJ}. 
We analyze the Rc-, i'-, H- and K- bands separately to determine whether there is any variation between observational bands, and indicate the number of data points in each region, summarized in Table~\ref{tab:diff_param_points}.

\begin{table}
\hspace{-0.5cm}\resizebox{0.80\textwidth}{!}{\begin{minipage}{\textwidth}
\begin{tabular}{|l|c|c|c|}
 \hline
 Region & RA & Dec & Number of points \\
 \hline  \hline
 A & $21^{\rm h} \;\!53^{\rm m}\;\!0^{\rm s}$ & $47^{\rm o} \;\!14'\;\!0''$ & 34 (Rc), -- (i), 120 (H-), 15 (K-) \\
  B & $21^{\rm h} \;\!50^{\rm m}\;\!0^{\rm s}$ & $47^{\rm o} \;\!30'\;\!0''$ & 53 (Rc), -- (i), 87 (H-), 15 (K-) \\
  C & $21^{\rm h} \;\!45^{\rm m}\;\!0^{\rm s}$ & $47^{\rm o} \;\!40'\;\!0''$ & 48 (Rc), 41 (i), 165 (H-), 14 (K-) \\
    D & $21^{\rm h} \;\!48^{\rm m}\;\!0^{\rm s}$ & $48^{\rm o} \;\!10'\;\!0''$ & 70 (Rc), 92 (i), 187 (H-), 10 (K-) \\
 \hline
\end{tabular}
\end{minipage}}
\caption{Location and number of points in each analysis region. Each region selects all points within a radius of 10' from the center. Note that only regions C and D intersect the smaller observation window for the i-band data. Region A roughly corresponds to the Cocoon Nebula.}
\label{tab:diff_param_points}
\end{table}

\begin{figure*}
\vspace*{0.2cm}
    \centering
    \includegraphics[width=0.8\textwidth]{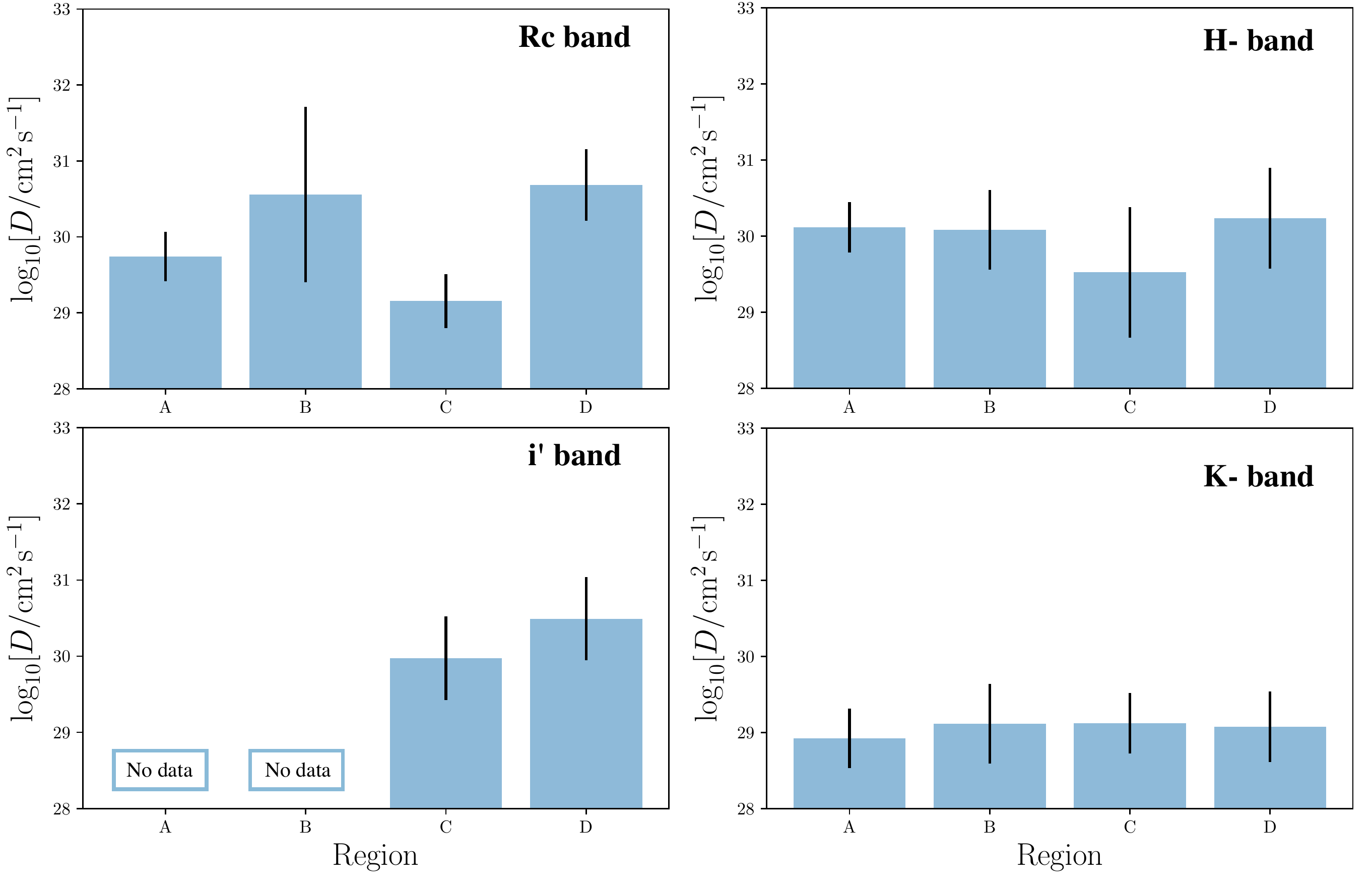}
    \caption{Estimated empirical diffusion parameter values for each of the regions A, B, C and D in each of the four bands. 1$\sigma$ error bars are shown.}
    \label{fig:diff_param_results}
\end{figure*}

The diffusion parameter is computed for each region and band according to section~\ref{sec:d_ang_sep_method}, for which a magnetic field strength on the plane of the sky is estimated from the density using equation~\ref{eq:bfield_ic_region} (a relation specifically for the IC 5146 region), and for a CR energy of 1 GeV. Given the large ranges in volume density throughout each region, we use a characteristic value of $10^3~\text{cm}^{-3}$ for this comparative estimate and note that future dedicated work should more carefully quantify the structures of magnetic and density fields with higher resolution. For our current approach, this is sufficient as our analysis in section~\ref{sec:application} separately accounts for variation in density/magnetic fields through the IC 5146 filamentary structures. We present our comparative results in Fig.~\ref{fig:diff_param_results}, where error bars are at 1$\sigma$ confidence level. 
There is a slight tension
in 
the Rc-band data 
between 
the values derived for the four regions, however there is no evidence of variation in any of the other bands. Moreover, the K-band yields slightly lower values for the diffusion coefficient, which presumably results from each of the bands being sensitive to slightly different scales~\citep[see also][]{Wang2019ApJ} 
and subjected to different opacities into the cloud. We argue there is insufficient evidence to motivate the use of region-specific computations of the diffusion coefficient, but the tension between the bands does merit further work and comparison with simulation results to assess the suitability of each observational band in estimating a local diffusion parameter.

\section{Abundance ratios}
\label{sec:appendixd}

CR ionization rates can be inferred from the astrochemistry of molecular clouds.
The derived rates account for direct CR ionizations together with 
the knock-on events caused by electrons being 
released in the ionization.
The chemical processes can be split into three categories: (1) {\it initiation} steps arise when a species ionized by a CR subsequently reacts with species expected to be present in abundance in the MC environment (e.g. H, ${\rm H}_2$ or C); (2) {\it propagation} steps yield the formation of readily observable species, or facilitate their destruction in a way which would impact on the chemical balance/abundance of that species, and (3) {\it termination} steps result in the neutralization of an ionized species within the chain, without leading to the production of a further species of interest -- 
the `products' in Table~\ref{tab:astrochemistry_rates}, where
the key reactions resulting from a CR ionization event in a MC are listed together with their associated rate coefficient.
CR ionization in a MC proceeds as 
\begin{align*}  
   {\rm H} + {\rm CR} &  \rightarrow {\rm H}^+ + {\rm CR'} 
    \ ;  \\ 
    {\rm H_2} + {\rm CR} &  \rightarrow {\rm H_2}^+ + {\rm CR'} \ ; \\ 
{\rm C} + {\rm CR} & \rightarrow {\rm C}^+ + {\rm CR'}  \ , 
\end{align*}
which lead directly to steps I1, I2 and I3, respectively (see ID keys in Table~\ref{tab:astrochemistry_rates}).


In a steady state the reactions in Table~\ref{tab:astrochemistry_rates} 
  give the abundance ratios for the species of interest, OH$^+$, CO$^+$ and C$^+$ (see section~\ref{sec:cr_ionization}). The key reactions yield
\begin{align}
    \label{eq:ohplus}
    n({\rm OH}^+) &= \frac{n({\rm O}^+) n({\rm H}_2)\;\!k_{\rm P1} + n({\rm H}_3^+) n({\rm O}) k_{\rm P3}}{n({\rm H}_2)\;\!k_{\rm P4} + n_{\rm e}\;\!k_{\rm T1}} \ ; \\
    \label{eq:coplus}
    n({\rm CO}^+) &= \frac{n({\rm C}^+)n({\rm OH})k_{\rm I3}}{n({\rm H}) k_{\rm P5} + n({\rm H}_2)k_{\rm P6} + n_{\rm e}k_{\rm T5}} \ ;  \\
    \label{eq:cplus}
    n({\rm C}^+) &= \frac{n({\rm C}) \zeta^{\rm H}}{n({\rm OH}) k_{\rm I3}} \ ,  
\end{align}
which may be used together with
\begin{align}
    \label{eq:h3plus}
    n({\rm H}_3^+) &= \frac{n({\rm H}_2)\zeta^{\rm H}}{n({\rm CO})k_{\rm P2} + n({\rm O})k_{\rm P3} + n_{\rm e}(k_{\rm T2} + k_{\rm T3} + k_{\rm T4})}
\end{align}
to assess the chemical balances of the required species. The rate coefficients for the relevant processes are given in Table~\ref{tab:astrochemistry_rates}.
These assume that atomic and molecular hydrogen and atomic carbon ionization rates by CRs are all equivalent.
Equation~\ref{eq:ohplus} can be reduced to
\begin{align}
    n(&{\rm OH}^+) \approx \frac{\left[2 + f_{{\rm H}_2}\right] \zeta^{\rm H}}{k_{\rm P4}\;\!f_{{\rm H}_2} + 2 x_i k_{\rm T1}}  \ , 
    \label{eq:ohplus_reduced} 
\end{align}
assuming that the value of $n({\rm H}_3^+)n({\rm CO}) k_{\rm P2}$ (i.e. the rate of process P2) and the ${\rm H}_3^+$ dissociative recombination rate are negligible compared to the ionization rates of H and H$_2$ (which would presumably be valid in a neutral MC with low ionization fraction).
Moreover, equation~\ref{eq:coplus} becomes 
\begin{align}
    n(&{\rm CO}^+) = \frac{x_{\rm C}\zeta^{\rm H}}
    {\left({2k_{\rm P5}}/{f_{{\rm H}_2}} \right) 
    + k_{\rm P6} + 
    \left({2 x_i k_{\rm T5}}/{f_{{\rm H}_2}}\right) }  \ , 
    \label{eq:coplus_reduced} 
\end{align}
where $x_{\rm C} = n({\rm C})/n({\rm H}_2)$. These expressions, together with equation~\ref{eq:cplus}, then yield equations~\ref{eq:ohplus_ratio},~\ref{eq:coplus_ratio} and~\ref{eq:cplus_ratio} in section~\ref{sec:cr_ionization}, where ratios are taken to allow the abundances to be expressed as column densities.

\begin{table*}
\centering
\hspace{1.5cm} \resizebox{0.90\textwidth}{!}{\begin{minipage}{\textwidth}
\begin{tabular}{lllr} 
\hline
\hline
{\bf ID} & ~{\bf Reaction}
  & {\bf Rate coefficient}$^{\rm (a)}$ 
  ~\{\;\! $c_1$/\;\!{cm}$^{3}$\;{s}$^{-1}$, ~$c_2$, ~$c_3$ \;\!/\;\!{K} \;\!\} 
  & {\bf Reference(s)}$^{\rm (b)}$ \\ 
\hline 
\hline
\vspace{-0.2cm}
& & & \\  \vspace*{0.1cm}
{I1} & ~H$^+$ + O $\protect\rightarrow$ O$^+$ + H & 
 $k_{\rm I1} = \{7.0\times 10^{-10},\ 0.26,\ 224.3\}$ 
 & \protect\cite{Stancil1999} \\ \protect\vspace*{0.1cm}
{I2} & ~${{\rm H}_2}^+$ + H$_2$ $\protect\rightarrow$ ${{\rm H}_3}^+$ + H & 
 $k_{\rm I2} = \{2.1\times 10^{-9},\  0.0,\ 0.0\}$ 
 &  \protect\cite{Theard1974} \\  \protect\vspace*{0.1cm}
{I3} & ~{\bf C}$^+$ + {\bf OH} $\protect\rightarrow$ {\bf CO}$^+$ + H  & 
  $k_{\rm I3} = \{7.7\times 10^{-10},\  -0.5,\  0.0\}$ 
  & \protect\cite{Prasad1980} \\  \protect\vspace*{0.1cm}  
{P1} & ~O$^+$ + H$_2$ $\protect\rightarrow$ {\bf OH}$^+$ + H & 
  $k_{\rm P1} = \{1.7\times 10^{-9},\ 0.0,\ 0.0\}$ 
  & \protect\cite{Adams1980} \\  \protect\vspace*{0.1cm}
{P2} & ~${{\rm H}_3}^+$ + {\bf CO} $\protect\rightarrow$ HCO$^+$ + H$_2$ & 
  $k_{\rm P2} = \{1.4\times 10^{-9},\ -0.14,\  -3.4\}^{(d)}$ 
  & \protect\cite{Klippenstein2010} \\ \protect\vspace*{0.1cm}
{P3} & ~${{\rm H}_3}^+$ + O $\protect\rightarrow$ {\bf OH$^+$} + H$_2$ & 
  $k_{\rm P3} = \{8.0\times 10^{-10},\  -0.16,\  1.4\}$ 
  & \protect\cite{Bettens1999} \\ \protect\vspace*{0.1cm}
{P4} & ~{\bf OH}$^{+}$ + H$_2$ $\protect\rightarrow$ H$_2$O$^+$ + H & 
  $k_{\rm P4} = \{1.0\times 10^{-9},\ 0.0,\  0.0\}$ 
  & \protect\cite{Jones1981}\\ \protect\vspace*{0.1cm}
{P5} & ~{\bf CO}$^+$ + H $\protect\rightarrow$ {\bf CO} + H$^+$ & 
  $k_{\rm P5} = \{7.5\times 10^{-10},\  0.0,\  0.0\}$ 
  & \protect\cite{Federer1984}$^{\rm (c)}$ \\   \protect\vspace*{0.1cm}
{P6} & ~{\bf CO}$^+$ + H$_2$ $\protect\rightarrow$ HCO$^+$ + H & 
  $k_{\rm P6} = \{1.8\times 10^{-9},\ 0.0,\  0.0\}$ & 
  \protect\cite{Adams1978} \\   \protect\vspace*{0.1cm} 
{T1} & ~{\bf OH}$^+$ + e$^{-}$ $\protect\rightarrow$ O + H & 
  $k_{\rm T1} = \{3.8\times 10^{-8},\  -0.50, \  0.0\}$ 
  & \protect\cite{Mitchell1990} \\ \protect\vspace*{0.1cm}
{T2} & ~${{\rm H}_3}^+$ + e$^{-}$ $\protect\rightarrow$ H$_2$ + H & 
  $k_{\rm T2} = \{2.3\times 10^{-8},\ -0.52,\ 0.0\}$ 
  & \protect\cite{McCall2004} \\ \protect\vspace*{0.1cm}
{T3} & ~${{\rm H}_3}^+$ + e$^{-}$ $\protect\rightarrow$ 3H & 
  $k_{\rm T3} = \{4.4\times 10^{-8},\ -0.52,\ 0.0\}$ 
  & \protect\cite{McCall2004} \\ \protect\vspace*{0.1cm} 
{T4} & ~${{\rm H}_2}^+$ + e$^{-}$ $\protect\rightarrow$ 2H & 
  $k_{\rm T4} = \{1.6\times 10^{-8},\ -0.43,\  0.0\}$ 
  & \protect\cite{Mitchell1990} \\ \protect\vspace*{0.1cm}
{T5} & ~{\bf CO}$^+$ + e$^{-}$ $\protect\rightarrow$ C + O & 
  $k_{\rm T5} = \{1.0\times 10^{-7},\ -0.46,\  0.0\}$ 
  & \protect\cite{Mitchell1990} \\ 
\hline 
\end{tabular}
\end{minipage}}
\caption{Rate coefficients for principal formation/destruction reactions of chemical tracers in MCs as collated in the UMIST {\protect\citetalias{Millar1997}}~{\protect\citep{Millar1997}} 
  and updated UMIST {\protect\citetalias{McElroy2012}}~{\protect\citep{McElroy2012}} databases. 
Reaction IDs indicate initiation steps (I), propagation steps (P) and termination steps (T). \\
\textbf{Notes:} \\
$^{\rm (a)}$ 
Rate coefficients are given in terms of the parameters 
  $\{c_1, c_2, c_3\}$, 
  for $k_{\rm xx}= c_1 \;\! \left( T/300~{\rm K} \right)^{c_2}\;\!\exp({-c_3/T})$, where ${\rm xx}$ denotes the process ID. \\
$^{\rm (b)}$ Reference(s) of the original source of the rate coefficients 
 in the {\protect\citetalias{McElroy2012}} database~{\protect\citep{McElroy2012}}.
  \\
$^{\rm (c)}$ CO production via CH and CH$_2$ channels can also become important when the metallicity is high (i.e. large CH/OH ratio). 
We do not consider this channel in the reaction network as it is sub-dominant for the objects of interest in this work.
This should, however, be included in the analyses of the high metallicity (above solar) regions.
 \\
$^{\rm (d)}$ The sign of the parameter $c_3$ in process P2 differs 
from other processes due to the molecular geometry. The charge-dipole, charge-quadrupole, and the charge-induced-dipole interactions influence the rate coefficient differently at different temperatures for the capture of reacting species in forming the transition state -- see \protect\cite{Klippenstein2010} for details.
}
\label{tab:astrochemistry_rates} 
\end{table*}

\bibliography{references}{}

\begin{thebibliography}{}
\expandafter\ifx\csname natexlab\endcsname\relax\def\natexlab#1{#1}\fi
\providecommand{\url}[1]{\href{#1}{#1}}
\providecommand{\dodoi}[1]{doi:~\href{http://doi.org/#1}{\nolinkurl{#1}}}
\providecommand{\doeprint}[1]{\href{http://ascl.net/#1}{\nolinkurl{http://ascl.net/#1}}}
\providecommand{\doarXiv}[1]{\href{https://arxiv.org/abs/#1}{\nolinkurl{https://arxiv.org/abs/#1}}}

\bibitem[{{Abraham} {et~al.}(1966){Abraham}, {Brunstein}, \&
  {Cline}}]{Abraham1966}
{Abraham}, P.~B., {Brunstein}, K.~A., \& {Cline}, T.~L. 1966, Physical Review,
  150, 1088, \dodoi{10.1103/PhysRev.150.1088}

\bibitem[{{Adams} {et~al.}(1978){Adams}, {Smith}, \& {Grief}}]{Adams1978}
{Adams}, N.~G., {Smith}, D., \& {Grief}, D. 1978, International Journal of Mass
  Spectrometry and Ion Processes, 26, 405, \dodoi{10.1016/0020-7381(78)80059-X}

\bibitem[{Adams {et~al.}(1980)Adams, Smith, \& Paulson}]{Adams1980}
Adams, N.~G., Smith, D., \& Paulson, J.~F. 1980, The Journal of Chemical
  Physics, 72, 288, \dodoi{10.1063/1.438893}

\bibitem[{Aharonian {et~al.}(2012)Aharonian, Bykov, Parizot, Ptuskin, \&
  Watson}]{Aharonian2012}
Aharonian, F., Bykov, A., Parizot, E., Ptuskin, V., \& Watson, A. 2012, \ssr,
  166, 97, \dodoi{10.1007/s11214-011-9770-3}

\bibitem[{{Aharonian} {et~al.}(2006){Aharonian}, {Akhperjanian}, {Bazer-Bachi},
  {Beilicke}, {Benbow}, {Berge}, {Bernl{\"o}hr}, {Boisson}, {Bolz}, {Borrel},
  {Braun}, {Breitling}, {Brown}, {Chadwick}, {Chounet}, {Cornils},
  {Costamante}, {Degrange}, {Dickinson}, {Djannati-Ata{\"\i}}, {Drury},
  {Dubus}, {Emmanoulopoulos}, {Espigat}, {Feinstein}, {Fontaine}, {Fuchs},
  {Funk}, {Gallant}, {Giebels}, {Gillessen}, {Glicenstein}, {Goret},
  {Hadjichristidis}, {Hauser}, {Hauser}, {Heinzelmann}, {Henri}, {Hermann},
  {Hinton}, {Hofmann}, {Holleran}, {Horns}, {Jacholkowska}, {de Jager},
  {Kh{\'e}lifi}, {Klages}, {Komin}, {Konopelko}, {Latham}, {Le Gallou},
  {Lemi{\`e}re}, {Lemoine-Goumard}, {Leroy}, {Lohse}, {Marcowith}, {Martin},
  {Martineau-Huynh}, {Masterson}, {McComb}, {de Naurois}, {Nolan}, {Noutsos},
  {Orford}, {Osborne}, {Ouchrif}, {Panter}, {Pelletier}, {Pita},
  {P{\"u}hlhofer}, {Punch}, {Raubenheimer}, {Raue}, {Raux}, {Rayner}, {Reimer},
  {Reimer}, {Ripken}, {Rob}, {Rolland}, {Rowell}, {Sahakian}, {Saug{\'e}},
  {Schlenker}, {Schlickeiser}, {Schuster}, {Schwanke}, {Siewert}, {Sol},
  {Spangler}, {Steenkamp}, {Stegmann}, {Tavernet}, {Terrier}, {Th{\'e}oret},
  {Tluczykont}, {van Eldik}, {Vasileiadis}, {Venter}, {Vincent}, {V{\"o}lk}, \&
  {Wagner}}]{Aharonian2006}
{Aharonian}, F., {Akhperjanian}, A.~G., {Bazer-Bachi}, A.~R., {et~al.} 2006,
  \nat, 439, 695, \dodoi{10.1038/nature04467}

\bibitem[{{Albertsson} {et~al.}(2018){Albertsson}, {Kauffmann}, \&
  {Menten}}]{Albertsson2018}
{Albertsson}, T., {Kauffmann}, J., \& {Menten}, K.~M. 2018, \apj, 868, 40,
  \dodoi{10.3847/1538-4357/aae775}

\bibitem[{Almeida {et~al.}(1968)Almeida, Rushbrooke, Scharenguivel, Behrens,
  Blobel, Borecka, Dehne, Dfaz, Knies, Schmitt, Str\"omer, \&
  Swanson}]{Almeida1968PR}
Almeida, S.~P., Rushbrooke, J.~G., Scharenguivel, J.~H., {et~al.} 1968, Phys.
  Rev., 174, 1638, \dodoi{10.1103/PhysRev.174.1638}

\bibitem[{{Arzoumanian} {et~al.}(2011){Arzoumanian}, {Andr{\'e}}, {Didelon},
  {K{\"o}nyves}, {Schneider}, {Men'shchikov}, {Sousbie}, {Zavagno}, {Bontemps},
  {di Francesco}, {Griffin}, {Hennemann}, {Hill}, {Kirk}, {Martin}, {Minier},
  {Molinari}, {Motte}, {Peretto}, {Pezzuto}, {Spinoglio}, {Ward-Thompson},
  {White}, \& {Wilson}}]{Arzoumanian2011}
{Arzoumanian}, D., {Andr{\'e}}, P., {Didelon}, P., {et~al.} 2011, \aap, 529,
  L6, \dodoi{10.1051/0004-6361/201116596}

\bibitem[{{Axford} {et~al.}(1977){Axford}, {Leer}, \&
  {Skadron}}]{Axford1977ICRC}
{Axford}, W.~I., {Leer}, E., \& {Skadron}, G. 1977, in International Cosmic Ray
  Conference, Vol.~11, International Cosmic Ray Conference, 132

\bibitem[{{Basu}(2000)}]{Basu2000}
{Basu}, S. 2000, \apj, 540, L103, \dodoi{10.1086/312885}

\bibitem[{{Basu} {et~al.}(2009){Basu}, {Ciolek}, {Dapp}, \&
  {Wurster}}]{Basu2009}
{Basu}, S., {Ciolek}, G.~E., {Dapp}, W.~B., \& {Wurster}, J. 2009, \na, 14,
  483, \dodoi{10.1016/j.newast.2009.01.004}

\bibitem[{{Bell}(1978{\natexlab{a}})}]{Bell1978MNRASI}
{Bell}, A.~R. 1978{\natexlab{a}}, \mnras, 182, 147,
  \dodoi{10.1093/mnras/182.2.147}

\bibitem[{{Bell}(1978{\natexlab{b}})}]{Bell1978MNRASII}
---. 1978{\natexlab{b}}, \mnras, 182, 443, \dodoi{10.1093/mnras/182.3.443}

\bibitem[{{Berezinskii} {et~al.}(1990){Berezinskii}, {Bulanov}, {Dogiel}, \&
  {Ptuskin}}]{Berezinskii1990}
{Berezinskii}, V.~S., {Bulanov}, S.~V., {Dogiel}, V.~A., \& {Ptuskin}, V.~S.
  1990, {Astrophysics of cosmic rays} (Amsterdam: North-Holland)

\bibitem[{{Bergin} \& {Tafalla}(2007)}]{Bergin2007}
{Bergin}, E.~A., \& {Tafalla}, M. 2007, \araa, 45, 339,
  \dodoi{10.1146/annurev.astro.45.071206.100404}

\bibitem[{{Berrington} \& {Dermer}(2003)}]{Berrington2003ApJ}
{Berrington}, R.~C., \& {Dermer}, C.~D. 2003, \apj, 594, 709,
  \dodoi{10.1086/376981}

\bibitem[{Bettens {et~al.}(1999)Bettens, Hansen, \& Collins}]{Bettens1999}
Bettens, R. P.~A., Hansen, T.~A., \& Collins, M.~A. 1999, The Journal of
  Chemical Physics, 111, 6322, \dodoi{10.1063/1.479937}

\bibitem[{{Bisbas} {et~al.}(2017){Bisbas}, {van Dishoeck}, {Papadopoulos},
  {Sz{\H u}cs}, {Bialy}, \& {Zhang}}]{Bisbas2017}
{Bisbas}, T.~G., {van Dishoeck}, E.~F., {Papadopoulos}, P.~P., {et~al.} 2017,
  \apj, 839, 90, \dodoi{10.3847/1538-4357/aa696d}

\bibitem[{{Black} {et~al.}(1978){Black}, {Hartquist}, \&
  {Dalgarno}}]{Black1978ApJ}
{Black}, J.~H., {Hartquist}, T.~W., \& {Dalgarno}, A. 1978, \apj, 224, 448,
  \dodoi{10.1086/156392}

\bibitem[{{Blandford} \& {Ostriker}(1978)}]{Blandford1978ApJ}
{Blandford}, R.~D., \& {Ostriker}, J.~P. 1978, \apjl, 221, L29,
  \dodoi{10.1086/182658}

\bibitem[{{Blasi}(2011)}]{Blasi2011crpa}
{Blasi}, P. 2011, in Cosmic Rays for Particle and Astroparticle Physics, ed.
  S.~{Giani}, C.~{Leroy}, \& P.~G. {Rancoita}, 493--506,
  \dodoi{10.1142/9789814329033_0061}

\bibitem[{{Blattnig} {et~al.}(2000){Blattnig}, {Swaminathan}, {Kruger}, {Ngom},
  {Norbury}, \& {Tripathi}}]{Blattnig2000}
{Blattnig}, S.~R., {Swaminathan}, S.~R., {Kruger}, A.~T., {et~al.} 2000,
  {Parameterized Cross Sections for Pion Production in Proton-Proton
  Collisions}, Tech. rep.

\bibitem[{{Blumenthal}(1970)}]{Blumenthal1970PRD}
{Blumenthal}, G.~R. 1970, \prd, 1, 1596, \dodoi{10.1103/PhysRevD.1.1596}

\bibitem[{{Brown} \& {Marscher}(1977)}]{Brown1977ApJ}
{Brown}, R.~L., \& {Marscher}, A.~P. 1977, \apj, 212, 659,
  \dodoi{10.1086/155088}

\bibitem[{{Brunstein}(1965)}]{Brunstein1965}
{Brunstein}, K.~A. 1965, Physical Review, 137, 757,
  \dodoi{10.1103/PhysRev.137.B757}

\bibitem[{{Bykov} {et~al.}(2020){Bykov}, {Marcowith}, {Amato}, {Kalyashova},
  {Kruijssen}, \& {Waxman}}]{Bykov2020SSRv}
{Bykov}, A.~M., {Marcowith}, A., {Amato}, E., {et~al.} 2020, \ssr, 216, 42,
  \dodoi{10.1007/s11214-020-00663-0}

\bibitem[{{Carlberg} \& {Pudritz}(1990)}]{Carlberg1990}
{Carlberg}, R.~G., \& {Pudritz}, R.~E. 1990, \mnras, 247, 353

\bibitem[{{Casanova} {et~al.}(2010){Casanova}, {Aharonian}, {Fukui}, {Gabici},
  {Jones}, {Kawamura}, {Onishi}, {Rowell}, {Sano}, {Torii}, \&
  {Yamamoto}}]{Casanova2010PASJ}
{Casanova}, S., {Aharonian}, F.~A., {Fukui}, Y., {et~al.} 2010, \pasj, 62, 769,
  \dodoi{10.1093/pasj/62.3.769}

\bibitem[{{Caselli} {et~al.}(1998){Caselli}, {Walmsley}, {Terzieva}, \&
  {Herbst}}]{Caselli1998}
{Caselli}, P., {Walmsley}, C.~M., {Terzieva}, R., \& {Herbst}, E. 1998, \apj,
  499, 234, \dodoi{10.1086/305624}

\bibitem[{{Cashman} \& {Clemens}(2014)}]{Cashman2014ApJ}
{Cashman}, L.~R., \& {Clemens}, D.~P. 2014, \apj, 793, 126,
  \dodoi{10.1088/0004-637X/793/2/126}

\bibitem[{{Ceccarelli} {et~al.}(1998){Ceccarelli}, {Caux}, {Wolfire},
  {Rudolph}, {Nisini}, {Saraceno}, \& {White}}]{Ceccarelli1998}
{Ceccarelli}, C., {Caux}, E., {Wolfire}, M., {et~al.} 1998, \aap, 331, L17

\bibitem[{{C{\'e}cere} {et~al.}(2016){C{\'e}cere}, {Vel{\'a}zquez}, {Araudo},
  {De Colle}, {Esquivel}, {Carrasco-Gonz{\'a}lez}, \&
  {Rodr{\'\i}guez}}]{Cecere2016ApJ}
{C{\'e}cere}, M., {Vel{\'a}zquez}, P.~F., {Araudo}, A.~T., {et~al.} 2016, \apj,
  816, 64, \dodoi{10.3847/0004-637X/816/2/64}

\bibitem[{{Cesarsky} \& {Volk}(1978)}]{Cesarsky1978}
{Cesarsky}, C.~J., \& {Volk}, H.~J. 1978, \aap, 70, 367

\bibitem[{{Chandran}(2000)}]{Chandran2000}
{Chandran}, B.~D.~G. 2000, \apj, 529, 513, \dodoi{10.1086/308232}

\bibitem[{{Chandrasekhar} \& {Fermi}(1953)}]{Chandrasekhar1953ApJ}
{Chandrasekhar}, S., \& {Fermi}, E. 1953, \apj, 118, 113,
  \dodoi{10.1086/145731}

\bibitem[{{Chernyshov} {et~al.}(2018){Chernyshov}, {Caselli}, {Cheng},
  {Dogiel}, {Ivlev}, \& {Ko}}]{Chernyshov2018NPPP}
{Chernyshov}, D.~O., {Caselli}, P., {Cheng}, K.~S., {et~al.} 2018, Nuclear and
  Particle Physics Proceedings, 297-299, 80,
  \dodoi{10.1016/j.nuclphysbps.2018.07.012}

\bibitem[{{Ching} {et~al.}(2017){Ching}, {Lai}, {Zhang}, {Girart}, {Qiu}, \&
  {Liu}}]{Ching2017}
{Ching}, T.-C., {Lai}, S.-P., {Zhang}, Q., {et~al.} 2017, \apj, 838, 121,
  \dodoi{10.3847/1538-4357/aa65cc}

\bibitem[{{Colafrancesco} \& {Marchegiani}(2008)}]{Colafrancesco2008}
{Colafrancesco}, S., \& {Marchegiani}, P. 2008, \aap, 484, 51,
  \dodoi{10.1051/0004-6361:20078428}

\bibitem[{{Commer{\c{c}}on} {et~al.}(2019){Commer{\c{c}}on}, {Marcowith}, \&
  {Dubois}}]{Commercon2019AA}
{Commer{\c{c}}on}, B., {Marcowith}, A., \& {Dubois}, Y. 2019, \aap, 622, A143,
  \dodoi{10.1051/0004-6361/201833809}

\bibitem[{Cooley \& Tukey(1965)}]{Cooley1965_book}
Cooley, J.~W., \& Tukey, J.~W. 1965, Mathematics of Computation, 19, 297

\bibitem[{{Coud{\'e}} {et~al.}(2019){Coud{\'e}}, {Bastien}, {Houde}, {Sadavoy},
  {Friesen}, {Di Francesco}, {Johnstone}, {Mairs}, {Hasegawa}, \&
  {Kwon}}]{Coude2019}
{Coud{\'e}}, S., {Bastien}, P., {Houde}, M., {et~al.} 2019, \apj, 877, 88,
  \dodoi{10.3847/1538-4357/ab1b23}

\bibitem[{{Crutcher}(1999)}]{Crutcher1999}
{Crutcher}, R.~M. 1999, \apj, 520, 706, \dodoi{10.1086/307483}

\bibitem[{{Crutcher}(2012)}]{Crutcher2012ARAA}
---. 2012, \araa, 50, 29, \dodoi{10.1146/annurev-astro-081811-125514}

\bibitem[{{Crutcher} {et~al.}(1993){Crutcher}, {Troland}, {Goodman}, {Heiles},
  {Kazes}, \& {Myers}}]{Crutcher1993ApJ}
{Crutcher}, R.~M., {Troland}, T.~H., {Goodman}, A.~A., {et~al.} 1993, \apj,
  407, 175, \dodoi{10.1086/172503}

\bibitem[{{Crutcher} {et~al.}(2010){Crutcher}, {Wandelt}, {Heiles},
  {Falgarone}, \& {Troland}}]{Crutcher2010}
{Crutcher}, R.~M., {Wandelt}, B., {Heiles}, C., {Falgarone}, E., \& {Troland},
  T.~H. 2010, \apj, 725, 466, \dodoi{10.1088/0004-637X/725/1/466}

\bibitem[{{Cummings} {et~al.}(2016){Cummings}, {Stone}, {Heikkila}, {Lal},
  {Webber}, {J{\'o}hannesson}, {Moskalenko}, {Orlando}, \&
  {Porter}}]{Cummings2016ApJ}
{Cummings}, A.~C., {Stone}, E.~C., {Heikkila}, B.~C., {et~al.} 2016, \apj, 831,
  18, \dodoi{10.3847/0004-637X/831/1/18}

\bibitem[{{Dalgarno}(2006)}]{Dalgarno2006PNAS}
{Dalgarno}, A. 2006, Proceedings of the National Academy of Science, 103,
  12269, \dodoi{10.1073/pnas.0602117103}

\bibitem[{{Dalgarno} \& {McCray}(1972)}]{Dalgarno1972}
{Dalgarno}, A., \& {McCray}, R.~A. 1972, \araa, 10, 375,
  \dodoi{10.1146/annurev.aa.10.090172.002111}

\bibitem[{{Davis}(1951)}]{Davis1951PhRv}
{Davis}, L. 1951, Physical Review, 81, 890, \dodoi{10.1103/PhysRev.81.890.2}

\bibitem[{{Dermer} \& {Menon}(2009)}]{Dermer2009book}
{Dermer}, C.~D., \& {Menon}, G. 2009, {High Energy Radiation from Black Holes:
  Gamma Rays, Cosmic Rays, and Neutrinos}

\bibitem[{{Desch} {et~al.}(2004){Desch}, {Connolly}, \&
  {Srinivasan}}]{Desch2004ApJ}
{Desch}, S.~J., {Connolly}, Jr., H.~C., \& {Srinivasan}, G. 2004, \apj, 602,
  528, \dodoi{10.1086/380831}

\bibitem[{{Dib} {et~al.}(2010){Dib}, {Shadmehri}, {Padoan}, {Maheswar}, {Ojha},
  \& {Khajenabi}}]{Dib2010MNRAS}
{Dib}, S., {Shadmehri}, M., {Padoan}, P., {et~al.} 2010, \mnras, 405, 401,
  \dodoi{10.1111/j.1365-2966.2010.16451.x}

\bibitem[{{Dogel} \& {Sharov}(1990)}]{Dogel1990AA}
{Dogel}, V.~A., \& {Sharov}, G.~S. 1990, \aap, 229, 259

\bibitem[{{Dogiel} {et~al.}(2018){Dogiel}, {Chernyshov}, {Ivlev}, {Malyshev},
  {Strong}, \& {Cheng}}]{Dogiel2018ApJ}
{Dogiel}, V.~A., {Chernyshov}, D.~O., {Ivlev}, A.~V., {et~al.} 2018, \apj, 868,
  114, \dodoi{10.3847/1538-4357/aae827}

\bibitem[{{Dogiel} {et~al.}(2015){Dogiel}, {Chernyshov}, {Kiselev}, {Nobukawa},
  {Cheng}, {Hui}, {Ko}, {Nobukawa}, \& {Tsuru}}]{Dogiel2015ApJ}
{Dogiel}, V.~A., {Chernyshov}, D.~O., {Kiselev}, A.~M., {et~al.} 2015, \apj,
  809, 48, \dodoi{10.1088/0004-637X/809/1/48}

\bibitem[{{Dolginov} \& {Mitrofanov}(1976)}]{Dolginov1976ApSS}
{Dolginov}, A.~Z., \& {Mitrofanov}, I.~G. 1976, \apss, 43, 291,
  \dodoi{10.1007/BF00640010}

\bibitem[{{Doty} {et~al.}(2002){Doty}, {van Dishoeck}, {van der Tak}, \&
  {Boonman}}]{Doty2002}
{Doty}, S.~D., {van Dishoeck}, E.~F., {van der Tak}, F.~F.~S., \& {Boonman},
  A.~M.~S. 2002, \aap, 389, 446, \dodoi{10.1051/0004-6361:20020597}

\bibitem[{{Draine}(2011)}]{Draine2011Book}
{Draine}, B.~T. 2011, {Physics of the Interstellar and Intergalactic Medium}

\bibitem[{{Draine} \& {Weingartner}(1996)}]{Draine1996ApJ}
{Draine}, B.~T., \& {Weingartner}, J.~C. 1996, \apj, 470, 551,
  \dodoi{10.1086/177887}

\bibitem[{{Draine} \& {Weingartner}(1997)}]{Draine1997ApJ}
---. 1997, \apj, 480, 633, \dodoi{10.1086/304008}

\bibitem[{{Dunham} {et~al.}(2015){Dunham}, {Allen}, {Evans},
  {Broekhoven-Fiene}, {Cieza}, {Di Francesco}, {Gutermuth}, {Harvey},
  {Hatchell}, {Heiderman}, {Huard}, {Johnstone}, {Kirk}, {Matthews}, {Miller},
  {Peterson}, \& {Young}}]{Dunham2015ApJS}
{Dunham}, M.~M., {Allen}, L.~E., {Evans}, Neal~J., I., {et~al.} 2015, \apjs,
  220, 11, \dodoi{10.1088/0067-0049/220/1/11}

\bibitem[{{Dzib} {et~al.}(2018){Dzib}, {Loinard}, {Ortiz-Le{\'o}n},
  {Rodr{\'\i}guez}, \& {Galli}}]{Dzib2018ApJ}
{Dzib}, S.~A., {Loinard}, L., {Ortiz-Le{\'o}n}, G.~N., {Rodr{\'\i}guez}, L.~F.,
  \& {Galli}, P. A.~B. 2018, \apj, 867, 151, \dodoi{10.3847/1538-4357/aae687}

\bibitem[{{Elmegreen}(1979)}]{Elmegreen1979ApJ}
{Elmegreen}, B.~G. 1979, \apj, 232, 729, \dodoi{10.1086/157333}

\bibitem[{{Falceta-Gon{\c c}alves} {et~al.}(2003){Falceta-Gon{\c c}alves}, {de
  Juli}, \& {Jatenco-Pereira}}]{Falceta2003}
{Falceta-Gon{\c c}alves}, D., {de Juli}, M.~C., \& {Jatenco-Pereira}, V. 2003,
  \apj, 597, 970, \dodoi{10.1086/378584}

\bibitem[{{Farmer} \& {Goldreich}(2004)}]{Farmer2004}
{Farmer}, A.~J., \& {Goldreich}, P. 2004, \apj, 604, 671,
  \dodoi{10.1086/382040}

\bibitem[{{Federer} {et~al.}(1984){Federer}, {Villinger}, {Howorka},
  {Lindinger}, {Tosis}, {Bassi}, \& {Ferguson}}]{Federer1984}
{Federer}, W., {Villinger}, H., {Howorka}, F., {et~al.} 1984, Physical Review
  Letters, 52, 2084, \dodoi{10.1103/PhysRevLett.52.2084}

\bibitem[{{Federman} {et~al.}(1996){Federman}, {Weber}, \&
  {Lambert}}]{Federman1996ApJ}
{Federman}, S.~R., {Weber}, J., \& {Lambert}, D.~L. 1996, \apj, 463, 181,
  \dodoi{10.1086/177233}

\bibitem[{{Federrath}(2016)}]{Federrath2016}
{Federrath}, C. 2016, in Journal of Physics Conference Series, Vol. 719,
  Journal of Physics Conference Series, 012002,
  \dodoi{10.1088/1742-6596/719/1/012002}

\bibitem[{{Federrath} \& {Klessen}(2013)}]{Federrath2013ApJ}
{Federrath}, C., \& {Klessen}, R.~S. 2013, \apj, 763, 51,
  \dodoi{10.1088/0004-637X/763/1/51}

\bibitem[{{Felice} \& {Kulsrud}(2001)}]{Felice2001ApJ}
{Felice}, G.~M., \& {Kulsrud}, R.~M. 2001, \apj, 553, 198,
  \dodoi{10.1086/320651}

\bibitem[{{Fermi}(1949)}]{Fermi1949}
{Fermi}, E. 1949, Physical Review, 75, 1169, \dodoi{10.1103/PhysRev.75.1169}

\bibitem[{{Ferri{\`e}re}(2001)}]{Ferriere2001RvMP}
{Ferri{\`e}re}, K.~M. 2001, Reviews of Modern Physics, 73, 1031,
  \dodoi{10.1103/RevModPhys.73.1031}

\bibitem[{{Field} {et~al.}(1969){Field}, {Goldsmith}, \& {Habing}}]{Field1969}
{Field}, G.~B., {Goldsmith}, D.~W., \& {Habing}, H.~J. 1969, \apjl, 155, L149,
  \dodoi{10.1086/180324}

\bibitem[{{Fuente} \& {Mart{\'\i}n-Pintado}(1997)}]{Fuente1997}
{Fuente}, A., \& {Mart{\'\i}n-Pintado}, J. 1997, \apjl, 477, L107,
  \dodoi{10.1086/310532}

\bibitem[{{Fujita} {et~al.}(2013){Fujita}, {Kimura}, \& {Ohira}}]{Fujita2013}
{Fujita}, Y., {Kimura}, S., \& {Ohira}, Y. 2013, \mnras, 432, 1434,
  \dodoi{10.1093/mnras/stt563}

\bibitem[{{Fujita} \& {Ohira}(2011)}]{Fujita2011}
{Fujita}, Y., \& {Ohira}, Y. 2011, \apj, 738, 182,
  \dodoi{10.1088/0004-637X/738/2/182}

\bibitem[{{Gabici}(2011)}]{Gabici2011crpa}
{Gabici}, S. 2011, in Cosmic Rays for Particle and Astroparticle Physics, ed.
  S.~{Giani}, C.~{Leroy}, \& P.~G. {Rancoita}, 343--351,
  \dodoi{10.1142/9789814329033_0044}

\bibitem[{{Gabici} {et~al.}(2009){Gabici}, {Aharonian}, \&
  {Casanova}}]{Gabici2009MNRAS}
{Gabici}, S., {Aharonian}, F.~A., \& {Casanova}, S. 2009, \mnras, 396, 1629,
  \dodoi{10.1111/j.1365-2966.2009.14832.x}

\bibitem[{{Gaches} {et~al.}(2019){Gaches}, {Offner}, \& {Bisbas}}]{Gaches2019}
{Gaches}, B. A.~L., {Offner}, S. S.~R., \& {Bisbas}, T.~G. 2019, \apj, 878,
  105, \dodoi{10.3847/1538-4357/ab20c7}

\bibitem[{Gaggero(2012)}]{Gaggero2012}
Gaggero, D. 2012, Cosmic Ray Diffusion in the Galaxy and Diffuse Gamma
  Emission, Springer Theses (Springer: Berlin)

\bibitem[{{Gaggero} {et~al.}(2017){Gaggero}, {Grasso}, {Marinelli}, {Taoso}, \&
  {Urbano}}]{Gaggero2017}
{Gaggero}, D., {Grasso}, D., {Marinelli}, A., {Taoso}, M., \& {Urbano}, A.
  2017, Physical Review Letters, 119, 031101,
  \dodoi{10.1103/PhysRevLett.119.031101}

\bibitem[{{Gaia Collaboration} {et~al.}(2018){Gaia Collaboration}, {Brown},
  {Vallenari}, {Prusti}, {de Bruijne}, {Babusiaux}, {Bailer-Jones}, {Biermann},
  {Evans}, {Eyer}, {Jansen}, {Jordi}, {Klioner}, {Lammers}, {Lindegren},
  {Luri}, {Mignard}, {Panem}, {Pourbaix}, {Randich}, {Sartoretti}, {Siddiqui},
  {Soubiran}, {van Leeuwen}, {Walton}, {Arenou}, {Bastian}, {Cropper},
  {Drimmel}, {Katz}, {Lattanzi}, {Bakker}, {Cacciari}, {Casta{\~n}eda},
  {Chaoul}, {Cheek}, {De Angeli}, {Fabricius}, {Guerra}, {Holl}, {Masana},
  {Messineo}, {Mowlavi}, {Nienartowicz}, {Panuzzo}, {Portell}, {Riello},
  {Seabroke}, {Tanga}, {Th{\'e}venin}, {Gracia-Abril}, {Comoretto},
  {Garcia-Reinaldos}, {Teyssier}, {Altmann}, {Andrae}, {Audard},
  {Bellas-Velidis}, {Benson}, {Berthier}, {Blomme}, {Burgess}, {Busso},
  {Carry}, {Cellino}, {Clementini}, {Clotet}, {Creevey}, {Davidson}, {De
  Ridder}, {Delchambre}, {Dell'Oro}, {Ducourant},
  {Fern{\'a}ndez-Hern{\'a}ndez}, {Fouesneau}, {Fr{\'e}mat}, {Galluccio},
  {Garc{\'\i}a-Torres}, {Gonz{\'a}lez-N{\'u}{\~n}ez}, {Gonz{\'a}lez-Vidal},
  {Gosset}, {Guy}, {Halbwachs}, {Hambly}, {Harrison}, {Hern{\'a}ndez},
  {Hestroffer}, {Hodgkin}, {Hutton}, {Jasniewicz}, {Jean-Antoine-Piccolo},
  {Jordan}, {Korn}, {Krone-Martins}, {Lanzafame}, {Lebzelter}, {L{\"o}ffler},
  {Manteiga}, {Marrese}, {Mart{\'\i}n-Fleitas}, {Moitinho}, {Mora}, {Muinonen},
  {Osinde}, {Pancino}, {Pauwels}, {Petit}, {Recio-Blanco}, {Richards},
  {Rimoldini}, {Robin}, {Sarro}, {Siopis}, {Smith}, {Sozzetti}, {S{\"u}veges},
  {Torra}, {van Reeven}, {Abbas}, {Abreu Aramburu}, {Accart}, {Aerts},
  {Altavilla}, {{\'A}lvarez}, {Alvarez}, {Alves}, {Anderson}, {Andrei},
  {Anglada Varela}, {Antiche}, {Antoja}, {Arcay}, {Astraatmadja}, {Bach},
  {Baker}, {Balaguer-N{\'u}{\~n}ez}, {Balm}, {Barache}, {Barata}, {Barbato},
  {Barblan}, {Barklem}, {Barrado}, {Barros}, {Barstow}, {Bartholom{\'e}
  Mu{\~n}oz}, {Bassilana}, {Becciani}, {Bellazzini}, {Berihuete}, {Bertone},
  {Bianchi}, {Bienaym{\'e}}, {Blanco-Cuaresma}, {Boch}, {Boeche}, {Bombrun},
  {Borrachero}, {Bossini}, {Bouquillon}, {Bourda}, {Bragaglia}, {Bramante},
  {Breddels}, {Bressan}, {Brouillet}, {Br{\"u}semeister}, {Brugaletta},
  {Bucciarelli}, {Burlacu}, {Busonero}, {Butkevich}, {Buzzi}, {Caffau},
  {Cancelliere}, {Cannizzaro}, {Cantat-Gaudin}, {Carballo}, {Carlucci},
  {Carrasco}, {Casamiquela}, {Castellani}, {Castro-Ginard}, {Charlot},
  {Chemin}, {Chiavassa}, {Cocozza}, {Costigan}, {Cowell}, {Crifo}, {Crosta},
  {Crowley}, {Cuypers}, {Dafonte}, {Damerdji}, {Dapergolas}, {David}, {David},
  {de Laverny}, {De Luise}, {De March}, {de Martino}, {de Souza}, {de Torres},
  {Debosscher}, {del Pozo}, {Delbo}, {Delgado}, {Delgado}, {Di Matteo},
  {Diakite}, {Diener}, {Distefano}, {Dolding}, {Drazinos}, {Dur{\'a}n},
  {Edvardsson}, {Enke}, {Eriksson}, {Esquej}, {Eynard Bontemps}, {Fabre},
  {Fabrizio}, {Faigler}, {Falc{\~a}o}, {Farr{\`a}s Casas}, {Federici},
  {Fedorets}, {Fernique}, {Figueras}, {Filippi}, {Findeisen}, {Fonti},
  {Fraile}, {Fraser}, {Fr{\'e}zouls}, {Gai}, {Galleti}, {Garabato},
  {Garc{\'\i}a-Sedano}, {Garofalo}, {Garralda}, {Gavel}, {Gavras}, {Gerssen},
  {Geyer}, {Giacobbe}, {Gilmore}, {Girona}, {Giuffrida}, {Glass}, {Gomes},
  {Granvik}, {Gueguen}, {Guerrier}, {Guiraud}, {Guti{\'e}rrez-S{\'a}nchez},
  {Haigron}, {Hatzidimitriou}, {Hauser}, {Haywood}, {Heiter}, {Helmi}, {Heu},
  {Hilger}, {Hobbs}, {Hofmann}, {Holland}, {Huckle}, {Hypki}, {Icardi},
  {Jan{\ss}en}, {Jevardat de Fombelle}, {Jonker}, {Juh{\'a}sz}, {Julbe},
  {Karampelas}, {Kewley}, {Klar}, {Kochoska}, {Kohley}, {Kolenberg},
  {Kontizas}, {Kontizas}, {Koposov}, {Kordopatis}, {Kostrzewa-Rutkowska},
  {Koubsky}, {Lambert}, {Lanza}, {Lasne}, {Lavigne}, {Le Fustec}, {Le
  Poncin-Lafitte}, {Lebreton}, {Leccia}, {Leclerc}, {Lecoeur-Taibi},
  {Lenhardt}, {Leroux}, {Liao}, {Licata}, {Lindstr{\o}m}, {Lister}, {Livanou},
  {Lobel}, {L{\'o}pez}, {Managau}, {Mann}, {Mantelet}, {Marchal}, {Marchant},
  {Marconi}, {Marinoni}, {Marschalk{\'o}}, {Marshall}, {Martino}, {Marton},
  {Mary}, {Massari}, {Matijevi{\v{c}}}, {Mazeh}, {McMillan}, {Messina},
  {Michalik}, {Millar}, {Molina}, {Molinaro}, {Moln{\'a}r}, {Montegriffo},
  {Mor}, {Morbidelli}, {Morel}, {Morris}, {Mulone}, {Muraveva}, {Musella},
  {Nelemans}, {Nicastro}, {Noval}, {O'Mullane}, {Ord{\'e}novic},
  {Ord{\'o}{\~n}ez-Blanco}, {Osborne}, {Pagani}, {Pagano}, {Pailler},
  {Palacin}, {Palaversa}, {Panahi}, {Pawlak}, {Piersimoni}, {Pineau}, {Plachy},
  {Plum}, {Poggio}, {Poujoulet}, {Pr{\v{s}}a}, {Pulone}, {Racero}, {Ragaini},
  {Rambaux}, {Ramos-Lerate}, {Regibo}, {Reyl{\'e}}, {Riclet}, {Ripepi}, {Riva},
  {Rivard}, {Rixon}, {Roegiers}, {Roelens}, {Romero-G{\'o}mez}, {Rowell},
  {Royer}, {Ruiz-Dern}, {Sadowski}, {Sagrist{\`a} Sell{\'e}s}, {Sahlmann},
  {Salgado}, {Salguero}, {Sanna}, {Santana-Ros}, {Sarasso}, {Savietto},
  {Schultheis}, {Sciacca}, {Segol}, {Segovia}, {S{\'e}gransan}, {Shih},
  {Siltala}, {Silva}, {Smart}, {Smith}, {Solano}, {Solitro}, {Sordo}, {Soria
  Nieto}, {Souchay}, {Spagna}, {Spoto}, {Stampa}, {Steele},
  {Steidelm{\"u}ller}, {Stephenson}, {Stoev}, {Suess}, {Surdej}, {Szabados},
  {Szegedi-Elek}, {Tapiador}, {Taris}, {Tauran}, {Taylor}, {Teixeira},
  {Terrett}, {Teyssand ier}, {Thuillot}, {Titarenko}, {Torra Clotet}, {Turon},
  {Ulla}, {Utrilla}, {Uzzi}, {Vaillant}, {Valentini}, {Valette}, {van Elteren},
  {Van Hemelryck}, {van Leeuwen}, {Vaschetto}, {Vecchiato}, {Veljanoski},
  {Viala}, {Vicente}, {Vogt}, {von Essen}, {Voss}, {Votruba}, {Voutsinas},
  {Walmsley}, {Weiler}, {Wertz}, {Wevers}, {Wyrzykowski}, {Yoldas},
  {{\v{Z}}erjal}, {Ziaeepour}, {Zorec}, {Zschocke}, {Zucker}, {Zurbach}, \&
  {Zwitter}}]{Gaia2018AA}
{Gaia Collaboration}, {Brown}, A.~G.~A., {Vallenari}, A., {et~al.} 2018, \aap,
  616, A1, \dodoi{10.1051/0004-6361/201833051}

\bibitem[{{Galli} {et~al.}(2002){Galli}, {Walmsley}, \&
  {Gon{\c{c}}alves}}]{Galli2002AA}
{Galli}, D., {Walmsley}, M., \& {Gon{\c{c}}alves}, J. 2002, \aap, 394, 275,
  \dodoi{10.1051/0004-6361:20021125}

\bibitem[{{Gammie} \& {Ostriker}(1996)}]{Gammie1996}
{Gammie}, C.~F., \& {Ostriker}, E.~C. 1996, \apj, 466, 814,
  \dodoi{10.1086/177556}

\bibitem[{{Gao} {et~al.}(2015){Gao}, {Xu}, \& {Law}}]{Gao2015ApJ}
{Gao}, Y., {Xu}, H., \& {Law}, C.~K. 2015, \apj, 799, 227,
  \dodoi{10.1088/0004-637X/799/2/227}

\bibitem[{{Geballe} {et~al.}(2007){Geballe}, {Indriolo}, {McCall}, \&
  {Oka}}]{Geballe2007AAS}
{Geballe}, T.~R., {Indriolo}, N., {McCall}, B.~J., \& {Oka}, T. 2007, in
  Bulletin of the American Astronomical Society, Vol.~39, American Astronomical
  Society Meeting Abstracts, 985

\bibitem[{{Geballe} {et~al.}(1999){Geballe}, {McCall}, {Hinkle}, \&
  {Oka}}]{Geballe1999ApJ}
{Geballe}, T.~R., {McCall}, B.~J., {Hinkle}, K.~H., \& {Oka}, T. 1999, \apj,
  510, 251, \dodoi{10.1086/306580}

\bibitem[{{Ginzburg} \& {Syrovatskii}(1964)}]{Ginzburg1964}
{Ginzburg}, V.~L., \& {Syrovatskii}, S.~I. 1964, {The Origin of Cosmic Rays}

\bibitem[{{Girart} {et~al.}(2006){Girart}, {Rao}, \& {Marrone}}]{Girart2006}
{Girart}, J.~M., {Rao}, R., \& {Marrone}, D.~P. 2006, Science, 313, 812,
  \dodoi{10.1126/science.1129093}

\bibitem[{{Glassgold} \& {Langer}(1974)}]{Glassgold1974ApJ}
{Glassgold}, A.~E., \& {Langer}, W.~D. 1974, \apj, 193, 73,
  \dodoi{10.1086/153130}

\bibitem[{{Goldsmith}(2001)}]{Goldsmith2001ApJ}
{Goldsmith}, P.~F. 2001, \apj, 557, 736, \dodoi{10.1086/322255}

\bibitem[{{Goldsmith} \& {Langer}(1978)}]{Goldsmith1978}
{Goldsmith}, P.~F., \& {Langer}, W.~D. 1978, \apj, 222, 881,
  \dodoi{10.1086/156206}

\bibitem[{{H.~E.~S.~S. Collaboration} {et~al.}(2018{\natexlab{a}}){H.~E.~S.~S.
  Collaboration}, {Abdalla}, {Abramowski}, {Aharonian}, {Ait Benkhali},
  {Ang{\"u}ner}, {Arakawa}, {Arrieta}, {Aubert}, {Backes}, {Balzer}, {Barnard},
  {Becherini}, {Becker Tjus}, {Berge}, {Bernhard}, {Bernl{\"o}hr}, {Blackwell},
  {B{\"o}ttcher}, {Boisson}, {Bolmont}, {Bonnefoy}, {Bordas}, {Bregeon},
  {Brun}, {Brun}, {Bryan}, {B{\"u}chele}, {Bulik}, {Capasso}, {Carrigan},
  {Caroff}, {Carosi}, {Casanova}, {Cerruti}, {Chakraborty}, {Chaves}, {Chen},
  {Chevalier}, {Colafrancesco}, {Condon}, {Conrad}, {Davids}, {Decock}, {Deil},
  {Devin}, {deWilt}, {Dirson}, {Djannati-Ata{\"\i}}, {Domainko}, {Donath},
  {Drury}, {Dutson}, {Dyks}, {Edwards}, {Egberts}, {Eger}, {Emery},
  {Ernenwein}, {Eschbach}, {Farnier}, {Fegan}, {Fernand es}, {Fiasson},
  {Fontaine}, {F{\"o}rster}, {Funk}, {F{\"u}{\ss}ling}, {Gabici}, {Gallant},
  {Garrigoux}, {Gast}, {Gat{\'e}}, {Giavitto}, {Giebels}, {Glawion},
  {Glicenstein}, {Gottschall}, {Grondin}, {Hahn}, {Haupt}, {Hawkes},
  {Heinzelmann}, {Henri}, {Hermann}, {Hinton}, {Hofmann}, {Hoischen}, {Holch},
  {Holler}, {Horns}, {Ivascenko}, {Iwasaki}, {Jacholkowska}, {Jamrozy},
  {Jankowsky}, {Jankowsky}, {Jingo}, {Jouvin}, {Jung-Richardt}, {Kastendieck},
  {Katarzy{\'n}ski}, {Katsuragawa}, {Katz}, {Kerszberg}, {Khangulyan},
  {Kh{\'e}lifi}, {King}, {Klepser}, {Klochkov}, {Klu{\'z}niak}, {Komin},
  {Kosack}, {Krakau}, {Kraus}, {Kr{\"u}ger}, {Laffon}, {Lamanna}, {Lau},
  {Lees}, {Lefaucheur}, {Lemi{\`e}re}, {Lemoine-Goumard}, {Lenain}, {Leser},
  {Lohse}, {Lorentz}, {Liu}, {L{\'o}pez-Coto}, {Lypova}, {Marandon},
  {Malyshev}, {Marcowith}, {Mariaud}, {Marx}, {Maurin}, {Maxted}, {Mayer},
  {Meintjes}, {Meyer}, {Mitchell}, {Moderski}, {Mohamed}, {Mohrmann},
  {Mor{\r{a}}}, {Moulin}, {Murach}, {Nakashima}, {de Naurois}, {Ndiyavala},
  {Niederwanger}, {Niemiec}, {Oakes}, {O'Brien}, {Odaka}, {Ohm}, {Ostrowski},
  {Oya}, {Padovani}, {Panter}, {Parsons}, {Paz Arribas}, {Pekeur}, {Pelletier},
  {Perennes}, {Petrucci}, {Peyaud}, {Piel}, {Pita}, {Poireau}, {Poon},
  {Prokhorov}, {Prokoph}, {P{\"u}hlhofer}, {Punch}, {Quirrenbach}, {Raab},
  {Rauth}, {Reimer}, {Reimer}, {Renaud}, {de los Reyes}, {Rieger}, {Rinchiuso},
  {Romoli}, {Rowell}, {Rudak}, {Rulten}, {Safi-Harb}, {Sahakian}, {Saito},
  {Sanchez}, {Santangelo}, {Sasaki}, {Schand ri}, {Schlickeiser},
  {Sch{\"u}ssler}, {Schulz}, {Schwanke}, {Schwemmer}, {Seglar-Arroyo},
  {Settimo}, {Seyffert}, {Shafi}, {Shilon}, {Shiningayamwe}, {Simoni}, {Sol},
  {Spanier}, {Spir-Jacob}, {Stawarz}, {Steenkamp}, {Stegmann}, {Steppa},
  {Sushch}, {Takahashi}, {Tavernet}, {Tavernier}, {Taylor}, {Terrier},
  {Tibaldo}, {Tiziani}, {Tluczykont}, {Trichard}, {Tsirou}, {Tsuji}, {Tuffs},
  {Uchiyama}, {van der Walt}, {van Eldik}, {van Rensburg}, {van Soelen},
  {Vasileiadis}, {Veh}, {Venter}, {Viana}, {Vincent}, {Vink}, {Voisin},
  {V{\"o}lk}, {Vuillaume}, {Wadiasingh}, {Wagner}, {Wagner}, {Wagner}, {White},
  {Wierzcholska}, {Willmann}, {W{\"o}rnlein}, {Wouters}, {Yang}, {Zaborov},
  {Zacharias}, {Zanin}, {Zdziarski}, {Zech}, {Zefi}, {Ziegler}, {Zorn}, \&
  {{\.Z}ywucka}}]{HESS2018b}
{H.~E.~S.~S. Collaboration}, {Abdalla}, H., {Abramowski}, A., {et~al.}
  2018{\natexlab{a}}, \aap, 612, A1, \dodoi{10.1051/0004-6361/201732098}

\bibitem[{{H.~E.~S.~S. Collaboration} {et~al.}(2018{\natexlab{b}}){H.~E.~S.~S.
  Collaboration}, {Abdalla}, {Abramowski}, {Aharonian}, {Ait Benkhali},
  {Akhperjanian}, {Andersson}, {Ang{\"u}ner}, {Arakawa}, {Arrieta}, {Aubert},
  {Backes}, {Balzer}, {Barnard}, {Becherini}, {Becker Tjus}, {Berge},
  {Bernhard}, {Bernl{\"o}hr}, {Blackwell}, {B{\"o}ttcher}, {Boisson},
  {Bolmont}, {Bonnefoy}, {Bordas}, {Bregeon}, {Brun}, {Brun}, {Bryan},
  {B{\"u}chele}, {Bulik}, {Capasso}, {Carr}, {Casanova}, {Cerruti},
  {Chakraborty}, {Chaves}, {Chen}, {Chevalier}, {Coffaro}, {Colafrancesco},
  {Cologna}, {Condon}, {Conrad}, {Cui}, {Davids}, {Decock}, {Degrange}, {Deil},
  {Devin}, {deWilt}, {Dirson}, {Djannati-Ata{\"\i}}, {Domainko}, {Donath},
  {Drury}, {Dutson}, {Dyks}, {Edwards}, {Egberts}, {Eger}, {Ernenwein},
  {Eschbach}, {Farnier}, {Fegan}, {Fernand es}, {Fiasson}, {Fontaine},
  {F{\"o}rster}, {Funk}, {F{\"u}{\ss}ling}, {Gabici}, {Gallant}, {Garrigoux},
  {Giavitto}, {Giebels}, {Glicenstein}, {Gottschall}, {Goyal}, {Grondin},
  {Hahn}, {Haupt}, {Hawkes}, {Heinzelmann}, {Henri}, {Hermann}, {Hinton},
  {Hofmann}, {Hoischen}, {Holch}, {Holler}, {Horns}, {Ivascenko}, {Iwasaki},
  {Jacholkowska}, {Jamrozy}, {Janiak}, {Jankowsky}, {Jankowsky}, {Jingo},
  {Jogler}, {Jouvin}, {Jung-Richardt}, {Kastendieck}, {Katarzy{\'n}ski},
  {Katsuragawa}, {Katz}, {Kerszberg}, {Khangulyan}, {Kh{\'e}lifi}, {King},
  {Klepser}, {Klochkov}, {Klu{\'z}niak}, {Kolitzus}, {Komin}, {Kosack},
  {Krakau}, {Kraus}, {Kr{\"u}ger}, {Laffon}, {Lamanna}, {Lau}, {Lees},
  {Lefaucheur}, {Lefranc}, {Lemi{\`e}re}, {Lemoine-Goumard}, {Lenain}, {Leser},
  {Lohse}, {Lorentz}, {Liu}, {L{\'o}pez-Coto}, {Lypova}, {Marandon},
  {Marcowith}, {Mariaud}, {Marx}, {Maurin}, {Maxted}, {Mayer}, {Meintjes},
  {Meyer}, {Mitchell}, {Moderski}, {Mohamed}, {Mohrmann}, {Mor{\r{a}}},
  {Moulin}, {Murach}, {Nakashima}, {de Naurois}, {Niederwanger}, {Niemiec},
  {Oakes}, {O'Brien}, {Odaka}, {Ohm}, {Ostrowski}, {Oya}, {Padovani}, {Panter},
  {Parsons}, {Pekeur}, {Pelletier}, {Perennes}, {Petrucci}, {Peyaud}, {Piel},
  {Pita}, {Poon}, {Prokhorov}, {Prokoph}, {P{\"u}hlhofer}, {Punch},
  {Quirrenbach}, {Raab}, {Rauth}, {Reimer}, {Reimer}, {Renaud}, {de los Reyes},
  {Richter}, {Rieger}, {Romoli}, {Rowell}, {Rudak}, {Rulten}, {Sahakian},
  {Saito}, {Salek}, {Sanchez}, {Santangelo}, {Sasaki}, {Schlickeiser},
  {Sch{\"u}ssler}, {Schulz}, {Schwanke}, {Schwemmer}, {Seglar-Arroyo},
  {Settimo}, {Seyffert}, {Shafi}, {Shilon}, {Simoni}, {Sol}, {Spanier},
  {Spengler}, {Spies}, {Stawarz}, {Steenkamp}, {Stegmann}, {Stycz}, {Sushch},
  {Takahashi}, {Tavernet}, {Tavernier}, {Taylor}, {Terrier}, {Tibaldo},
  {Tiziani}, {Tluczykont}, {Trichard}, {Tsuji}, {Tuffs}, {Uchiyama}, {van der
  Walt}, {van Eldik}, {van Rensburg}, {van Soelen}, {Vasileiadis}, {Veh},
  {Venter}, {Viana}, {Vincent}, {Vink}, {Voisin}, {V{\"o}lk}, {Vuillaume},
  {Wadiasingh}, {Wagner}, {Wagner}, {Wagner}, {White}, {Wierzcholska},
  {Willmann}, {W{\"o}rnlein}, {Wouters}, {Yang}, {Zaborov}, {Zacharias},
  {Zanin}, {Zdziarski}, {Zech}, {Zefi}, {Ziegler}, \&
  {{\.Z}ywucka}}]{HESS2018a}
---. 2018{\natexlab{b}}, \aap, 612, A9, \dodoi{10.1051/0004-6361/201730824}

\bibitem[{Hairer {et~al.}(1993)Hairer, N{\o}rsett, \& Wanner}]{Hairer1993book}
Hairer, E., N{\o}rsett, S., \& Wanner, G. 1993, Solving Ordinary Differential
  Equations II: Stiff and Differential-Algebraic Problems, Lecture Notes in
  Economic and Mathematical Systems (Springer)

\bibitem[{{Harju} {et~al.}(2008){Harju}, {Juvela}, {Schlemmer}, {Haikala},
  {Lehtinen}, \& {Mattila}}]{Harju2008}
{Harju}, J., {Juvela}, M., {Schlemmer}, S., {et~al.} 2008, \aap, 482, 535,
  \dodoi{10.1051/0004-6361:20079259}

\bibitem[{Harris {et~al.}(2020)Harris, Millman, van~der Walt, Gommers,
  Virtanen, Cournapeau, Wieser, Taylor, Berg, Smith, Kern, Picus, Hoyer, van
  Kerkwijk, Brett, Haldane, del R{'{\i}}o, Wiebe, Peterson,
  G{'{e}}rard-Marchant, Sheppard, Reddy, Weckesser, Abbasi, Gohlke, \&
  Oliphant}]{harris2020}
Harris, C.~R., Millman, K.~J., van~der Walt, S.~J., {et~al.} 2020, Nature, 585,
  357, \dodoi{10.1038/s41586-020-2649-2}

\bibitem[{{Hartquist} {et~al.}(1978){Hartquist}, {Black}, \&
  {Dalgarno}}]{Hartquist1978MNRAS}
{Hartquist}, T.~W., {Black}, J.~H., \& {Dalgarno}, A. 1978, \mnras, 185, 643,
  \dodoi{10.1093/mnras/185.3.643}

\bibitem[{{Harvey} {et~al.}(2008){Harvey}, {Huard}, {J{\o}rgensen},
  {Gutermuth}, {Mamajek}, {Bourke}, {Mer{\'\i}n}, {Cieza}, {Brooke}, {Chapman},
  {Alcal{\'a}}, {Allen}, {Evans}, {Di Francesco}, \& {Kirk}}]{Harvey2008ApJ}
{Harvey}, P.~M., {Huard}, T.~L., {J{\o}rgensen}, J.~K., {et~al.} 2008, \apj,
  680, 495, \dodoi{10.1086/587687}

\bibitem[{{Hasselmann} \& {Wibberenz}(1968)}]{Hasselmann1968}
{Hasselmann}, K., \& {Wibberenz}, G. 1968, Z. Geophys, 328, 269

\bibitem[{{Hayakawa} {et~al.}(1961){Hayakawa}, {Nishimura}, \&
  {Takayanagi}}]{Hayakawa1961PASJ}
{Hayakawa}, S., {Nishimura}, S., \& {Takayanagi}, T. 1961, \pasj, 13, 184

\bibitem[{{Hildebrand} {et~al.}(2009){Hildebrand}, {Kirby}, {Dotson}, {Houde},
  \& {Vaillancourt}}]{Hildebrand2009ApJ}
{Hildebrand}, R.~H., {Kirby}, L., {Dotson}, J.~L., {Houde}, M., \&
  {Vaillancourt}, J.~E. 2009, \apj, 696, 567,
  \dodoi{10.1088/0004-637X/696/1/567}

\bibitem[{{Hillas}(2006)}]{Hillas2006astroph}
{Hillas}, A.~M. 2006, arXiv e-prints, astro.
\newblock \doarXiv{astro-ph/0607109}

\bibitem[{{Hollenbach} {et~al.}(2012){Hollenbach}, {Kaufman}, {Neufeld},
  {Wolfire}, \& {Goicoechea}}]{Hollenbach2012}
{Hollenbach}, D., {Kaufman}, M.~J., {Neufeld}, D., {Wolfire}, M., \&
  {Goicoechea}, J.~R. 2012, \apj, 754, 105, \dodoi{10.1088/0004-637X/754/2/105}

\bibitem[{{Houde} {et~al.}(2009){Houde}, {Vaillancourt}, {Hildebrand},
  {Chitsazzadeh}, \& {Kirby}}]{Houde2009ApJ}
{Houde}, M., {Vaillancourt}, J.~E., {Hildebrand}, R.~H., {Chitsazzadeh}, S., \&
  {Kirby}, L. 2009, \apj, 706, 1504, \dodoi{10.1088/0004-637X/706/2/1504}

\bibitem[{{Hull} {et~al.}(2017){Hull}, {Mocz}, {Burkhart}, {Goodman}, {Girart},
  {Cort{\'e}s}, {Hernquist}, {Springel}, {Li}, \& {Lai}}]{Hull2017}
{Hull}, C.~L.~H., {Mocz}, P., {Burkhart}, B., {et~al.} 2017, \apjl, 842, L9,
  \dodoi{10.3847/2041-8213/aa71b7}

\bibitem[{{Indriolo}(2012)}]{Indriolo2012RSPTA}
{Indriolo}, N. 2012, Philosophical Transactions of the Royal Society of London
  Series A, 370, 5142, \dodoi{10.1098/rsta.2012.0022}

\bibitem[{{Indriolo}(2013)}]{Indriolo2013ASSP}
{Indriolo}, N. 2013, in Cosmic Rays in Star-Forming Environments, ed. D.~F.
  {Torres} \& O.~{Reimer}, Vol.~34, 83, \dodoi{10.1007/978-3-642-35410-6_7}

\bibitem[{{Indriolo} {et~al.}(2007){Indriolo}, {Geballe}, {Oka}, \&
  {McCall}}]{Indriolo2007ApJ}
{Indriolo}, N., {Geballe}, T.~R., {Oka}, T., \& {McCall}, B.~J. 2007, \apj,
  671, 1736, \dodoi{10.1086/523036}

\bibitem[{{Indriolo} \& {McCall}(2012)}]{Indriolo2012}
{Indriolo}, N., \& {McCall}, B.~J. 2012, \apj, 745, 91,
  \dodoi{10.1088/0004-637X/745/1/91}

\bibitem[{{Indriolo} {et~al.}(2015){Indriolo}, {Neufeld}, {Gerin}, {Schilke},
  {Benz}, {Winkel}, {Menten}, {Chambers}, {Black}, {Bruderer}, {Falgarone},
  {Godard}, {Goicoechea}, {Gupta}, {Lis}, {Ossenkopf}, {Persson},
  {Sonnentrucker}, {van der Tak}, {van Dishoeck}, {Wolfire}, \&
  {Wyrowski}}]{Indriolo2015}
{Indriolo}, N., {Neufeld}, D.~A., {Gerin}, M., {et~al.} 2015, \apj, 800, 40,
  \dodoi{10.1088/0004-637X/800/1/40}

\bibitem[{{Ivlev} {et~al.}(2018){Ivlev}, {Dogiel}, {Chernyshov}, {Caselli},
  {Ko}, \& {Cheng}}]{Ivlev2018ApJ}
{Ivlev}, A.~V., {Dogiel}, V.~A., {Chernyshov}, D.~O., {et~al.} 2018, \apj, 855,
  23, \dodoi{10.3847/1538-4357/aaadb9}

\bibitem[{{Jaacks} {et~al.}(2019){Jaacks}, {Finkelstein}, \&
  {Bromm}}]{Jaacks2019MNRAS}
{Jaacks}, J., {Finkelstein}, S.~L., \& {Bromm}, V. 2019, \mnras, 488, 2202,
  \dodoi{10.1093/mnras/stz1529}

\bibitem[{{Jacob} \& {Pfrommer}(2017)}]{Jacob2017}
{Jacob}, S., \& {Pfrommer}, C. 2017, \mnras, 467, 1449,
  \dodoi{10.1093/mnras/stx131}

\bibitem[{{Jenkins} {et~al.}(1983){Jenkins}, {Jura}, \&
  {Loewenstein}}]{Jenkins1983}
{Jenkins}, E.~B., {Jura}, M., \& {Loewenstein}, M. 1983, \apj, 270, 88,
  \dodoi{10.1086/161100}

\bibitem[{{Jokipii}(1966)}]{Jokipii1966ApJ}
{Jokipii}, J.~R. 1966, \apj, 146, 480, \dodoi{10.1086/148912}

\bibitem[{{Jones}(2014)}]{Jones2014ApJL}
{Jones}, D.~I. 2014, \apjl, 792, L14, \dodoi{10.1088/2041-8205/792/1/L14}

\bibitem[{Jones {et~al.}(1981)Jones, Birkinshaw, \& Twiddy}]{Jones1981}
Jones, J., Birkinshaw, K., \& Twiddy, N. 1981, Chemical Physics Letters, 77,
  484 , \dodoi{https://doi.org/10.1016/0009-2614(81)85191-3}

\bibitem[{{Jura}(1975)}]{Jura1975}
{Jura}, M. 1975, \apj, 197, 581, \dodoi{10.1086/153546}

\bibitem[{{Juvela} \& {Ysard}(2011)}]{Juvella2011ApJ}
{Juvela}, M., \& {Ysard}, N. 2011, \apj, 739, 63,
  \dodoi{10.1088/0004-637X/739/2/63}

\bibitem[{{Kafexhiu} {et~al.}(2014){Kafexhiu}, {Aharonian}, {Taylor}, \&
  {Vila}}]{Kafexhiu2014}
{Kafexhiu}, E., {Aharonian}, F., {Taylor}, A.~M., \& {Vila}, G.~S. 2014, \prd,
  90, 123014, \dodoi{10.1103/PhysRevD.90.123014}

\bibitem[{{Kalv{\={a}}ns}(2018)}]{Kalvans2018ApJS}
{Kalv{\={a}}ns}, J. 2018, \apjs, 239, 6, \dodoi{10.3847/1538-4365/aae527}

\bibitem[{{Kirk} {et~al.}(1988){Kirk}, {Schlickeiser}, \&
  {Schneider}}]{Kirk1988}
{Kirk}, J.~G., {Schlickeiser}, R., \& {Schneider}, P. 1988, \apj, 328, 269,
  \dodoi{10.1086/166290}

\bibitem[{{Klippenstein} {et~al.}(2010){Klippenstein}, {Georgievskii}, \&
  {McCall}}]{Klippenstein2010}
{Klippenstein}, S.~J., {Georgievskii}, Y., \& {McCall}, B.~J. 2010, Journal of
  Physical Chemistry A, 114, 278, \dodoi{10.1021/jp908500h}

\bibitem[{{Ko}(1992)}]{Ko1992}
{Ko}, C.-M. 1992, \aap, 259, 377

\bibitem[{{Koch} {et~al.}(2014){Koch}, {Tang}, {Ho}, {Zhang}, {Girart}, {Chen},
  {Frau}, {Li}, {Li}, {Liu}, {Padovani}, {Qiu}, {Yen}, {Chen}, {Ching}, {Lai},
  \& {Rao}}]{Koch2014}
{Koch}, P.~M., {Tang}, Y.-W., {Ho}, P.~T.~P., {et~al.} 2014, \apj, 797, 99,
  \dodoi{10.1088/0004-637X/797/2/99}

\bibitem[{{Kossmann} {et~al.}(1990){Kossmann}, {Schwarzkopf}, \&
  {Schmidt}}]{Kossmann1990}
{Kossmann}, H., {Schwarzkopf}, O., \& {Schmidt}, V. 1990, Journal of Physics B
  Atomic Molecular Physics, 23, 301, \dodoi{10.1088/0953-4075/23/2/012}

\bibitem[{{Kotera} \& {Olinto}(2011)}]{Kotera2011ARAA}
{Kotera}, K., \& {Olinto}, A.~V. 2011, \araa, 49, 119,
  \dodoi{10.1146/annurev-astro-081710-102620}

\bibitem[{{Krause} {et~al.}(2015){Krause}, {Morlino}, \&
  {Gabici}}]{Krause2015ICRC}
{Krause}, J., {Morlino}, G., \& {Gabici}, S. 2015, arXiv e-prints,
  arXiv:1507.05127.
\newblock \doarXiv{1507.05127}

\bibitem[{{Krymskii}(1977)}]{Krymskii1977DoSSR}
{Krymskii}, G.~F. 1977, Akademiia Nauk SSSR Doklady, 234, 1306

\bibitem[{{Kudoh} \& {Basu}(2008)}]{Kudoh2008}
{Kudoh}, T., \& {Basu}, S. 2008, \apjl, 679, L97, \dodoi{10.1086/589618}

\bibitem[{{Kulsrud} \& {Pearce}(1969)}]{Kulsrud1969}
{Kulsrud}, R., \& {Pearce}, W.~P. 1969, \apj, 156, 445, \dodoi{10.1086/149981}

\bibitem[{{Kulsrud}(2005)}]{Kulsrud2005book}
{Kulsrud}, R.~M. 2005, {Plasma physics for astrophysics}

\bibitem[{{Kulsrud} \& {Cesarsky}(1971)}]{Kulsrud1971}
{Kulsrud}, R.~M., \& {Cesarsky}, C.~J. 1971, \aplett, 8, 189

\bibitem[{{Lacki} \& {Beck}(2013)}]{Lacki2013}
{Lacki}, B.~C., \& {Beck}, R. 2013, \mnras, 430, 3171,
  \dodoi{10.1093/mnras/stt122}

\bibitem[{{Lacki} {et~al.}(2010){Lacki}, {Thompson}, \& {Quataert}}]{Lacki2010}
{Lacki}, B.~C., {Thompson}, T.~A., \& {Quataert}, E. 2010, \apj, 717, 1,
  \dodoi{10.1088/0004-637X/717/1/1}

\bibitem[{{Lada} {et~al.}(1999){Lada}, {Alves}, \& {Lada}}]{Lada1999ApJ}
{Lada}, C.~J., {Alves}, J., \& {Lada}, E.~A. 1999, \apj, 512, 250,
  \dodoi{10.1086/306756}

\bibitem[{{Larson} {et~al.}(2015){Larson}, {Evans}, {Green}, \&
  {Yang}}]{Larson2015ApJ}
{Larson}, R.~L., {Evans}, Neal~J., I., {Green}, J.~D., \& {Yang}, Y.-L. 2015,
  \apj, 806, 70, \dodoi{10.1088/0004-637X/806/1/70}

\bibitem[{{Latter} {et~al.}(1993){Latter}, {Walker}, \& {Maloney}}]{Latter1993}
{Latter}, W.~B., {Walker}, C.~K., \& {Maloney}, P.~R. 1993, \apjl, 419, L97,
  \dodoi{10.1086/187146}

\bibitem[{{Lazarian} {et~al.}(1997){Lazarian}, {Goodman}, \&
  {Myers}}]{Lazarian1997ApJ}
{Lazarian}, A., {Goodman}, A.~A., \& {Myers}, P.~C. 1997, \apj, 490, 273,
  \dodoi{10.1086/304874}

\bibitem[{{Lazarian} \& {Hoang}(2007)}]{Lazarian2007MNRAS}
{Lazarian}, A., \& {Hoang}, T. 2007, \mnras, 378, 910,
  \dodoi{10.1111/j.1365-2966.2007.11817.x}

\bibitem[{{Lazarian} \& {Pogosyan}(2016)}]{Lazarian2016ApJ}
{Lazarian}, A., \& {Pogosyan}, D. 2016, \apj, 818, 178,
  \dodoi{10.3847/0004-637X/818/2/178}

\bibitem[{{Lee} {et~al.}(2003){Lee}, {Evans}, {Shirley}, \&
  {Tatematsu}}]{Lee2003ApJ}
{Lee}, J.-E., {Evans}, Neal~J., I., {Shirley}, Y.~L., \& {Tatematsu}, K. 2003,
  \apj, 583, 789, \dodoi{10.1086/345428}

\bibitem[{{Lequeux}(2005)}]{Lequeux2005}
{Lequeux}, J. 2005, {The Interstellar Medium}, \dodoi{10.1007/b137959}

\bibitem[{{Li} {et~al.}(2009){Li}, {Dowell}, {Goodman}, {Hildebrand}, \&
  {Novak}}]{Li2009}
{Li}, H.-b., {Dowell}, C.~D., {Goodman}, A., {Hildebrand}, R., \& {Novak}, G.
  2009, \apj, 704, 891, \dodoi{10.1088/0004-637X/704/2/891}

\bibitem[{{Li} {et~al.}(2015){Li}, {Yuen}, {Otto}, {Leung}, {Sridharan},
  {Zhang}, {Liu}, {Tang}, \& {Qiu}}]{Li2015Natur}
{Li}, H.-B., {Yuen}, K.~H., {Otto}, F., {et~al.} 2015, \nat, 520, 518,
  \dodoi{10.1038/nature14291}

\bibitem[{{Lin} {et~al.}(2020){Lin}, {Pagani}, {Lai}, {Lef{\`e}vre}, \&
  {Lique}}]{lin2020AA}
{Lin}, S.-J., {Pagani}, L., {Lai}, S.-P., {Lef{\`e}vre}, C., \& {Lique}, F.
  2020, \aap, 635, A188, \dodoi{10.1051/0004-6361/201936877}

\bibitem[{{Liu} \& {Shemansky}(2004)}]{Liu2004}
{Liu}, X., \& {Shemansky}, D.~E. 2004, \apj, 614, 1132, \dodoi{10.1086/423890}

\bibitem[{{Loewenstein} {et~al.}(1991){Loewenstein}, {Zweibel}, \&
  {Begelman}}]{Loewenstein1991}
{Loewenstein}, M., {Zweibel}, E.~G., \& {Begelman}, M.~C. 1991, \apj, 377, 392,
  \dodoi{10.1086/170369}

\bibitem[{{Mac Low} \& {Klessen}(2004)}]{MacLow2004}
{Mac Low}, M.-M., \& {Klessen}, R.~S. 2004, Reviews of Modern Physics, 76, 125,
  \dodoi{10.1103/RevModPhys.76.125}

\bibitem[{{Martin} {et~al.}(1997){Martin}, {Heyvaerts}, \&
  {Priest}}]{Martin1997}
{Martin}, C.~E., {Heyvaerts}, J., \& {Priest}, E.~R. 1997, \aap, 326, 1176

\bibitem[{{Mart{\'\i}nez-G{\'o}mez} {et~al.}(2018){Mart{\'\i}nez-G{\'o}mez},
  {Soler}, \& {Terradas}}]{Gomez2018ApJ}
{Mart{\'\i}nez-G{\'o}mez}, D., {Soler}, R., \& {Terradas}, J. 2018, \apj, 856,
  16, \dodoi{10.3847/1538-4357/aab156}

\bibitem[{{McCall} {et~al.}(2004){McCall}, {Huneycutt}, {Saykally}, {Djuric},
  {Dunn}, {Semaniak}, {Novotny}, {Al-Khalili}, {Ehlerding}, {Hellberg},
  {Kalhori}, {Neau}, {Thomas}, {Paal}, {{\"O}sterdahl}, \&
  {Larsson}}]{McCall2004}
{McCall}, B.~J., {Huneycutt}, A.~J., {Saykally}, R.~J., {et~al.} 2004, \pra,
  70, 052716, \dodoi{10.1103/PhysRevA.70.052716}

\bibitem[{{McElroy} {et~al.}(2013){McElroy}, {Walsh}, {Markwick}, {Cordiner},
  {Smith}, \& {Millar}}]{McElroy2012}
{McElroy}, D., {Walsh}, C., {Markwick}, A.~J., {et~al.} 2013, \aap, 550, A36,
  \dodoi{10.1051/0004-6361/201220465}

\bibitem[{{McKee} \& {Zweibel}(1995)}]{McKee1995}
{McKee}, C.~F., \& {Zweibel}, E.~G. 1995, \apj, 440, 686,
  \dodoi{10.1086/175306}

\bibitem[{{Mestel}(1966)}]{Mestel1966}
{Mestel}, L. 1966, \mnras, 133, 265, \dodoi{10.1093/mnras/133.2.265}

\bibitem[{{Mestel} \& {Spitzer}(1956)}]{Mestel1956}
{Mestel}, L., \& {Spitzer}, Jr., L. 1956, \mnras, 116, 503,
  \dodoi{10.1093/mnras/116.5.503}

\bibitem[{{Millar} {et~al.}(1997){Millar}, {Farquhar}, \&
  {Willacy}}]{Millar1997}
{Millar}, T.~J., {Farquhar}, P.~R.~A., \& {Willacy}, K. 1997, \aaps, 121, 139,
  \dodoi{10.1051/aas:1997118}

\bibitem[{{Minter} \& {Spangler}(1996)}]{Minter1996ApJ}
{Minter}, A.~H., \& {Spangler}, S.~R. 1996, \apj, 458, 194,
  \dodoi{10.1086/176803}

\bibitem[{{Mitchell}(1990)}]{Mitchell1990}
{Mitchell}, J.~B.~A. 1990, \physrep, 186, 215

\bibitem[{{Morfill}(1982{\natexlab{a}})}]{Morfill1982MNRAS}
{Morfill}, G.~E. 1982{\natexlab{a}}, \mnras, 198, 583,
  \dodoi{10.1093/mnras/198.2.583}

\bibitem[{{Morfill}(1982{\natexlab{b}})}]{Morfill1982ApJ}
---. 1982{\natexlab{b}}, \apj, 262, 749, \dodoi{10.1086/160470}

\bibitem[{{Morlino} \& {Gabici}(2015)}]{Morlino2015}
{Morlino}, G., \& {Gabici}, S. 2015, \mnras, 451, L100,
  \dodoi{10.1093/mnrasl/slv074}

\bibitem[{{Moskalenko} {et~al.}(2002){Moskalenko}, {Strong}, {Ormes}, \&
  {Potgieter}}]{Moskalenko2002ApJ}
{Moskalenko}, I.~V., {Strong}, A.~W., {Ormes}, J.~F., \& {Potgieter}, M.~S.
  2002, \apj, 565, 280, \dodoi{10.1086/324402}

\bibitem[{{Mouschovias}(1991)}]{Mouschovias1991}
{Mouschovias}, T.~C. 1991, in NATO Advanced Science Institutes (ASI) Series C,
  ed. C.~J. {Lada} \& N.~D. {Kylafis}, Vol. 342, 61

\bibitem[{{Mouschovias} \& {Ciolek}(1999)}]{Mouschovias1999}
{Mouschovias}, T.~C., \& {Ciolek}, G.~E. 1999, in NATO Advanced Science
  Institutes (ASI) Series C, ed. C.~J. {Lada} \& N.~D. {Kylafis}, Vol. 540, 305

\bibitem[{{Murphy} {et~al.}(1987){Murphy}, {Dermer}, \&
  {Ramaty}}]{Murphy1987ApJS}
{Murphy}, R.~J., {Dermer}, C.~D., \& {Ramaty}, R. 1987, \apjs, 63, 721,
  \dodoi{10.1086/191180}

\bibitem[{Myers(1995)}]{Myers1995}
Myers, P. 1995, Molecular Clouds and Star Formation, ed. C. Yuan \& J.-H. You,
  Singapore: World Scientific

\bibitem[{{Neufeld} \& {Wolfire}(2017)}]{Neufeld2017ApJ}
{Neufeld}, D.~A., \& {Wolfire}, M.~G. 2017, \apj, 845, 163,
  \dodoi{10.3847/1538-4357/aa6d68}

\bibitem[{{Neufeld} {et~al.}(2010){Neufeld}, {Goicoechea}, {Sonnentrucker},
  {Black}, {Pearson}, {Yu}, {Phillips}, {Lis}, {de Luca}, {Herbst}, {Rimmer},
  {Gerin}, {Bell}, {Boulanger}, {Cernicharo}, {Coutens}, {Dartois},
  {Kazmierczak}, {Encrenaz}, {Falgarone}, {Geballe}, {Giesen}, {Godard},
  {Goldsmith}, {Gry}, {Gupta}, {Hennebelle}, {Hily-Blant}, {Joblin},
  {Ko{\l}os}, {Kre{\l}owski}, {Mart{\'\i}n-Pintado}, {Menten}, {Monje},
  {Mookerjea}, {Perault}, {Persson}, {Plume}, {Salez}, {Schlemmer}, {Schmidt},
  {Stutzki}, {Teyssier}, {Vastel}, {Cros}, {Klein}, {Lorenzani}, {Philipp},
  {Samoska}, {Shipman}, {Tielens}, {Szczerba}, \& {Zmuidzinas}}]{Neufeld2010}
{Neufeld}, D.~A., {Goicoechea}, J.~R., {Sonnentrucker}, P., {et~al.} 2010,
  \aap, 521, L10, \dodoi{10.1051/0004-6361/201015077}

\bibitem[{{Osterbrock}(1989)}]{Osterbrock_book}
{Osterbrock}, D.~E. 1989, {Astrophysics of gaseous nebulae and active galactic
  nuclei}

\bibitem[{{Owen} {et~al.}(2018){Owen}, {Jacobsen}, {Wu}, \&
  {Surajbali}}]{Owen2018}
{Owen}, E.~R., {Jacobsen}, I.~B., {Wu}, K., \& {Surajbali}, P. 2018, \mnras,
  481, 666, \dodoi{10.1093/mnras/sty2279}

\bibitem[{{Owen} {et~al.}(2019{\natexlab{a}}){Owen}, {Jin}, {Wu}, \&
  {Chan}}]{Owen2019MNRAS}
{Owen}, E.~R., {Jin}, X., {Wu}, K., \& {Chan}, S. 2019{\natexlab{a}}, \mnras,
  484, 1645, \dodoi{10.1093/mnras/stz060}

\bibitem[{{Owen} {et~al.}(2019{\natexlab{b}}){Owen}, {Wu}, {Jin}, {Surajbali},
  \& {Kataoka}}]{Owen2019AA}
{Owen}, E.~R., {Wu}, K., {Jin}, X., {Surajbali}, P., \& {Kataoka}, N.
  2019{\natexlab{b}}, \aap, 626, A85, \dodoi{10.1051/0004-6361/201834350}

\bibitem[{{Padoan} \& {Nordlund}(1999)}]{Padoan1999}
{Padoan}, P., \& {Nordlund}, {\AA}. 1999, \apj, 526, 279,
  \dodoi{10.1086/307956}

\bibitem[{{Padoan} \& {Scalo}(2005)}]{Padoan2005ApJ}
{Padoan}, P., \& {Scalo}, J. 2005, \apjl, 624, L97, \dodoi{10.1086/430598}

\bibitem[{{Padovani} \& {Galli}(2011)}]{Padovani2011}
{Padovani}, M., \& {Galli}, D. 2011, \aap, 530, A109,
  \dodoi{10.1051/0004-6361/201116853}

\bibitem[{{Padovani} \& {Galli}(2018)}]{Padovani2018Lett}
---. 2018, \aap, 620, L4, \dodoi{10.1051/0004-6361/201834222}

\bibitem[{{Padovani} {et~al.}(2009){Padovani}, {Galli}, \&
  {Glassgold}}]{Padovani2009AA}
{Padovani}, M., {Galli}, D., \& {Glassgold}, A.~E. 2009, \aap, 501, 619,
  \dodoi{10.1051/0004-6361/200911794}

\bibitem[{{Padovani} {et~al.}(2018){Padovani}, {Galli}, {Ivlev}, {Caselli}, \&
  {Ferrara}}]{Padovani2018}
{Padovani}, M., {Galli}, D., {Ivlev}, A.~V., {Caselli}, P., \& {Ferrara}, A.
  2018, \aap, 619, A144, \dodoi{10.1051/0004-6361/201834008}

\bibitem[{{Padovani} {et~al.}(2013){Padovani}, {Hennebelle}, \&
  {Galli}}]{Padovani2013}
{Padovani}, M., {Hennebelle}, P., \& {Galli}, D. 2013, \aap, 560, A114,
  \dodoi{10.1051/0004-6361/201322407}

\bibitem[{{Padovani} {et~al.}(2015){Padovani}, {Hennebelle}, {Marcowith}, \&
  {Ferri{\`e}re}}]{Padovani2015AA}
{Padovani}, M., {Hennebelle}, P., {Marcowith}, A., \& {Ferri{\`e}re}, K. 2015,
  \aap, 582, L13, \dodoi{10.1051/0004-6361/201526874}

\bibitem[{{Padovani} {et~al.}(2016){Padovani}, {Marcowith}, {Hennebelle}, \&
  {Ferri{\`e}re}}]{Padovani2016AA}
{Padovani}, M., {Marcowith}, A., {Hennebelle}, P., \& {Ferri{\`e}re}, K. 2016,
  \aap, 590, A8, \dodoi{10.1051/0004-6361/201628221}

\bibitem[{{Padovani} {et~al.}(2020){Padovani}, {Ivlev}, {Galli}, {Offner},
  {Indriolo}, {Rodgers-Lee}, {Marcowith}, {Girichidis}, {Bykov}, \&
  {Kruijssen}}]{Padovani2020SSRv}
{Padovani}, M., {Ivlev}, A.~V., {Galli}, D., {et~al.} 2020, \ssr, 216, 29,
  \dodoi{10.1007/s11214-020-00654-1}

\bibitem[{{Pagani} {et~al.}(2007){Pagani}, {Bacmann}, {Cabrit}, \&
  {Vastel}}]{Pagani2007AA}
{Pagani}, L., {Bacmann}, A., {Cabrit}, S., \& {Vastel}, C. 2007, \aap, 467,
  179, \dodoi{10.1051/0004-6361:20066670}

\bibitem[{{Pagani} {et~al.}(2015){Pagani}, {Lef{\`e}vre}, {Juvela}, {Pelkonen},
  \& {Schuller}}]{Pagani2015AA}
{Pagani}, L., {Lef{\`e}vre}, C., {Juvela}, M., {Pelkonen}, V.~M., \&
  {Schuller}, F. 2015, \aap, 574, L5, \dodoi{10.1051/0004-6361/201425095}

\bibitem[{{Pagani} {et~al.}(2009){Pagani}, {Vastel}, {Hugo}, {Kokoouline},
  {Greene}, {Bacmann}, {Bayet}, {Ceccarelli}, {Peng}, \&
  {Schlemmer}}]{Pagani2009AA}
{Pagani}, L., {Vastel}, C., {Hugo}, E., {et~al.} 2009, \aap, 494, 623,
  \dodoi{10.1051/0004-6361:200810587}

\bibitem[{{Pan} \& {Padoan}(2009)}]{Pan2009ApJ}
{Pan}, L., \& {Padoan}, P. 2009, \apj, 692, 594,
  \dodoi{10.1088/0004-637X/692/1/594}

\bibitem[{Patrignani {et~al.}(2016)}]{Patrignani2016ChPh}
Patrignani, C., {et~al.} 2016, Chin. Phys., C40, 100001,
  \dodoi{10.1088/1674-1137/40/10/100001}

\bibitem[{{Percival} \& {Walden}(1993)}]{Percival1995_book}
{Percival}, D.~B., \& {Walden}, A.~T. 1993, {Spectral analysis for physical
  applications : multitaper and conventional univariate techniques}

\bibitem[{{Phan} {et~al.}(2018){Phan}, {Morlino}, \& {Gabici}}]{Phan2018MNRAS}
{Phan}, V.~H.~M., {Morlino}, G., \& {Gabici}, S. 2018, \mnras, 480, 5167,
  \dodoi{10.1093/mnras/sty2235}

\bibitem[{{Planck Collaboration} {et~al.}(2016){Planck Collaboration}, {Ade},
  {Aghanim}, {Alves}, {Arnaud}, {Arzoumanian}, {Ashdown}, {Aumont},
  {Baccigalupi}, \& {Band ay}}]{Planck2016_mag}
{Planck Collaboration}, {Ade}, P.~A.~R., {Aghanim}, N., {et~al.} 2016, \aap,
  586, A138, \dodoi{10.1051/0004-6361/201525896}

\bibitem[{{Prasad} \& {Huntress}(1980)}]{Prasad1980}
{Prasad}, S.~S., \& {Huntress}, Jr., W.~T. 1980, \apjs, 43, 1,
  \dodoi{10.1086/190665}

\bibitem[{Press {et~al.}(2007)Press, Teukolsky, Vetterling, \&
  Flannery}]{NRbook3_2007}
Press, W., Teukolsky, S., Vetterling, W., \& Flannery, B. 2007, Numerical
  Recipes 3rd Edition: The Art of Scientific Computing (Cambridge University
  Press)

\bibitem[{{Press} {et~al.}(1992){Press}, {Teukolsky}, {Vetterling}, \&
  {Flannery}}]{Press1992book}
{Press}, W.~H., {Teukolsky}, S.~A., {Vetterling}, W.~T., \& {Flannery}, B.~P.
  1992, {Numerical recipes in FORTRAN. The art of scientific computing}

\bibitem[{{Price} \& {Bate}(2008)}]{Price2008}
{Price}, D.~J., \& {Bate}, M.~R. 2008, \mnras, 385, 1820,
  \dodoi{10.1111/j.1365-2966.2008.12976.x}

\bibitem[{{Protheroe} {et~al.}(2008){Protheroe}, {Ott}, {Ekers}, {Jones}, \&
  {Crocker}}]{Protheroe2008MNRAS}
{Protheroe}, R.~J., {Ott}, J., {Ekers}, R.~D., {Jones}, D.~I., \& {Crocker},
  R.~M. 2008, \mnras, 390, 683, \dodoi{10.1111/j.1365-2966.2008.13752.x}

\bibitem[{{Rao} {et~al.}(2009){Rao}, {Girart}, {Marrone}, {Lai}, \&
  {Schnee}}]{Rao2009}
{Rao}, R., {Girart}, J.~M., {Marrone}, D.~P., {Lai}, S.-P., \& {Schnee}, S.
  2009, \apj, 707, 921, \dodoi{10.1088/0004-637X/707/2/921}

\bibitem[{{Redaelli} {et~al.}(2019){Redaelli}, {Alves}, {Santos}, \&
  {Caselli}}]{Redaelli2019}
{Redaelli}, E., {Alves}, F.~O., {Santos}, F.~P., \& {Caselli}, P. 2019, \aap,
  631, A154, \dodoi{10.1051/0004-6361/201936271}

\bibitem[{{Rodr{\'{\i}}guez}(2005)}]{Rodriguez2005}
{Rodr{\'{\i}}guez}, L.~F.~R. 2005, in Astronomical Society of the Pacific
  Conference Series, Vol. 344, The Cool Universe: Observing Cosmic Dawn, ed.
  C.~{Lidman} \& D.~{Alloin}, 146

\bibitem[{{Rudd}(1991)}]{Rudd1991}
{Rudd}, M.~E. 1991, \pra, 44, 1644, \dodoi{10.1103/PhysRevA.44.1644}

\bibitem[{{Rudd} {et~al.}(1983){Rudd}, {Goffe}, {Dubois}, {Toburen}, \&
  {Ratcliffe}}]{Rudd1983}
{Rudd}, M.~E., {Goffe}, T.~V., {Dubois}, R.~D., {Toburen}, L.~H., \&
  {Ratcliffe}, C.~A. 1983, \pra, 28, 3244, \dodoi{10.1103/PhysRevA.28.3244}

\bibitem[{Rudd {et~al.}(1985)Rudd, Kim, Madison, \& Gallagher}]{Rudd1985}
Rudd, M.~E., Kim, Y.~K., Madison, D.~H., \& Gallagher, J.~W. 1985, Rev. Mod.
  Phys., 57, 965, \dodoi{10.1103/RevModPhys.57.965}

\bibitem[{{Ruszkowski} {et~al.}(2017){Ruszkowski}, {Yang}, \&
  {Reynolds}}]{Ruszkowski2017}
{Ruszkowski}, M., {Yang}, H.-Y.~K., \& {Reynolds}, C.~S. 2017, \apj, 844, 13,
  \dodoi{10.3847/1538-4357/aa79f8}

\bibitem[{{Salem} {et~al.}(2016){Salem}, {Bryan}, \& {Corlies}}]{Salem2016}
{Salem}, M., {Bryan}, G.~L., \& {Corlies}, L. 2016, \mnras, 456, 582,
  \dodoi{10.1093/mnras/stv2641}

\bibitem[{{Schlickeiser}(2002)}]{Schlickeiser2002_book}
{Schlickeiser}, R. 2002, {Cosmic Ray Astrophysics}

\bibitem[{{Schlickeiser} \& {Achatz}(1993{\natexlab{a}})}]{SA1993_1}
{Schlickeiser}, R., \& {Achatz}, U. 1993{\natexlab{a}}, Journal of Plasma
  Physics, 49, 63, \dodoi{10.1017/S0022377800016822}

\bibitem[{{Schlickeiser} \& {Achatz}(1993{\natexlab{b}})}]{SA1993_2}
---. 1993{\natexlab{b}}, Journal of Plasma Physics, 50, 85,
  \dodoi{10.1017/S0022377800026933}

\bibitem[{{Schulz-Dubois} \& {Rehberg}(1981)}]{SchulzDubois1981ApPhy}
{Schulz-Dubois}, E.~O., \& {Rehberg}, I. 1981, Applied Physics, 24, 323,
  \dodoi{10.1007/BF00899730}

\bibitem[{{Seifried} \& {Walch}(2015)}]{Seifried2015}
{Seifried}, D., \& {Walch}, S. 2015, \mnras, 452, 2410,
  \dodoi{10.1093/mnras/stv1458}

\bibitem[{{Sikora} {et~al.}(1987){Sikora}, {Kirk}, {Begelman}, \&
  {Schneider}}]{Sikora1987ApJ}
{Sikora}, M., {Kirk}, J.~G., {Begelman}, M.~C., \& {Schneider}, P. 1987, \apjl,
  320, L81, \dodoi{10.1086/184980}

\bibitem[{{Silsbee} \& {Ivlev}(2019)}]{Silsbee2019ApJ}
{Silsbee}, K., \& {Ivlev}, A.~V. 2019, \apj, 879, 14,
  \dodoi{10.3847/1538-4357/ab22b4}

\bibitem[{{Silsbee} {et~al.}(2018){Silsbee}, {Ivlev}, {Padovani}, \&
  {Caselli}}]{Silsbee2018ApJ}
{Silsbee}, K., {Ivlev}, A.~V., {Padovani}, M., \& {Caselli}, P. 2018, \apj,
  863, 188, \dodoi{10.3847/1538-4357/aad3cf}

\bibitem[{{Skorodko} {et~al.}(2008){Skorodko}, {Bashkanov}, {Bogoslawsky},
  {Calen}, {Cappellaro}, {Clement}, {Demiroers}, {Doroshkevich}, {Duniec},
  {Ekstr{\"o}m}, {Franssen}, {Gustafsson}, {H{\"o}istad}, {Ivanov}, {Jacewicz},
  {Jiganov}, {Johansson}, {Khakimova}, {Kaskulov}, {Keleta}, {Koch}, {Kren},
  {Kullander}, {Kup{\'s}{\'c}}, {Kuznetsov}, {Marciniewski}, {Meier},
  {Morosov}, {Pauly}, {Petterson}, {Petukhov}, {Povtorejko}, {Sch{\"o}nning},
  {Scobel}, {Shwartz}, {Sopov}, {Stepeniak}, {Th{\"o}rngren-Engblom},
  {Tikhomirov}, {Wagner}, {Wolke}, {Yamamoto}, {Zabierowski}, \&
  {Z{\l}omanczuk}}]{Skorodko2008EPJA}
{Skorodko}, T., {Bashkanov}, M., {Bogoslawsky}, D., {et~al.} 2008, Eur. Phys.
  J. A, 35, 317, \dodoi{10.1140/epja/i2008-10569-6}

\bibitem[{{Skrutskie} {et~al.}(2006){Skrutskie}, {Cutri}, {Stiening},
  {Weinberg}, {Schneider}, {Carpenter}, {Beichman}, {Capps}, {Chester},
  {Elias}, {Huchra}, {Liebert}, {Lonsdale}, {Monet}, {Price}, {Seitzer},
  {Jarrett}, {Kirkpatrick}, {Gizis}, {Howard}, {Evans}, {Fowler}, {Fullmer},
  {Hurt}, {Light}, {Kopan}, {Marsh}, {McCallon}, {Tam}, {Van Dyk}, \&
  {Wheelock}}]{Skrutskie2006AJ}
{Skrutskie}, M.~F., {Cutri}, R.~M., {Stiening}, R., {et~al.} 2006, \aj, 131,
  1163, \dodoi{10.1086/498708}

\bibitem[{{Sofia} {et~al.}(2004){Sofia}, {Lauroesch}, {Meyer}, \&
  {Cartledge}}]{Sofia2004}
{Sofia}, U.~J., {Lauroesch}, J.~T., {Meyer}, D.~M., \& {Cartledge}, S. I.~B.
  2004, \apj, 605, 272, \dodoi{10.1086/382592}

\bibitem[{{Sonnentrucker} {et~al.}(2007){Sonnentrucker}, {Welty}, {Thorburn},
  \& {York}}]{Sonnentrucker2007}
{Sonnentrucker}, P., {Welty}, D.~E., {Thorburn}, J.~A., \& {York}, D.~G. 2007,
  \apjs, 168, 58, \dodoi{10.1086/508687}

\bibitem[{{Spitzer} \& {Tomasko}(1968)}]{Spitzer1968ApJ}
{Spitzer}, Jr., L., \& {Tomasko}, M.~G. 1968, \apj, 152, 971,
  \dodoi{10.1086/149610}

\bibitem[{{Stancil} {et~al.}(1999){Stancil}, {Schultz}, {Kimura}, {Gu},
  {Hirsch}, \& {Buenker}}]{Stancil1999}
{Stancil}, P.~C., {Schultz}, D.~R., {Kimura}, M., {et~al.} 1999, \aaps, 140,
  225, \dodoi{10.1051/aas:1999419}

\bibitem[{{Straub} {et~al.}(1996){Straub}, {Renault}, {Lindsay}, {Smith}, \&
  {Stebbings}}]{Straub1996}
{Straub}, H.~C., {Renault}, P., {Lindsay}, B.~G., {Smith}, K.~A., \&
  {Stebbings}, R.~F. 1996, \pra, 54, 2146, \dodoi{10.1103/PhysRevA.54.2146}

\bibitem[{{Strong} {et~al.}(2014){Strong}, {Dickinson}, \&
  {Murphy}}]{Strong2014arXiv}
{Strong}, A.~W., {Dickinson}, C., \& {Murphy}, E.~J. 2014, arXiv e-prints,
  arXiv:1412.4500.
\newblock \doarXiv{1412.4500}

\bibitem[{Strong {et~al.}(2007)Strong, Moskalenko, \& Ptuskin}]{Strong2007}
Strong, A.~W., Moskalenko, I.~V., \& Ptuskin, V.~S. 2007, Ann. Rev. Nucl. and
  Part. Sci., 57, 285, \dodoi{10.1146/annurev.nucl.57.090506.123011}

\bibitem[{{Strong} {et~al.}(2000){Strong}, {Moskalenko}, \&
  {Reimer}}]{Strong2000ApJ}
{Strong}, A.~W., {Moskalenko}, I.~V., \& {Reimer}, O. 2000, \apj, 537, 763,
  \dodoi{10.1086/309038}

\bibitem[{{Tang} {et~al.}(2009){Tang}, {Ho}, {Koch}, {Girart}, {Lai}, \&
  {Rao}}]{Tang2009}
{Tang}, Y.-W., {Ho}, P.~T.~P., {Koch}, P.~M., {et~al.} 2009, \apj, 700, 251,
  \dodoi{10.1088/0004-637X/700/1/251}

\bibitem[{{Tang} {et~al.}(2019){Tang}, {Koch}, {Peretto}, {Novak},
  {Duarte-Cabral}, {Chapman}, {Hsieh}, \& {Yen}}]{Tang2019}
{Tang}, Y.-W., {Koch}, P.~M., {Peretto}, N., {et~al.} 2019, \apj, 878, 10,
  \dodoi{10.3847/1538-4357/ab1484}

\bibitem[{{Theard} \& {Huntress}(1974)}]{Theard1974}
{Theard}, L.~P., \& {Huntress}, W.~T. 1974, \jcp, 60, 2840,
  \dodoi{10.1063/1.1681453}

\bibitem[{{Thompson} {et~al.}(2019){Thompson}, {Troland}, \&
  {Heiles}}]{Thompson2019ApJ}
{Thompson}, K.~L., {Troland}, T.~H., \& {Heiles}, C. 2019, \apj, 884, 49,
  \dodoi{10.3847/1538-4357/ab364e}

\bibitem[{{Thompson} {et~al.}(2007){Thompson}, {Quataert}, \&
  {Waxman}}]{Thompson2007}
{Thompson}, T.~A., {Quataert}, E., \& {Waxman}, E. 2007, \apj, 654, 219,
  \dodoi{10.1086/509068}

\bibitem[{{Torres}(2004)}]{Torres2004}
{Torres}, D.~F. 2004, \apj, 617, 966, \dodoi{10.1086/425415}

\bibitem[{{van der Tak} \& {van Dishoeck}(2000)}]{Tak2000}
{van der Tak}, F.~F.~S., \& {van Dishoeck}, E.~F. 2000, \aap, 358, L79

\bibitem[{{van Dishoeck} \& {Black}(1986)}]{vanDishoeck1986ApJS}
{van Dishoeck}, E.~F., \& {Black}, J.~H. 1986, \apjs, 62, 109,
  \dodoi{10.1086/191135}

\bibitem[{{V{\'a}zquez-Semadeni} {et~al.}(2011){V{\'a}zquez-Semadeni},
  {Banerjee}, {G{\'o}mez}, {Hennebelle}, {Duffin}, \& {Klessen}}]{Vazquez2011}
{V{\'a}zquez-Semadeni}, E., {Banerjee}, R., {G{\'o}mez}, G.~C., {et~al.} 2011,
  \mnras, 414, 2511, \dodoi{10.1111/j.1365-2966.2011.18569.x}

\bibitem[{{Virtanen} {et~al.}(2020){Virtanen}, {Gommers}, {Oliphant},
  {Haberland}, {Reddy}, {Cournapeau}, {Burovski}, {Peterson}, {Weckesser},
  {Bright}, {van der Walt}, {Brett}, {Wilson}, {Millman}, {Mayorov}, {Nelson},
  {Jones}, {Kern}, {Larson}, {Carey}, {Polat}, {Feng}, {Moore}, {VanderPlas},
  {Laxalde}, {Perktold}, {Cimrman}, {Henriksen}, {Quintero}, {Harris},
  {Archibald}, {Ribeiro}, {Pedregosa}, {van Mulbregt}, \& {SciPy 1. 0
  Contributors}}]{SciPy2019}
{Virtanen}, P., {Gommers}, R., {Oliphant}, T.~E., {et~al.} 2020, Nature
  Methods, 17, 261, \dodoi{10.1038/s41592-019-0686-2}

\bibitem[{{Walker}(2016)}]{Walker2016}
{Walker}, M.~A. 2016, \apj, 818, 23, \dodoi{10.3847/0004-637X/818/1/23}

\bibitem[{{Wang} {et~al.}(2020){Wang}, {Lai}, {Clemens}, {Koch}, {Eswaraiah},
  {Chen}, \& {Pand ey}}]{Wang2020ApJ}
{Wang}, J.-W., {Lai}, S.-P., {Clemens}, D.~P., {et~al.} 2020, \apj, 888, 13,
  \dodoi{10.3847/1538-4357/ab5c1c}

\bibitem[{{Wang} {et~al.}(2017){Wang}, {Lai}, {Eswaraiah}, {Clemens}, {Chen},
  \& {Pandey}}]{Wang2017ApJ}
{Wang}, J.-W., {Lai}, S.-P., {Eswaraiah}, C., {et~al.} 2017, \apj, 849, 157,
  \dodoi{10.3847/1538-4357/aa937f}

\bibitem[{{Wang} {et~al.}(2019){Wang}, {Lai}, {Eswaraiah}, {Pattle}, {Di
  Francesco}, {Johnstone}, {Koch}, {Liu}, {Tamura}, {Furuya}, {Onaka},
  {Ward-Thompson}, {Soam}, {Kim}, {Lee}, {Lee}, {Mairs}, {Arzoumanian}, {Kim},
  {Hoang}, {Hwang}, {Liu}, {Berry}, {Bastien}, {Hasegawa}, {Kwon}, {Qiu},
  {Andr{\'e}}, {Aso}, {Byun}, {Chen}, {Chen}, {Chen}, {Ching}, {Cho}, {Choi},
  {Chrysostomou}, {Chung}, {Coud{\'e}}, {Doi}, {Dowell}, {Drabek-Maunder},
  {Duan}, {Eyres}, {Falle}, {Fanciullo}, {Fiege}, {Franzmann}, {Friberg},
  {Friesen}, {Fuller}, {Gledhill}, {Graves}, {Greaves}, {Griffin}, {Gu}, {Han},
  {Hatchell}, {Hayashi}, {Holland}, {Houde}, {Inoue}, {Inutsuka}, {Iwasaki},
  {Jeong}, {Kanamori}, {Kang}, {Kang}, {Kang}, {Kataoka}, {Kawabata}, {Kemper},
  {Kim}, {Kim}, {Kim}, {Kim}, {Kirk}, {Kobayashi}, {Konyves}, {Kwon},
  {Lacaille}, {Lee}, {Lee}, {Lee}, {Lee}, {Li}, {Li}, {Li}, {Liu}, {Liu},
  {Lyo}, {Matsumura}, {Matthews}, {Moriarty-Schieven}, {Nagata}, {Nakamura},
  {Nakanishi}, {Ohashi}, {Park}, {Parsons}, {Pascale}, {Peretto}, {Pon}, {Pyo},
  {Qian}, {Rao}, {Rawlings}, {Retter}, {Richer}, {Rigby}, {Robitaille},
  {Sadavoy}, {Saito}, {Savini}, {Scaife}, {Seta}, {Shinnaga}, {Tang},
  {Tomisaka}, {Tsukamoto}, {van Loo}, {Wang}, {Whitworth}, {Yen}, {Yoo},
  {Yuan}, {Yun}, {Zenko}, {Zhang}, {Zhang}, {Zhang}, {Zhou}, \&
  {Zhu}}]{Wang2019ApJ}
---. 2019, \apj, 876, 42, \dodoi{10.3847/1538-4357/ab13a2}

\bibitem[{{Webber}(1998)}]{Webber1998ApJ}
{Webber}, W.~R. 1998, \apj, 506, 329, \dodoi{10.1086/306222}

\bibitem[{{Wentzel}(1969)}]{Wentzel1969}
{Wentzel}, D.~G. 1969, \apj, 156, 303, \dodoi{10.1086/149965}

\bibitem[{{Wentzel}(1971)}]{Wentzel1971}
---. 1971, \apj, 163, 503, \dodoi{10.1086/150794}

\bibitem[{{Wentzel}(1974)}]{Wentzel1974ARAA}
---. 1974, \araa, 12, 71, \dodoi{10.1146/annurev.aa.12.090174.000443}

\bibitem[{{Whittet} {et~al.}(2008){Whittet}, {Hough}, {Lazarian}, \&
  {Hoang}}]{Whittet2008ApJ}
{Whittet}, D.~C.~B., {Hough}, J.~H., {Lazarian}, A., \& {Hoang}, T. 2008, \apj,
  674, 304, \dodoi{10.1086/525040}

\bibitem[{{Whitworth} \& {Jaffa}(2018)}]{Whitworth2018AA}
{Whitworth}, A.~P., \& {Jaffa}, S.~E. 2018, \aap, 611, A20,
  \dodoi{10.1051/0004-6361/201731871}

\bibitem[{{Whitworth} \& {Ward-Thompson}(2001)}]{Whitworth2001ApJ}
{Whitworth}, A.~P., \& {Ward-Thompson}, D. 2001, \apj, 547, 317,
  \dodoi{10.1086/318373}

\bibitem[{{Wiener} {et~al.}(2013{\natexlab{a}}){Wiener}, {Oh}, \&
  {Guo}}]{Wiener2013}
{Wiener}, J., {Oh}, S.~P., \& {Guo}, F. 2013{\natexlab{a}}, \mnras, 434, 2209,
  \dodoi{10.1093/mnras/stt1163}

\bibitem[{{Wiener} {et~al.}(2013{\natexlab{b}}){Wiener}, {Zweibel}, \&
  {Oh}}]{Wiener2013_b}
{Wiener}, J., {Zweibel}, E.~G., \& {Oh}, S.~P. 2013{\natexlab{b}}, \apj, 767,
  87, \dodoi{10.1088/0004-637X/767/1/87}

\bibitem[{Wiener(1930)}]{Wiener1930_book}
Wiener, N. 1930, Acta Math., 55, 117, \dodoi{10.1007/BF02546511}

\bibitem[{{Xu} \& {Zhang}(2016)}]{Xu2016ApJ}
{Xu}, S., \& {Zhang}, B. 2016, \apj, 824, 113,
  \dodoi{10.3847/0004-637X/824/2/113}

\bibitem[{{Yamamoto}(2017)}]{Yamamoto2017}
{Yamamoto}, S. 2017, {Introduction to Astrochemistry: Chemical Evolution from
  Interstellar Clouds to Star and Planet Formation},
  \dodoi{10.1007/978-4-431-54171-4}

\bibitem[{{Yoast-Hull} {et~al.}(2015){Yoast-Hull}, {Gallagher}, \&
  {Zweibel}}]{Yoast-Hull2015MNRAS}
{Yoast-Hull}, T.~M., {Gallagher}, J.~S., \& {Zweibel}, E.~G. 2015, \mnras, 453,
  222, \dodoi{10.1093/mnras/stv1525}

\bibitem[{{Yoast-Hull} {et~al.}(2016){Yoast-Hull}, {Gallagher}, \&
  {Zweibel}}]{Yoast-Hull2016MNRAS}
---. 2016, \mnras, 457, L29, \dodoi{10.1093/mnrasl/slv195}

\bibitem[{{Yusef-Zadeh} {et~al.}(2002){Yusef-Zadeh}, {Law}, \&
  {Wardle}}]{YusefZadeh2002ApJ}
{Yusef-Zadeh}, F., {Law}, C., \& {Wardle}, M. 2002, \apjl, 568, L121,
  \dodoi{10.1086/340379}

\bibitem[{{Yusef-Zadeh} {et~al.}(2013){Yusef-Zadeh}, {Hewitt}, {Wardle},
  {Tatischeff}, {Roberts}, {Cotton}, {Uchiyama}, {Nobukawa}, {Tsuru}, {Heinke},
  \& {Royster}}]{YusefZadeh2013ApJ}
{Yusef-Zadeh}, F., {Hewitt}, J.~W., {Wardle}, M., {et~al.} 2013, \apj, 762, 33,
  \dodoi{10.1088/0004-637X/762/1/33}

\bibitem[{{Zhang} \& {Li}(2017)}]{Zhang2017}
{Zhang}, C.-P., \& {Li}, G.-X. 2017, \mnras, 469, 2286,
  \dodoi{10.1093/mnras/stx954}

\bibitem[{{Zhang} {et~al.}(2014){Zhang}, {Qiu}, {Girart}, {Liu}, {Tang},
  {Koch}, {Li}, {Keto}, {Ho}, {Rao}, {Lai}, {Ching}, {Frau}, {Chen}, {Li},
  {Padovani}, {Bontemps}, {Csengeri}, \& {Ju{\'a}rez}}]{Zhang2014}
{Zhang}, Q., {Qiu}, K., {Girart}, J.~M., {et~al.} 2014, \apj, 792, 116,
  \dodoi{10.1088/0004-637X/792/2/116}

\bibitem[{{Zhang} {et~al.}(2019){Zhang}, {Guo}, {Wang}, \& {Li}}]{Zhang2019}
{Zhang}, Y., {Guo}, Z., {Wang}, H.~H., \& {Li}, H.~b. 2019, \apj, 871, 98,
  \dodoi{10.3847/1538-4357/aaf57c}

\bibitem[{{Zweibel} \& {Shull}(1982)}]{Zweibel1982}
{Zweibel}, E.~G., \& {Shull}, J.~M. 1982, \apj, 259, 859,
  \dodoi{10.1086/160220}

\end{thebibliography}
\bibliographystyle{aasjournal}



\end{document}